\documentclass[12pt]{article}

\usepackage{amsmath,amssymb,amsfonts,amsthm}
\usepackage{graphicx}
\usepackage[all]{xy}
\usepackage[paper=letterpaper,margin=0.9in]{geometry}
\input epsf.sty

\begin{document}

\def\CA{{\cal A}}
\def\CB{{\cal B}}
\def\CC{{\cal C}}
\def\CD{{\cal D}}
\def\CE{{\cal E}}
\def\CF{{\cal F}}
\def\CG{{\cal G}}
\def\CH{{\cal H}}
\def\CI{{\cal I}}
\def\CJ{{\cal J}}
\def\CK{{\cal K}}
\def\CL{{\cal L}}
\def\CM{{\cal M}}
\def\CN{{\cal N}}
\def\CO{{\cal O}}
\def\CQ{{\cal Q}}
\def\CR{{\cal R}}
\def\CS{{\cal S}}
\def\CT{{\cal T}}
\def\CU{{\cal U}}
\def\CV{{\cal V}}
\def\CW{{\cal W}}
\def\CX{{\cal X}}
\def\CY{{\cal Y}}
\def\CZ{{\cal Z}}
\def\T{{\cal T}}

\newcommand{\comment}[1]{}
\newcommand{\todo}[1]{{\em \small {#1}}\marginpar{$\Longleftarrow$}}
\newcommand{\labell}[1]{\label{#1}\qquad_{#1}} %
\newcommand{\bbibitem}[1]{\bibitem{#1}\marginpar{#1}}
\newcommand{\llabel}[1]{\label{#1}\marginpar{#1}}
\newtheorem*{BFSS}{BFSS Conjecture}

\newcommand{\sphere}[0]{{\rm S}^3}
\newcommand{\su}[0]{{\rm SU(2)}}
\newcommand{\so}[0]{{\rm SO(4)}}
\newcommand{\bK}[0]{{\bf K}}
\newcommand{\bL}[0]{{\bf L}}
\newcommand{\bR}[0]{{\bf R}}
\newcommand{\tK}[0]{\tilde{K}}
\newcommand{\tL}[0]{\bar{L}}
\newcommand{\tR}[0]{\tilde{R}}

\newcommand{\btzm}[0]{BTZ$_{\rm M}$}
\newcommand{\ads}[1]{{\rm AdS}_{#1}}
\newcommand{\ds}[1]{{\rm dS}_{#1}}
\newcommand{\dS}[1]{{\rm dS}_{#1}}
\newcommand{\eds}[1]{{\rm EdS}_{#1}}
\newcommand{\sph}[1]{{\rm S}^{#1}}
\newcommand{\gn}[0]{G_N}
\newcommand{\SL}[0]{{\rm SL}(2,R)}
\newcommand{\cosm}[0]{R}
\newcommand{\hdim}[0]{\bar{h}}
\newcommand{\bw}[0]{\bar{w}}
\newcommand{\bz}[0]{\bar{z}}
\newcommand{\be}{\begin{equation}}
\newcommand{\ee}{\end{equation}}
\newcommand{\bea}{\begin{eqnarray}}
\newcommand{\eea}{\end{eqnarray}}
\newcommand{\pat}{\partial}
\newcommand{\lp}{\lambda_+}
\newcommand{\bx}{ {\bf x}}
\newcommand{\bk}{{\bf k}}
\newcommand{\bb}{{\bf b}}
\newcommand{\BB}{{\bf B}}
\newcommand{\tp}{\tilde{\phi}}
\hyphenation{Min-kow-ski}

\def\apr{\alpha'}
\def\str{{str}}
\def\lstr{\ell_\str}
\def\gstr{g_\str}
\def\Mstr{M_\str}
\def\lpl{\ell_{pl}}
\def\Mpl{M_{pl}}
\def\varep{\varepsilon}
\def\del{\nabla}
\def\grad{\nabla}
\def\perp{\bot}
\def\half{\frac{1}{2}}
\def\p{\partial}
\def\perp{\bot}
\def\eps{\epsilon}

\newcommand{\cf}{{\em cf.}\ }
\newcommand{\ie}{{\em i.e.}\ }
\newcommand{\eg}{{\em e.g.},}
\newcommand{\viz}{{\em viz.}\ }
\newcommand{\nb}{{\em N.B.}\ }
\newcommand{\etal}{{\em et al.}\ }

\newcommand{\ba}{\begin{array}}
\newcommand{\ea}{\end{array}}
\newcommand{\ben}{\begin{enumerate}}
\newcommand{\een}{\end{enumerate}}
\newcommand{\bi}{\begin{itemize}}
\newcommand{\ei}{\end{itemize}}
\newcommand{\bc}{\begin{center}}
\newcommand{\ec}{\end{center}}
\newcommand{\bt}{\begin{table}}
\newcommand{\et}{\end{table}}
\newcommand{\btab}{\begin{tabular}}
\newcommand{\etab}{\end{tabular}}

\newcommand{\nn}{\nonumber}
\newcommand{\eref}[1]{(\ref{#1})}

\newcommand{\BC}{{\mathbb C}}
\newcommand{\CP}[1]{{\mathbb P}^{#1}}
\newcommand{\BR}{{\mathbb R}}
\newcommand{\BZ}{{\mathbb Z}}

\newcommand{\AdS}[1]{{\rm AdS}_{#1}}

\newcommand{\dee}{{\rm d}}
\newcommand{\Diff}{{\rm Diff}}
\newcommand{\pa}{\partial}
\newcommand{\tr}[1]{{\rm tr}\, {#1}}
\newcommand{\bra}[1]{\langle{#1}|}
\newcommand{\ket}[1]{|{#1}\rangle}
\newcommand{\ip}[2]{\langle{#1}|{#2}\rangle}
\newcommand{\vev}[1]{\langle{#1}\rangle}

\newcommand{\ack}[1]{[{\bf Ack!: {#1}}]}

\newcommand{\Xtr}{X_{\rm tr}}
\newcommand{\Xlong}{X_{\rm long}}
\newcommand{\Seff}{S_{\rm eff}}
\newcommand{\msusy}{m_{\rm susy}}

\newcommand{\Real}{\mathfrak{Re}}
\newcommand{\Imag}{\mathfrak{Im}}

\def\NPB{{\it Nucl. Phys. }{\bf B}}
\def\PL{{\it Phys. Lett. }}
\def\PRL{{\it Phys. Rev. Lett. }}
\def\PRD{{\it Phys. Rev. }{\bf D}}
\def\CQG{{\it Class. Quantum Grav. }}
\def\JMP{{\it J. Math. Phys. }}
\def\SJNP{{\it Sov. J. Nucl. Phys. }}
\def\SPJ{{\it Sov. Phys. J. }}
\def\JETPL{{\it JETP Lett. }}
\def\TMP{{\it Theor. Math. Phys. }}
\def\IJMPA{{\it Int. J. Mod. Phys. }{\bf A}}
\def\MPL{{\it Mod. Phys. Lett. }}
\def\CMP{{\it Commun. Math. Phys. }}
\def\AP{{\it Ann. Phys. }}
\def\PR{{\it Phys. Rep. }}

\renewcommand{\thepage}{\arabic{page}}
\setcounter{page}{1}

\rightline{\tt hep-th/0706.2252}
\rightline{DCPT-07/31}
\rightline{VPI-IPNAS-07-05}

\vskip 2.00 cm
\renewcommand{\thefootnote}{\fnsymbol{footnote}}

\centerline{\Large \bf Time and M-theory}
\vskip 0.75 cm

\centerline{{\bf
Vishnu Jejjala,${}^{1}$\footnote{\tt vishnu.jejjala@durham.ac.uk}
Michael Kavic,${}^{2}$\footnote{\tt kavic@vt.edu}
Djordje Minic${}^{2}$\footnote{\tt dminic@vt.edu}
}}
\vskip .5cm
\centerline{${}^1$\it Department of Mathematical Sciences,}
\centerline{\it Durham University,}
\centerline{\it South Road, Durham DH1 3LE, U.K.}
\vskip .5cm
\centerline{${}^2$\it Institute for Particle, Nuclear and Astronomical Sciences,}
\centerline{\it Department of Physics, Virginia Tech}
\centerline{\it Blacksburg, VA 24061, U.S.A.}
\vskip .5cm

\setcounter{footnote}{0}
\renewcommand{\thefootnote}{\arabic{footnote}}

\begin{abstract}
We review our recent proposal for a background independent formulation of a holographic theory of quantum gravity.
The present review incorporates the necessary background material on geometry of canonical quantum theory, holography and spacetime thermodynamics, Matrix theory, as well as our specific proposal for a dynamical theory of geometric quantum mechanics, as applied to Matrix theory.
At the heart of this review is a new analysis of the conceptual problem of time and the closely related and phenomenologically relevant problem of vacuum energy in quantum gravity.
We also present a discussion of some observational implications of this new viewpoint on the problem of vacuum energy.
\end{abstract}

\newpage

\section{Introduction and Overview}

While thinking about conceptual problems in fundamental physics, it is illuminating to begin with a comparison of two {\em fin de si\`ecle} periods:
the end of the nineteenth/beginning of the twentieth century and the end of the twentieth/beginning of the twenty-first century.
The central puzzles of the two epochs possess key parallels.
We have
(a) the black-body radiation problem exemplified in the classical sum over the mean energy $\frac12 k_B T$ per degree of freedom, which led to the ultraviolet catastrophe; and
(a${}'$) the vacuum energy problem exemplified in the sum over the quantum zero point energy $\frac12 \hbar \omega$ per oscillator degree of freedom, which leads to the still unresolved cosmological constant catastrophe and the related question of the origin of ``dark energy.''
We have
(b) the fundamental (non)existence of the {\ae}ther; and
(b${}'$) the fundamental (non)existence (or emergence) of spacetime.
Related to these are the origin of inertial frames and masses and the origin of spacetime and inertial mass.
We have also
(c) the problem of missing mass: the ``missing mass'' that explains the precession of Mercury's perihelion; and
(c${}'$) the modern cosmological problem of missing mass: the missing mass in galaxies and clusters of galaxies, the so-called ``dark matter.''
Finally, there is
(d) the need for a fundamental explanation of the periodic table of elements; and
(d${}'$) the need for a fundamental explanation/derivation of around 30-35 dimensionless numbers that go into the formulation of the Standard Model of particle physics (25 numbers) and the Standard Model of cosmology (5-10 numbers).

The questions (a), (b), (c), and (d) had ``unreasonable'' answers from the point of view of late nineteenth century physics.
The ``unreasonable'' answers were provided by revolutionary new physics: the Special and General Theories of Relativity and the quantum theory.
In light of this historical metaphor (which is only meant as a motivational tool!), perhaps it is only natural to expect that the true quantum theory of gravity and matter
(for which string theory is a promising theoretical candidate)
will provide equally ``unreasonable'' and currently unforeseen answers to the questions (a${}'$), (b${}'$), (c${}'$) and (d${}'$) as viewed through the lens of late twentieth century physics.\footnote{
Many other puzzles can be easily enumerated (see, for example, the top ten list from the Strings 2000 conference \cite{strings2000}).
We view these as important problems:
(1) What is string theory/quantum gravity?
Is it a quantum mechanical theory, and if so, what are its degrees of freedom and observable quantities, and if not, how does it go beyond the quantum?
(2) Are dimensionful parameters/coupling constants computable in principle or are they historical/quantum/statistical accidents?
(3) Can string theory/quantum gravity explain the origin of the Universe?
(4) Can it explain/rationalize the Standard Model of particle physics?
(5) Does it predict supersymmetry, if there is supersymmetry, and the specific breaking of supersymmetry?
Does it predict proton decay?
Does it explain the hierarchy of scales?
(6) Does it explain why our Universe looks $(3+1)$-dimensional?
(7) Does it explain/rationalize quantum mechanics?
(This is Wheeler's ``why the quantum?''\ question.)}

At the moment there remains a deep conceptual problem in string theory after two sweeping revolutions (1984/5, 1995):
the problem of a non-perturbative, background independent formulation,\footnote{
By {\em background independence} we mean that no {\em a priori} choice of a consistent background for string propagation is made.
The usage is as in string field theory.}
which would answer the question ``What is string theory?''\ and the related problem of dynamical vacuum selection based on a fundamentally new formulation, which is necessary for understanding how the real world fits within the framework of string theory.
In particular, the current technology of string theory, even in its most spectacular non-perturbative advancements such as the AdS/CFT correspondence, still relies on many low-energy artifacts such as effective field theory techniques, and, what is even more constraining, the requirement of specified asymptotic spacetime data.
On the other hand, there is a widespread conviction that the very notion of spacetime will need to be dramatically revised in some more fundamental formulation.

In some respects %
the present formulation of string theory may be compared to Yang--Mills theory in the 1950s.
A beautiful mathematical structure made an ``obvious'' wrong prediction:
a massless color particle, together with an associated (non-existent) long distance force.
Is the analogous ``obvious'' wrong prediction of string theory the ``eternal'' asymptotically ten-dimensional Minkowski vacuum that no known perturbative or non-perturbative device is able to eliminate in favor of the observed four-dimensional apparently de Sitter (dS) asymptotic background?

This question, we feel, may be profitably phrased in a different manner.
There are certainly many consistent backgrounds for string propagation ---
consider, for example, a favorite Calabi--Yau compactification of the heterotic string ---
and many perfectly respectable sigma-models (CFTs) on the string worldsheet that have no interpretation in terms of a background spacetime.
Thus, we would argue that there is nothing inherently special about ten-dimensional Minkowski space that makes it the central wrong prediction of string theory.
Instead we believe that the analogous wrong prediction of quantum gravity incorporating Standard Model-like matter
is a cosmological constant that scales neither as a power of the ultraviolet cutoff (as would be expected in a conventional field theory) nor is exactly vanishing (as would be expected if it were protected by a deep symmetry).
Rather the measured vacuum energy is a small and positive dimensionful number.
This is unnatural (\`a la 't Hooft \cite{nat}).
Moreover, the cosmology in the Universe we see is dynamic rather than static, and currently we have little control over such situations in string theory.
That is why in order to understand de Sitter-like spaces, given the present conceptual foundations of string theory, one must appeal to a KKLT-type of mechanism (see, for example, \cite{DouglasKachruReview}) which treats de Sitter-like backgrounds as metastable vacua and employs statistical analysis of string compactifications to explain small numbers \cite{svp}.
These ideas revise the traditional conception of naturalness.
While this remains an obviously fruitful and important avenue of research,
the present understanding of string theory does not allow us to answer difficult foundational questions involving the origins of space, time, and matter.
For this, we seemingly need to confront the hard question ``What is string theory?''\ head on.
It is one of our aims in this paper to review a novel approach to this question.

The historical development of Yang--Mills theory, from mathematically beautiful structures
into physical theories ({\em i.e.}, phenomenologically relevant models for physics),
came about with the development of the new concepts of spontaneous symmetry breaking and confinement.
The ultimate outcome was the phenomenally successful structure of the Standard Model of elementary particle physics.
A crucial question, arising from this particular historical comparison, is:
What is the non-perturbative formulation of string theory and what conceptually new insight does such a formulation offer about the structure of the ``vacuum,'' provided this latter concept makes sense?

Considering the more foundational questions of string theory concerning the nature and origins of space, time, and matter,
a still more ``revolutionary'' point of view may be in order. This would treat the
current deepest understanding of string theory
(as exemplified for the case of asymptotically anti-de Sitter backgrounds by the remarkable AdS/CFT correspondence \cite{ads})
as a WKB-like version ({\em cf.}, old quantum theory) of some conceptually deeper theory ({\em cf.},
quantum mechanics) in which the very notions of space and, in particular, time would be drastically modified.
That such a radically new physics might be needed for the non-perturbative background independent formulation of string theory has been advocated in our recent papers \cite{tzeminic}.
Of course, if this proposal has a bit of truth in it, especially with regards to its essential claim {\it that non-perturbative string theory is truly a fundamentally new domain of physics, namely a generalized quantum theory}, we are at the cusp of something spectacular.

It is our goal in this review to expand upon the scenario presented in \cite{tzeminic} and developed in our more recent papers \cite{jmt, jkm}.
In particular we wish to bring together the material presented over the last years in one place and in a pedagogical form, so that the main logic of our argument can be followed in detail.
Because our proposal combines information from different fields of theoretical physics (foundations of quantum theory, General Relativity, and string theory) and because we wish to make this review understandable to physicists working in these different fields, we have structured this article as follows.
In Sec. 2 we present a self-contained review of a geometric approach to canonical quantum theory.
An interesting, and experimentally testable generalization of canonical quantum theory, treated geometrically, is summarized in App. A.
In App. B we also discuss some foundational aspects of quantum theory.
In Sec. 3 we review an approach to a holographic description of classical Einstein gravity based on spacetime thermodynamics.
In Sec. 4 we review what is known about Matrix theory: a holographic quantum theory of Minkowski space.
Finally, in Sec. 5 we put together the material from the preceding sections.
These ideas are unified by a new physical principle:
a quantum version of an equivalence principle, which then leads us to an abstract formulation of a background independent Matrix theory.
The physical application of this abstract formulation is discussed in Sec. 6, where we consider the cosmological constant problem from a fundamentally new viewpoint.
This section is accompanied by App. C at the end of the paper.
We summarize various avenues for future research in the concluding Sec. 7 of this review.

Before we embark on the technical matters of the review, we would like to address what we think is the fundamental conceptual question on the nature of quantum gravity.

\subsection{Is Quantum Gravity a Canonical Quantum Theory?}

One interesting theoretical aspect of fundamental physical theories is presented by the way their mathematical structures get ``deformed'' in terms of new fundamental physical constants.
One way to summarize this is to consider the famous Planck cube (See Figure~\ref{fig:cube} in Sec.~\ref{sec:acv}.)
By glancing at the Planck cube (defined by the fundamental constants $c$, $\hbar$, and $G_N$), one sees at every corner something radically different (either classical, or relativistic, or quantum, or gravitational physics, and their respective combinations).
This might na\"{\i}vely suggest that the quantum theory of gravity is a profoundly new theory.
Nevertheless, the usual claim is that quantum gravity is a canonical quantum theory.
But within its architecture, there are any number of outstanding issues:
\begin{itemize}
\item The problem of time, or the problem of isolating a quantum evolution parameter in a background independent quantum theory of gravity.
\item The problem of identifying observables (or ``beables'') in a quantum theory of gravity consistent with spacetime diffeomorphism invariance.
\item The problem of defining the ``vacuum'' (and vacuum energy) in a background independent theory in the quantum mechanical, and not the classical sense.
\item The problem of recovering background dependent (that is, everyday) physics from a background independent formulation, by which we refer to the questions of
the emergence of spacetime background,
the emergence of causal structure,
the emergence of Standard Model-like matter,
the emergence of a realistic cosmology with dark energy and dark matter, and
the emergence of a successful low-energy description of physics via an effective quantum field theory of matter and gravity.
\item The problem of making sense of quantum cosmology:
the specification of initial conditions,
the question of the universality of inflation,
the resolution of the low entropy puzzle associated with the initial state to account for the observed degrees of freedom in the Universe and the associated ``cosmological origin'' of the arrow of time.
\item The resolution of ``classic'' quantum gravity puzzles:
horizons and quantum physics,
a microscopic explanation of gravitational entropy,
the black hole information paradox, and
the resolution of cosmological and black hole singularities.
\item The isolation of the true degrees of freedom of a quantum theory of gravitation consistent with the principle of holography.
\end{itemize}

We think that this list collapses in its essence to the following two fundamental conceptual issues:

\noindent (a) the problem of time in a quantum theory of gravity.
Associated to this is the problem of local vs.\ global observables and the question of how to describe a local observer without appealing to asymptotic spacetime data.
The non-decoupling between ultraviolet and infrared physics is central here.

\noindent (b) the related problem of vacuum energy (and the physical meaning of a vacuum in quantum gravity).
This is the familiar cosmological constant problem, and here, once again, we confront the issue of non-decoupling between short-distance and long-distance physics.
The persistence of large-scale structure in the Universe as well demands an explanation.
Why is the Universe not Planckian?
Why is it inflating?
And why is it stable (long-lived)?

In both of these puzzles the crucial missing piece is how to incorporate {\em time} (and the associated causal structure) in a purely quantum way and how to understand the dynamical evolution of the Universe within quantum theory.
What is the problem with time?\footnote{See \cite{Isham, Kuchar} for detailed reviews.}
Time is not an observable in quantum theory in the sense that there is no associated ``clock'' operator.
Time evolution, on the other hand, is driven by the dynamics of the Hamiltonian operator $\hat{H}$.
In the Schr\"odinger equation
\be
i\hbar \frac{d}{dt} | \psi(t) \rangle = \hat{H} | \psi(t) \rangle,
\ee
time enters as a parameter.
The energy-time uncertainty relation in quantum mechanics is intrinsically different in character than the position-momentum uncertainty relation.
This conception of the r\^ole of time as a Newtonian construct in a post-Newtonian theory persists even when we promote quantum mechanics to relativistic quantum field theory and constrain field operators to obey the causality condition,
\be
[\hat\phi(X), \hat\phi(Y)] = 0,
\ee
whenever the points $X$ and $Y$ are spacelike separated.

A theory of gravity must be diffeomorphism invariant.
In such a theory, time and spatial position are simply labels assigned to a point on the spacetime manifold that have no privileged meaning of their own.
Observables in General Relativity must also be diffeomorphism invariant, which typically means they are non-local (integrals of curvature invariants over spacetime, for example).
In a quantum theory of gravity, however, the situation is considerably more subtle.
As the spacetime metric becomes subject to quantum fluctuations, notions such as whether $X$ and $Y$ are spacelike separated become blurred.
Indeed, Lorentzian metrics exist for almost all pairs of points on the spacetime manifold such that the metric distance $g(X,Y)$ is not spacelike \cite{FredenhagenHaag}.
Clearly the notion of time, even locally, becomes problematic in such situations.

In thinking about these deep and complex issues, the core idea that we advocate is to take the structural lessons of General Relativity and emulate them in a quantum theoretic way in order to construct a theory that transcends the above problems.
In some sense, this is a repetition of what happened with the advent of quantum theory in the 1920s:
the structure of the classical theory (say its Hamiltonian or Hamilton--Jacobi formulation) was kept intact, but the kinematics were drastically altered.
The new theory thus developed was deeper in ways that classical physicists could not imagine.
The generalization of the structure of the classical theory makes sense of the success of the WKB limit in appropriate situations via the correspondence principle.
We likewise advocate making the quantum theory generally relativistic and constructing general backgrounds of string theory
from a fully holographic formulation of local Minkowski patches.
Holography is a crucial feature of non-perturbative string theory, as well as quantum field theory in fixed curved spacetime backgrounds, that we retain.
Minkowski space is fundamental in string theory, mainly due to supersymmetry.
In order to accomplish this synthesis, we must extend the usual framework of quantum theory.
This is the working point of view that we present throughout this review.

The natural uncertainty we must address at this point is whether there is any compelling theoretical (or even better, experimental) justification for such an outrageous proposal.
We consider the most compelling evidence to be the puzzle of ``dark energy,'' which, it seems, is nicely modeled by a cosmological constant in an effective Lagrangian approach.
(A purely theoretical test of our approach to building general string backgrounds would be in trying to recover what is known about asymptotically AdS backgrounds, or in other words to rederive the AdS/CFT correspondence as a WKB-like limit of our general philosophy.
We discuss this important ongoing project in the conclusion to this review.
In some sense, we would advocate that AdS space is like a ``hydrogen atom'' of the Born--Sommerfeld old quantum theory.
This means that for reasons of the symmetry of this particular background, a deeper formulation has to reproduce the successes of the existing AdS/CFT approach.)

Let us turn now to the central question.
{\it Is quantum gravity a canonical quantum theory?}
By canonical quantum theory, we refer to one that is formulated using ``canonical'' tools: path integrals, Hilbert spaces, etc., and whatever appropriate interpretation is necessary to address the questions posed by quantum cosmology using the ordinary quantum theory.
If it is not, why is it not, and what kind of new theory then is quantum gravity?
If quantum gravity is not an ordinary quantum theory, this must be immensely important for the foundations of physics, and it must have shattering observational consequences.\footnote{
There are some historical precedents regarding the generalization of quantum theory within string theory:
\begin{enumerate}
\item some general ideas related to the issue of non-linear wave equation in string field theory \cite{gsw} (See Vol.\ 1, Sec. 3.2);
\item the appearance of non-associative structures, which are seemingly incompatible with the canonical formulation of quantum theory \cite{stromer};
\item weaving the string background in the approaches based on abstract conformal field theory \cite{friedan1} and the general
sigma-model/non-perturbative renormalization group approaches to non-perturbative string theory \cite{friedan2},
the non-perturbative renormalization group looking like a non-linear wave equation \cite{ployakov};
\item third quantization and quantum cosmology \cite{thirdq}.
This was motivated by both string field theory and Euclidean quantum cosmology with topology change;
more recently there has been discussion of the ``multiverse'' \cite{susskind} in the landscape approach to the problem of string vacua.
\end{enumerate}}

As we have stated, the crucial reason why we think quantum gravity is not an ordinary quantum theory is precisely the way {\it time}
is treated by ordinary quantum theory, and the way {\it time} is supposed to ``emerge'' or be ``quantized'' in the beginning (at the Big Bang).
Space is treated differently from time in a very radical way within quantum theory, and this manifests in every approach to quantum
gravity based on the usual quantum mechanics, including string theory as it is currently understood.
This dichotomy is at the root of the problems of quantum gravity, and its explanation ultimately communicates what string theory is.
We will take the vacuum energy problem as a springboard for our specific proposal regarding the formulation of the quantum theory of
gravity and matter as a general geometric quantum theory.

\subsection{Vacuum Energy in Quantum Gravity}
As is well known, recent cosmological observations suggest that we live in an accelerating Universe \cite{sn, wmap3}.
One possible engine for late time acceleration is an unseen ``dark energy'' that comprises 74\% of the total energy density in the Universe.
The leading candidate for dark energy is the energy in the vacuum itself, and the data suggest a small, positive cosmological constant.
This leads to a two-fold cosmological constant problem \cite{weinb}:
\begin{enumerate}
\item Why is the energy density in the vacuum so small compared to the expectation of effective field theory?
\item Why are the energy densities of vacuum and matter comparable in the present epoch?
\end{enumerate}

The first cosmological constant problem concerns both ultraviolet and infrared physics.
In quantum field theory, the cosmological constant counts the degrees of freedom in the vacuum.
Heuristically, we sum the zero-point energies of harmonic oscillators and write
$
E_{\rm vac} = \sum_{\vec{k}} \left( \frac{1}{2}\hbar \omega_{\vec{k}} \right).
$
The sum is manifestly divergent.
Because quantum field theories are effective descriptions of Nature, we expect their validity to break down beyond a certain regime and be subsumed by more fundamental physics.
We may introduce a high-energy cutoff to regulate the sum, but $E_{\rm vac}$ will then scale with the cutoff.
The natural cutoff to impose on a quantum theory of gravity is the Planck energy $M_{\rm Pl}$.
This prescription yields an ultraviolet enumeration of the zero-point energy.

In the infrared, the cosmological constant feeds into Einstein's equations for gravity:
\be
R_{\mu \nu} - \frac{1}{2} g_{\mu \nu} R = 8\pi G_N \left( - \Lambda g_{\mu \nu} + T_{\mu \nu} \right).
\label{eq:einstein}
\ee
Present theories of quantum gravity are unable to deal with the cosmological constant problem.
Here we are concentrating on string theory, as the only known example of a consistent theory of
perturbative quantum gravity and Standard Model-like matter.
In perturbative string theory, the dynamics of the background spacetime are determined by the vanishing of the $\beta$-functional associated to the Weyl invariance of the worldsheet quantum theory;
the Einstein equation (\ref{eq:einstein}) is then corrected in an $\alpha'$ expansion.
In either case, the vacuum is just any solution to these equations.
We compute the vacuum energy in quantum field theory, include it on the right hand side of the gravitational field equation, and find that spacetime is not Minkowski.
Although the Einstein equations are {\em local} differential equations, the cosmological constant sets a {\em global} scale and determines the overall dynamics of spacetime.
Our Universe is approximately four-dimensional de Sitter space with $\Lambda \approx 10^{-123} M_{\rm Pl}^4$.

Quantum theory (as presently understood) therefore grossly over-counts the number of vacuum degrees of freedom.
There is no obvious way to reconcile the generic prediction of effective field theory that the vacuum energy density
should be $M_{\rm Pl}^4$ with the empirical observation that it spectacularly is not.
Note that the question of defining a vacuum in gravity is always tied to asymptotic conditions.
We are solving \eref{eq:einstein}, which is a differential equation, and boundary conditions are an input.
This methodology is imported to quantum gravity.
For example, in target space, string theory is formulated as an $S$-matrix theory whose long wavelength behavior is consistent with an effective field theory for gravity.
But the effective field theory in the infrared, in all cases that have been even partially understood, is particularly simple:
infinity is asymptotically flat or it is asymptotically anti-de Sitter space or it is a plane wave limit.
The formulation of string theory as a consistent theory of quantum gravity on de Sitter space \cite{dsh}, which the present second inflationary phase of our
Universe resembles, or in more general curved (time-dependent) backgrounds with (at best) approximate isometries is not at all understood.

The vacuum energy problem can also be couched in the following way.
In quantum field theory in flat space the vacuum is clearly defined as in quantum mechanics.
Vacuum energy is just the expectation value of the Hamiltonian in its ground state.
But what is the energy or Hamiltonian in a quantum theory of gravity, and what is the vacuum or lowest energy state?
Such concepts must be defined without invoking asymptopia as inputs, for clearly as local observers in spacetime, we cannot know what we are evolving towards.
How then is vacuum energy to be determined when there are no {\em a priori} fixed asymptotics and there is no Hamiltonian?

We find a clue in the equivalence principle.
In the classical theory of gravity, the nearness of a body's inertial mass to its gravitational mass is explained because an observer cannot distinguish between gravitation and acceleration.
Spacetime is locally indistinguishable from flat space (zero cosmological constant).
Globally it can be any solution at all to Einstein's equations.
It is the equivalence principle that is responsible for the dual nature of energy and the concept of a vacuum (the actual geometry of spacetime).
Because this is the root of the problem, we wish to implement the equivalence principle at the quantum mechanical level and see what light this throws on the vacuum energy problem.

In essence, we are turning the cosmological constant problem around, to argue that its natural solution
({\em i.e.}, natural adjustment to an almost zero value)
requires a major shift in the foundations of fundamental physics.
The new fundamental postulate needed is a quantum equivalence principle which demands a consistent gauging of the geometric structure of canonical quantum theory.
This, we believe, is the missing key element in present formulations of consistent quantum theory of gravity and matter.

Therefore we think that the case for a new radical approach to quantum gravity to be explained in this paper is strong and thus proceed with the review of the necessary background needed for its formulation.

\section{Background 1: Geometric Quantum Mechanics}

In this section we will review how standard quantum mechanics may be recast in a geometric framework \cite{kibble}.
There are several excellent reviews on this topic \cite{weinberg, geomqm, Bloch, anandan, provost}, and in particular we shall draw from \cite{as}.
The geometric formulation can be used to derive all standard results without reference to the underlying linear structure and as we shall see lends itself much more readily to generalization.

This section is organized as follows.
We begin by discussing how a Hilbert space may be given a phase space interpretation.
We will then reduce this phase space in an appropriate fashion in order to obey constraints provided by quantum mechanics.
The classical and quantum mechanical aspects of this reduced phase space are then outlined.
We next analyze the kinematic structure of quantum mechanics in this formalism, and finally we discuss possible generalizations of quantum mechanics from a geometric perspective.
The central aim of this section is to emphasize the quantum theoretic concepts underlying the measurement of time intervals as opposed to spatial distances.
This is where the geometric structure of quantum theory ties with the geometry of spacetime physics.
We append to this section a discussion of an experimentally testable generalization of quantum theory in geometric framework, as formulated by Weinberg (App. A), together with a discussion of what the geometric framework means for the foundations of quantum theory (App. B).

\subsection{From Hilbert Space to K\"ahler Space}

We will begin by developing the idea of a Hilbert space as a K\"ahler space.
The symplectic structure and Riemannian metric associated with a K\"ahler space will allow us to more clearly elucidate the relationship between classical mechanics and quantum mechanics.
Consider the Hilbert space ${\cal{H}}$.
We may choose to view ${\cal{H}}$ as a real vector space with a complex structure $J$.
The Hermitian inner product of two states may then be decomposed into its real and imaginary parts,
\begin{equation}
\langle \Psi|\Phi \rangle = \frac{1}{2\hbar} G(\Psi,\Phi)+\frac{i}{2\hbar}\Omega(\Psi,\Phi).
\end{equation}
The real part $G(\Psi, \Phi)$ is the Riemannian metric.
The imaginary part $\Omega$ is a non-degenerate symplectic two-form.
The two are related by
\begin{equation}
G(\Psi,\Phi)=\Omega(\Psi,J\Phi).
\end{equation}
The triad ($G,\Omega,J)$ endows ${\cal{H}}$ the structure of a K\"ahler space.

Let us explore the consequences of this.
First, the existence of the symplectic structure means that ${\cal{H}}$ is a symplectic manifold, namely a phase space.
Having already identified $\Omega(\Psi,J\Phi)$ as a symplectic two-form we may use it and its inverse $\Omega^{ab}$ to define a Poisson bracket
\begin{equation}
\Omega(X_A,X_B)=\{A, B\}={\partial A \over \partial p_a}{\partial B \over \partial q^a} - {\partial A \over \partial q^a}{\partial B \over \partial p_a}
\equiv \Omega^{ab} {\partial A \over \partial X^a}{\partial B \over \partial X^b}.
\end{equation}
In this and subsequent expressions $A\equiv\vev{\hat{A}}$ and $B\equiv\vev{\hat{B}}$ and $X_{A}$ and $X_{B}$ are the Hamiltonian vector fields generated by the expectation values of the operators.
In addition
$X^a= (p_a, q^a)$ are a set of canonical coordinates with $q^a =
\sqrt{2\hbar} Re \Psi_a$ and
$p_a = \sqrt{2\hbar} Im \Psi_a$.
Defining the Poisson bracket in this way is similar to the classical case except that instead of the observables being real functions they can be thought of as vector fields.
The Schr\"{o}dinger equation may be expressed as
\be
\dot{\Psi}=-\frac{1}{\hbar}J\hat{H}\Psi,
\ee
and we may associate a Schr\"{o}dinger vector field with each observable \cite{as},
\begin{equation}
W_{\hat{A}}(\Psi)=-\frac{1}{\hbar}J\hat{A}\Psi.
\label{sch}
\end{equation}
The Schr\"{o}dinger vector field preserves both the metric and the two-form, and due to the linear nature of ${\cal{H}}$, it is locally and globally Hamiltonian.
The Schr\"{o}dinger vector field generated by a given operator is equivalent to the Hamiltonian vector field which is generated by taking the expectation value of that same operator.
This implies that the time evolution of a quantum mechanical system may be described by Hamilton's equations, or put another way Schr\"{o}dinger's equation is an alternative expression of Hamilton's equations \cite{as}.
The Lie bracket of two operators $\hat{A}$ and $\hat{B}$ likewise has a correspondence to the Poisson brackets of two expectation values,
\begin{equation}
\{A,B\}=\bigg<\frac{1}{i\hbar}[\hat{A},\hat{B}]\bigg>.  
\end{equation}
It is important to note that this is not the correspondence between classical mechanics and quantum mechanics in the $\hbar\rightarrow0$ limit.
This is an equivalent formulation of quantum mechanics in the language of classical physics.

While analysis of the symplectic structure of the K\"ahler space emphasizes the similarities between quantum and classical mechanics, analysis of the Riemannian metric yields the differences.
The Riemannian metric is not present in classical phase space and as we shall see encodes purely quantum mechanical properties such as uncertainty relations.
We must first define what we will call the Riemannian bracket in terms of the Riemannian metric,
\begin{equation}
\{A,B\}_+\equiv\frac{\hbar}{2}G(X_{A}, X_{B}).
\end{equation}
As with the symplectic structure there is a correspondence but now with the Jordan product as opposed to the Lie bracket,
\begin{equation}
\bigg<\frac{1}{2}[\hat{A},\hat{B}]_+\bigg>=\{A,B\}_{+}.
\end{equation}
Now note that the uncertainty of an observable can be expressed as,
\begin{equation}
(\Delta\hat{A})^2=\vev{\hat{A}^2}-\vev{\hat{A}}^2=\{A,A\}_+ -A^2.
\end{equation}
We may also write the uncertainty relation in a nice form involving both the Poisson and Riemannian brackets \cite{as}:
\begin{equation}
(\Delta\hat{A})^2(\Delta\hat{B})^2\geq\bigg(\frac{\hbar}{2}\{A,B\}\bigg)^2+ (\{A,B\}_+ -AB)^2.
\end{equation}

Consideration of a Hilbert space as a K\"ahler space is the first step to a geometric reformulation of quantum mechanics.
The K\"ahler space has an associated symplectic structure that leads to a phase space interpretation reminiscent of classical mechanics.
It also has a Riemannian metric which endows it with purely quantum mechanical properties.

\subsection{Phase Space Reduction and Symplectic Structure}

The prior discussion contains an important subtlety.
In the Hilbert space ${\cal{H}}$, a given state is defined by more than one state vector.
The true space of physical states in quantum mechanics
is the space of rays in the Hilbert space, or the projective Hilbert space ${\cal{P}}$.
This is the space of pure states.
Therefore, we must reduce the %
phase space in the appropriate way.

We begin by noting that ${\cal{P}}$ is also a K\"ahler space.
This is true for the infinite dimensional case but can be seen most clearly by considering finite dimension.
For finite dimension ${\cal{H}}=\BC^{n+1}$ and $\cal{P}$ is the complex projective space $\CP{n}$.
This is clearly K\"ahler and in addition is the Hopf line bundle of the sphere $S^{2n+1}$ over $\CP{n}$.
The complex projective space is thus
\begin{equation}
\CP{n}=\frac{U(n+1)}{U(n)\times U(1)}.
\end{equation}
In this expression $U(1)$, the fiber, is the group of complex phases in quantum mechanics.
We shall see that the phase space reduction is directly related to the invariance under the choice of phase.
The geometric nature $\CP{n}$ can be seen more clearly by considering a specific quantum system.
If we simply considered a spin-$\frac12$ particle this would correspond to taking $n=1$ with $\CP{1}=S^2$.
This is the Bloch sphere which is the pure state space of a 1 qubit quantum register \cite{ing}.

We wish to perform the phase space reduction for the infinite dimensional case.
To do this we must first deal with the ambiguous nature of the state vectors.
We begin by implementing the Born rule,
\begin{equation}
\langle \Psi|\Psi \rangle={1 \over 2\hbar} \sum_a [ (p^a)^2 + (q_a)^2 ] =1.
\label{born}
\end{equation}
This implies that the states are normalized to unity and that $\Psi$ and $e^{i\alpha}\Psi$ should be identified.
We may express this as the following constraint function,
\begin{equation}
c(\Psi)\equiv\langle\Psi|\Psi\rangle-1=\frac{1}{2\hbar}G(\Psi,\Psi)-1=0.
\end{equation}
We may equally take the more geometric point of view and regard $c(\Psi)$ as defining the constraint surface $S$, which in this case is the unit sphere with
regard to the Hermitian inner product.
Thus we are isolating as physically relevant only the portion of $\cal{H}$ constrained to $S$.
The possible r\^{o}le of the remaining portion of $\cal{H}$ shall be addressed in subsequent sections.

For each constraint function on a Hamiltonian system there must be a corresponding gauge invariance.
The associated gauge transformation translates to the flow along the Hamiltonian vector field.
We denote the generator of these transformations as
\begin{equation}
{\cal{J}} = -J^{a}_{b}\Psi^{b}\arrowvert_{S}.
\end{equation}
This is the generator of phase rotations on $S$.
Thus the gauge transformation in question corresponds to phase invariance.
This is exactly what we would expect based on our implementation of the Born rule. %
The result of constraining our system in this way is to isolate the physical portion of the phase space.
We have done this by taking the quotient of our constraining surface with the action of the gauge transformation.
We are left with a gauge reduced phased space which is the projective Hilbert Space, $\cal{P}$.
We label this as the {\em quantum phase space}.
As in the case of the full Hilbert space, we must also consider the geometric structure of the quantum phase space.
We have already established that it is K\"ahler and, as before, therefore endowed with a symplectic structure and a Riemannian metric.

Recalling that the symplectic structure of the full Hilbert space $\cal{H}$ is encoded in the two-form $\Omega$, we constrain this to the unit sphere as $\Omega\arrowvert_{S}$.
We may then define a new symplectic two-form $\omega$ whose pull-back is equivalent to $\Omega\arrowvert_{S}$,
\begin{equation}
\pi^{*}\omega=\Omega\arrowvert_{S}
\end{equation}
where $\pi$ is the projection mapping $\pi: S \rightarrow \cal{P}$.
To define the Poisson bracket using $\omega$, we must first define the observables in the quantum phase space.
We begin with an operator $\hat{A}$ on $\cal{H}$.
We will take the expectation value of this operator and as with the two-form constrain it to the unit sphere as $A\arrowvert_{S}$.
This is a gauge invariant function, and thus we may define a new observable $a$ whose pull-back is equivalent to $A\arrowvert_{S}$,
\begin{equation}
\pi^{*}a= A\arrowvert_{S}.
\end{equation}
Next we wish to obtain the relationship between Hamiltonian vector fields, $X_{A}$ on $\cal{H}$ and $X_{a}$ on $\cal{P}$.
Because $A$ is gauge invariant, $X_{A}$ is constant along integral curves of $\cal{J}$.
Thus we may push-forward the vector field at every point $\Psi\in S$ and equate it to $X_{a}$ at that point,
\begin{equation}
\pi_{*}X_{A}|_{\Psi}=X_{a}.
\end{equation}
Now we are ready to consider Poisson brackets defined by $\omega$.
We consider the expectations values $A$ and $B$ on $\cal{H}$ and the expectations values $a$ and $b$ on $\cal{P}$.
Based on the results stated, we can derive a relationship to define a Poisson bracket for the reduced phase space:
\begin{equation}
\pi^{*}\{a,b\}_{\omega}=\{A,B\}_{\Omega}\arrowvert_{S}.
\end{equation}
There is a one-to-one mapping of the operators on $\cal{H}$ to the observable functions on $\cal{P}$.
The flow on $\cal{P}$ generated by the Schr\"{o}dinger vector field of $\hat{A}$ on $\cal{H}$ is equal to the flow of the Hamiltonian
vector field determined by $a$ on $\cal{P}$ \cite{as}.
We may, as previously indicated, express the Schr\"{o}dinger equation in the form of Hamilton's equations,
\begin{equation}
{d p_a \over dt} = \{h, p_a \}_{\omega}, \quad {d q^a \over dt} = \{h, q^a\}_{\omega}.
\end{equation}
In this expression $h =\langle \hat{H}\rangle|_S={1 \over 2} \sum_a [ (p^a)^2 + (q_a)^2 ] \lambda_a$, where $\lambda_a$ are the energy eigenvalues.
An observable $o$ will then evolve as
\begin{equation}
{ d o \over dt}= \{h, o \}_{\omega}.
\end{equation}
Thus the symplectic structure is carried over to the quantum phase space.

It is worth noting that in adopting this framework we have already generalized quantum mechanics in a certain sense.
We considered our system in boarder terms before constraining it because of physical considerations.
Specifically, we have constrained the expectation values of the operators on $\cal{H}$ to $S$.
We could, however, extend those expectation values off $S$ without disturbing the flow on $\cal{P}$.
We will return to this point when attempting to construct a generalized dynamical structure. \label{gd}

\subsection{Riemannian Geometry and Quantum Mechanics}

As in the case of the full Hilbert space, $\cal{P}$ is endowed with an almost complex structure $j$ and a Riemannian metric $g$.
These structures are inherited from the corresponding structures on the full Hilbert space.
We obtain the Riemannian metric on $\cal{P}$ by restricting the Riemannian metric on $\cal{H}$ to the constraint surface.
The metric obtained in this manner, however, is degenerate.
In order to correct this we must subtract off components in the direction of $\cal{J}$ \cite{as},
\begin{equation}
g=[G-\frac{1}{2\hbar}(\Psi\otimes\Psi+\cal{J}\otimes\cal{J})]|_{S}.
\end{equation}
Thus the metric will only be degenerate in that direction.
This is the metric on the complex Hopf bundle.
By requiring that the projection map associated to this bundle be Riemannian, the form of the metric can be determined.
It is the Cayley--Fubini--Study metric, which we may express for nearby states as
\begin{equation}
ds_{12}^2 = 4(\cos^{-1}{|\langle \psi_1|\psi_2 \rangle|})^2=4(1 -|\langle \psi_1|\psi_2 \rangle|^2). \label{fub}
\end{equation}
Clearly, this vanishes for $\ket{\psi_1} = \ket{\psi_2}$.
Suppose that $\ket{\psi_2} = \ket{\psi_1} + \ket{d\psi}$ and that both $\ket{\psi_i}$ are canonically normalized so that $\langle \psi_i \ket{\psi_i} = 1$.
The infinitesimal distance between $\ket{\psi_1}$ and $\ket{\psi_2}$ in the quantum configuration space is
\be
ds_{12}^2 = 4(\langle d\psi|d\psi\rangle - \langle d\psi|\psi_1\rangle\langle \psi_1|d\psi\rangle).
\ee

Before we address the relationship of the Riemannian metric to quantum mechanics,
let us first consider the nature of observables on the quantum phase space.
We have previously defined observable functions in terms of self-adjoint operators on $\cal{H}$.
However, it is preferable to define observables only in terms of $\cal{P}$, without any reference to the underlying linear structure of the Hilbert space.
In order to do this consider the Hamiltonian vector field, $X_A$.
It preserves $\Omega$ and  $G$ and is also a Killing vector field on $(G,\cal{H})$.
Now consider the corresponding observable function $a$ and the associated vector field, $X_a$.
It is a Killing vector field as well, but on $(g,\cal{P})$.
It is this property which we shall use to define $a$ as an observable.
A smooth function on $\cal{P}$ is an {\em observable} if and only if its Hamiltonian vector field is also Killing \cite{as}.

Thus the space of observables is isomorphic to the space of functions whose Hamiltonian vector fields are infinitesimal symmetries of the available structure.
This is exactly the same as in classical mechanics.
However, now the structure is far richer.
It contains not only the usual symplectic structure but the Riemannian metric as well.
In contrast to the classical case there is a small subset of smooth functions on $\cal{P}$ which qualify as observables.
These are K\"ahlerian functions.

Using $g$ we may define the connection $\Gamma^{a}_{bc}$ on the bundle by requiring that the fiber be orthogonal to the metric.
By parallel transport around a closed curve, we may determine the geometric phase factor. This is the Berry phase factor. A 
remarkable strength of this geometric approach to quantum mechanics is the natural understanding of Berry's phase that it provides. 
Also, with the triad $(j,\omega,g)$, it is easy to confirm that $\cal{P}$ is a K\"ahler manifold.

We may define the bracket Riemannian for the quantum phase space in terms of the metric on $\cal{P}$ without reference to the full phase space
\begin{equation}
(a,b)\equiv\frac{\hbar}{2}g(X_a,X_b).
\end{equation}
This can, however, be related to the Riemannian bracket for the full Hilbert space,
\be
\{A,B\}_+=\pi^{*}\big((a,b)+ab\big).
\ee
Note that unlike the symplectic bracket on $\cal{P}$, we cannot simply equate the pull-back of the new Riemannian bracket to the Riemannian bracket for full Hilbert space constrained on $S$.
Thus, we define a new bracket for which this is the case,
\begin{equation}
\{a,b\}_+\equiv(a,b)+ab.
\end{equation}
We will call this the {\em symmetric bracket} \cite{as}.
We may express the standard uncertainty relation in terms of the Poisson bracket and the Riemannian bracket,
\begin{equation}
\Delta a\Delta b\geq\bigg(\frac{\hbar}{2}\{a,b\}_{\omega}\bigg)^2+(a,b)^2.
\end{equation}
However we may use the Riemannian bracket alone to define the squared uncertainty for a given state $\psi$ \cite{as},
\begin{equation}
(\Delta a)^2(\psi)=(a,a)(\psi).
\end{equation}
If we now consider the time evolution of a system, we see that it is related to the metrical structure.
The Schr\"{o}dinger's equation plays the r\^{o}le of a geodesic equation on $\CP{n}$:
\begin{equation}
{d u^a \over d s} + \Gamma^{a}_{bc} u^b u^c = \frac{1}{2\Delta E}Tr(H F^a_b) u^b
\end{equation}
for the Fubini--Study metric $g_{ab}^{FS}$ with the canonical curvature two-form $F_{ab}$ valued in the holonomy gauge group $U(n)\times U(1)$.
Also, $u^a = \frac{d {\psi}^a}{ds}=\frac{d z^a}{ds}$, where $z^a$ denote the complex coordinates on $\CP{n}$.
The affine parameter $s$ is determined by the metric on $\CP{n}$.
This leads us directly to an important result given by Anandan and Aharanov \cite{anandan}.
Consider the uncertainty of the observable function associated with the Hamiltonian,
\begin{equation}
(\Delta h)^2=\frac{\hbar}{2}g(X_h,X_h)=(\Delta E)^2.
\end{equation}
The uncertainty in the energy is the length of the Hamiltonian vector field which generates the time evolution.
This can be interpreted as the speed with which the system moves through phase space.
In the above geodesic Schr\"{o}dinger equation, the Hamiltonian appears as the ``charge'' of an effective particle moving with a ``velocity'' $u^a$ in the background
of the ``Yang--Mills'' field $F_{ab}$.
The system passes quickly through the parts of phase space where the uncertainty is large and more slowly through the parts where the uncertainty is small.

The main features of quantum mechanics are embodied in the geometry of $\CP{n}$ and in the evolution equation.
The superposition principle is tied to viewing $\CP{n}$ as a collection of complex lines passing through the origin.
Entanglement arises from the embeddings of the products of two complex projective spaces within a higher dimensional one.
The geometric phase stems from the symplectic structure on $\CP{n}$.

\label{here}
Next we address the issue of coherent states, which are natural to consider in this formalism as they admit a straightforward
phase space description. These type of quantum states are the closest to being classical in nature in that they minimize the standard uncertainty relation
between position and momentum.
In fact the space of coherent states may be thought of a classical phase
space embedded in a quantum phase space with each point in the classical phase space corresponding to a
to a coherent state. It is also worth noting that coherent states are complete in that any state can be represented
as a superposition of coherent states \cite{ing}. In configuration space a coherent state
has the form of a Gaussian displaced a distance from the origin, with origin being the vacuum state,
\begin{equation}
\psi_l(x) \sim \exp\bigg(- \frac{{({\vec{x}}-{\vec{l}})}^2}{\delta l^2}\bigg).
\end{equation}
We may calculate $ds^2$ for two nearby coherent states making use of (\ref{fub}) and the convolution property of Gaussian integrals which gives the
overlap of the two states.
Interestingly, this yields the natural metric in the configuration space, namely
\begin{equation}
ds^2 = \frac{d {\vec{l}}^2}{\delta l^2}. \label{m}
\end{equation}
So, wherever the configuration space coincides with {\it space}, the natural metric on $\CP{n}$ in the $\hbar \to 0$ limit gives a spatial metric \cite{anandan}.
We shall make use of this important insight when we attempt to generalize to a background independent formulation of quantum mechanics.
For a generalized coherent state, the Fubini--Study metric reduces to the metric on the corresponding group manifold \cite{provost}.

In order to understand the measurement process in this new framework we require a better understanding of the nature eigenstates in the quantum phase space, $\cal{P}$.
Consider an operator $\hat{A}$.
A state $\Psi$ is an eigenstate provided $\hat{A}\Psi=\lambda\Psi$, where $\lambda$ is real.
Recalling our expression for the Schr\"{o}dinger vector field \eref{sch}, we can deduce that the Hamiltonian vector field $X_A$ must be pure gauge.
Consider now the projection of $\Psi$ on $\cal{P}$ which we denote $\psi$.
In order for $X_a$ to be pure gauge the Hamiltonian vector field associated with the observable function $a$ must vanish at $\psi$.
That is to say, $\psi$ is an eigenstate of $a$ if it is a critical point $p\in\cal{P}$ and the critical value $\lambda$ is the corresponding eigenvalue.

Now we wish to consider the measurement process from a geometric point of view.
We begin by investigating the probability distribution.
Each point in $\cal{P}$ is a particular state.
There is a geodesic that passes between any two given points which relates the transition amplitude between those two states.
The transition amplitude function is given by,
\begin{equation}
|\langle\psi_i|\psi_f\rangle|^2=\cos^2\bigg(\frac{\theta(p_i,p_f)}{\sqrt{2\hbar}}\bigg)
\end{equation}
where $\theta$ is the minimal geodesic distance between points $p_i$ and $p_f$, as measured by the Fubini--Study metric.
Thus the transition probability between two states is determine by the distance between the two corresponding points in $\cal{P}$.
Next consider a generic measurement.
Suppose a system is initially in state $\psi_0$ at point $p_0$.
Then an ideal measurement of an observable function $a$ is performed.
The system will then collapse into one of the available eigenstates, $\psi_f$ at point $p_f$.
There is a geodesic that passes from $p_0$ to each $p_f$.
Because the transition amplitude is governed by the distance between the various states the system is more likely to collapse to a nearby eigenstate than a more distant one.\footnote{
In the case of a degenerate eigenvalues, the eigenspace has an associated eigenmanifold.
The system will return the degenerate eigenvalue and collapse to the point in the eigenmanifold closest to $p_0$ \cite{as}.}

Armed with our newfound geometric view of probability, we return to the case of time evolution.
We observe (as underscored by Aharonov and Anandan \cite{anandan}), that time measurement in the evolution of a given system should reduce to that of distance on $\CP{n}$.
In particular we may rephrase our previous result as
\begin{equation}
\hbar\, ds = 2 \Delta E\, dt.
\label{anaeq}
\end{equation}
Such an expression naturally invokes a {\it{relational}} interpretation of time in quantum mechanics.
Even more striking is the fact that the geometric interpretation of probability as the geodesic distance on $\CP{n}$ is {\it{directly}} related to the definition of the evolution parameter $t$!
Moreover, the expression \eref{anaeq} relating time intervals to intervals in the projective space of the quantum theory is {\it exact}.
Now recall (\ref{m}) which is a relation between the spatial distances and geometric intervals.
This is in fact the most crucial difference between temporal and spatial geometry from the point of view of quantum geometry.
In essence the above exact relation underlies the way we build physical clocks based on quantum theory.
{\it This is the crucial conceptual statement of quantum theory, as illuminated by its
geometric formulation, that has to be re-examined when thinking about the problem of time
in quantum gravity.} For example, the gravitational redshift of weakly quantized
quantum gravity, immediately follows by changing the Hamiltonian in the above formula
by adding the gravitational potential.
Note also that in a general relativistic context, spacetime measurements can be viewed as measurements of time \cite{mtw}.
The tension between canonical quantum theory and background independent classical spacetime physics is precisely in the way the two treat measurements of time (and the corresponding canonically conjugate variable, energy).

It is now clear that the Riemannian metric enriches our phase space by providing it with quantum mechanical structure.
This includes reformulations of the standard uncertainty relations and the quantum measurement process.
We have recast the uncertainty relation in terms of the metrical structure and presented a novel way of viewing measurement theory.

\subsection{Geometric Quantum Kinematics}
We now wish to investigate the kinematics structure of quantum mechanics from a geometric perspective.
With an eye towards generalization, we will consider an arbitrary K\"ahler manifold $\cal{M}$.
We shall attempt to determine which characteristics impart the standard kinematics of quantum mechanics to this manifold.

We have already established that in the case of standard quantum mechanics $\cal{M}$ should be a projective Hilbert space.
Thus we begin by considering some important properties of projective Hilbert spaces.
The Riemann curvature tensor for a projective Hilbert space is of the form,
\begin{equation}
R_{\alpha\beta\gamma\delta}=\frac{\hbar}{4}\big[ g_{\gamma[\alpha}g_{\beta]\delta}+\omega_{\alpha\beta}\omega_{\delta\gamma}-\omega_{\gamma[\alpha}\omega_{\beta]\delta}\big].
\end{equation}
Because our quantum phase space has this type of curvature tensor, it also has {\em constant holomorphic sectional curvature}.
Holomorphic curvature plays the same r\^{o}le for complex manifolds that scalar curvature plays for real manifolds.
In the case of real manifolds, the number of independent Killing vectors is closely related to the form of the curvature tensor.
Thus, one could also expect the number of observables could also be related to the form of the curvature tensor.
In fact, as discussed by Ashtekar and Schilling in \cite{as}, a manifold with constant holomorphic sectional curvature has a maximal number of observables.
In addition we can only properly define a Lie algebra on the observables if the manifold is of constant holomorphic sectional curvature.
The value of the constant holomorphic sectional curvature is equal to $\frac{2}{\hbar}$.
This is determined by the need for the algebra of the observables to close under the previously defined symmetric bracket.

Now consider a finite dimensional complex projective space.
We can determine the characteristics that our manifold must have in order to possess standard quantum mechanical kinematics.
A theorem of Hawley \cite{geometry} and Igusa \cite{geometry} states that, for finite $n$, the projective spaces are up to isomorphism the only connected, simply connected, and complete K\"ahler manifolds of constant and positive holomorphic sectional curvature.
Thus they are isomorphic to $\CP{n}$.
Moreover, a more recent %
result of Siu and Yau and of Mori \cite{geometry} shows that the requirement of positive bisectional curvature {\em alone} necessarily implies that the underlying manifold is $\CP{n}$.

The infinite dimensional projective space is more problematic.
It is an open question whether the quoted theorem extends to the infinite dimensional case.
However, this is strongly hinted at by a theorem of Bessega \cite{geometry}, which suggests that every infinite dimensional Hilbert space is diffeomorphic with its unit sphere.
If so, there is no other infinite dimensional connected, simply connected, homogeneous, and isotropic K\"ahler manifold except $\CP{\infty}$.
In sharp contrast to the arbitrariness in the topology and geometry of the classical phase space and its symplectic structure, is the striking universality of the $\CP{n}$ of quantum mechanics.
The metric, symplectic and complex structures are so closely interlocked that the only freedoms left are the values of $n$ and $\hbar$.

We have determined to a certain extent what sets apart a given K\"ahler manifold from a manifold endowed with standard quantum mechanical kinematics.
As we shall see shortly this will allow us to more clearly see a path toward a generalization of quantum mechanics.

\subsection{Towards a Generalization of Quantum Mechanics}

The framework which we have developed readily lends itself to a generalization of the dynamical and kinematic
structures.\footnote{
Generalizations of quantum mechanics have a long history \cite{tzeminic, geomqm, anandan, generalqm}.
Most recently, such generalizations were discussed, for example, in \cite{generalqm}.}
Generalizing the dynamics involves, as indicated at the end of Sec. (\ref{gd}), extending the expectation values of an operators on $\cal{H}$ off the constraint surface $S$.
Generalizing the kinematics entails expanding the state space and/or the algebra of the observables.
There are direct extensions available for each of these.
However, the difficulty is in extending each in such a way as to create a self consistent generalization.
The generalization we will advocate is tied to a {\em quantum equivalence principle}.

We begin, however, by generalizing the dynamics.
As is the case with classical mechanics the dynamics preserve only the symplectic structure.
Thus we may also require that the dynamical flow preserve only the symplectic structure and not necessarily the Fubini--Study metric on the configuration space.
While the quantum phase space remains $\cal{P}$, the dynamical flow we seek to extend resides in $\cal{H}$.
This is because in terms of the quantum phase space the extension of the dynamical flow in the Hilbert space is arbitrary.

In order to see this more clearly consider $A$ the expectation value of an operator on $\cal{H}$.
We have up until this point restricted  such expectation value to the unit sphere.
However, suppose we extend off the unit sphere in some arbitrary way.
We can construct a Hamiltonian vector field generated by $A$ extended in $\cal{H}$, which we denote $X_A$.
As before, we project this vector field on to $\cal{P}$.
The part of the restriction of $X_A$ on the unit sphere, which is orthogonal to $\cal{J}$, is insensitive to the extension in $\cal{H}$.
Thus we are free to choose any extension we wish, and we shall always arrive at the same Hamiltonian vector field in the reduced quantum phase space.

There are many extensions from which to choose, the most obvious of which is to extend $A$ to the full Hilbert space and simply define it as,
\begin{equation}
A_{ext}(\Psi)\equiv\big<\Psi,\hat{A}\Psi\big>.
\label{ext}
\end{equation}
Extending $A$ off of $S$ in such a way we may recover the generalization of quantum mechanics given by Weinberg \cite{weinberg}.
Weinberg's formalism and its connection to geometric quantum mechanics are discussed in detail in App. A.

Turning now to constructing a generalized kinematic structure, we see immediately that several possibilities present themselves.
Recalling our previous analysis of the standard kinematic structure, we could consider expanding the quantum phase space to include all K\"ahler manifolds.
We might also consider expanding the class of observable beyond K\"ahler functions to include all smooth, real valued functions.
However, we will present a somewhat novel generalization that is motivated by our discussion concerning the nature of time in the geometric quantum framework as well as our intuition from General Relativity.

We begin by focusing on the observation that the geometry of $\CP{n}$ is closely tied to the interplay of the triadic structure $(g,\omega,j)$.
The Riemannian metric has its generic holonomy or stabilizer group $O(2n)$.
The symplectic structure has its stabilizer group $Sp(4n,\BR)$, and the almost complex structure $j$ has its group $GL(n,\BC)$.
The intersection of the these three associated Lie groups results in a subgroup of $O(2n)$, the unitary group $U(n,\BC)$.
This implies the unitarity in quantum mechanics, the Hermiticity of the observables and the Hermitian geometry of $\CP{n}$.
Note that any two elements of this triad plus their mutual compatibility condition imply the third.

The physical underpinnings of the triadic interplay was addressed by Gibbons and Pohle \cite{geomqm}.
They noted that observables in quantum mechanics play a {\em dual} r\^{o}le as providers of outcomes of measurements and generators of canonical transformations.
Specifically, the almost complex structure $j$ is dual to $g$ in that it generates canonical transformations corresponding to this metric, namely time evolution.
Thus time in quantum mechanics is tied in a one-to-one manner to $g$ as well as $j$.
We have already encountered this connection in \eref{anaeq}, the uncertainty in the energy (see \cite{anandan}).
As we previously noted, this linear relation between the metric and time shows the probabilistic nature of time and time as a correlator between statistical distances measured by different systems.

Now note that simply giving a manifold $\cal{M}$ a complex structure does not imply that $\cal{M}$ is complex.
The complex structure must be global.
For this to be the case the almost complex structure on a given manifold must be integrable.
The necessary and sufficient condition for integrability is given by the Newlander--Nirenberg theorem \cite{nn}, which states that an almost complex structure is integrable if the Nijenhuis torsion tensor vanishes.
Now note that $j$ is integrable on $\CP{n}$ for any $n$.
Thus standard quantum mechanics possesses absolute global time.
However, our intuition from General Relativity indicates a more provincial, local notion of time.
Therefore, we choose a generalized framework in which $j$ on a state space {\em fails} to be integrable.
Our generalized quantum mechanical structure would possess a local, relational time.
As a result there is a relativity among observers of the very notion of a quantum event.\footnote{This possibility was also discussed in \cite{isidro}.}

\section{Background 2: How to View Spacetime}
It is crucial to understand the microscopic degrees of freedom of quantum gravity.
The semiclassical limit, General Relativity, is one place to start addressing this question.
In what follows, we emphasize the thermodynamic nature of spacetime.
Then, inspired by the geometric formalism outlined above, we examine the quantum nature of spacetime.
Specifically, we will focus on comparison of the nature of observables and measurements in quantum field theory and General Relativity.
This will provide an entry point to a discussion of background independent quantum gravity.
Note that our crucial concern here is with a ``quasi-local'' (as opposed to
global) understanding of
holography (as implied by the second law of black hole thermodynamics)
from the underlying thermodynamic nature of General Theory of Relativity.

\subsection{Thermodynamics of Spacetime}
We have explored a method to reformulate quantum mechanics in a geometric formalism.
We have also discussed a possible generalization of quantum mechanics guided by intuition gained from General Relativity.
Now we would like to review a method for analyzing General Relativity from the perspective of thermodynamics.
This method is due to Jacobson and was applied in \cite{ted1} to systems
that possess local thermal equilibrium and extended to non-equilibrium systems in \cite{ted2}.
We will review and summarize both sets of results.

Black holes provide ideal laboratories for studying quantum gravity.
In particular, black holes are thermodynamic objects.
The familiar laws of thermodynamics have analogues for black holes \cite{bek, hawk1, hawk2}:
\begin{enumerate}
\item[0.] The surface gravity at the horizon of a stationary black hole is constant.
\item[1.] The infinitesimal change in mass is given by
$$
dM = \frac{\kappa}{8\pi G_N} dA + \Omega\ dJ + \Phi\ dQ,
$$
where $\kappa$ is the surface gravity, $A$ is the area of the horizon, $\Omega$ is the angular velocity, $J$ is the angular momentum, $\Phi$ is the electrostatic potential, and $Q$ is the electric charge.
\item[2.] The weak-energy condition implies that the surface area is a non-decreasing function in time: $dA \ge 0$.
\item[3.] It is not possible to have a black hole with zero surface gravity.
\end{enumerate}
The surface gravity specifies a black hole temperature $T_H = \hbar\kappa/2\pi$.
There is as well a notion of entropy associated to the area of the black hole horizon:
\be
S_{BH} = \frac{A}{4G_N\hbar}.
\ee
This is a quantum mechanical entropy as evinced by the $\hbar$ in the denominator.

The discovery of black hole entropy \cite{bek} and the four laws of classical black hole mechanics \cite{hawk1} therefore argue a connection between thermodynamics and gravitation.
This connection was put on a much firmer footing with the seminal work of Hawking \cite{hawk2} which established that black holes emit thermal radiation at the temperature $T_H$.
These advances in understanding black hole solutions were made by deriving thermodynamic quantities from gravitational considerations.
We now turn this method on its head and derive gravitational results from thermodynamic calculations.
Specifically, we will derive Einstein's equations as an equation of state by locally applying the Clausius relation, $dS=\frac{\delta Q}{T}$ in conjunction with the proportionality between the entropy of a system and the area of a causal horizon.

In order to do this we must first formalize the nature of heat flow and temperature in this setting, and we must also define exactly
which type of causal horizon we are considering. In traditional thermodynamics heat flow is the
transfer of energy between microscopic degrees of freedom which are unobservable at macroscopic scales.
We can draw an analogy between this phenomena and heat flow across a causal horizon. The gravitational field
created by this heat flow can be felt, although once across the horizon it can not be observed.
This horizon need not be the event horizon of a black hole. We
may simply consider the past causal boundary of an
observer as our horizon. These type of boundaries conceal information, and it is this property that
allows us to relate them to entropy calculations \cite{bek}. We would like to consider
equilibrium thermodynamics at each point along the horizon. Consider a spacetime point
$p$. By invoking the equivalence principle we can consider the neighborhood
about $p$ to be flat spacetime. In addition consider a small spacelike two-surface element $\cal{P}$
with its past directed null normal congruence to one side, which we will define as inside.
We require each spacetime point $p$ along the
horizon to be in equilibrium in the sense there is no shearing $\sigma$ or expansion $\theta$ of $\cal{P}$
and that Einstein's equations hold. To this
end we restrict ourselves to considering a local Rindler horizon which is the boundary of
a Rindler wedge. Most of the information stored beyond
the horizon is
stored in the correlations between the vacuum fluctuations
just inside and outside the horizon \cite{tangle}.
We may also use the vacuum fluctuations at the boundary to understand the system temperature.
Due to the Unruh effect \cite{ruh} these fluctuations become a thermal bath from the reference
frame of an uniformly accelerated observer. Thus we may define our system temperature as the
Unruh temperature of a uniformly accelerated observer just inside the horizon.
Restricted to Rindler wedge the vacuum density matrix for a relativistic
quantum field theory has the form of a canonical ensemble (Gibbs state), $\rho=Z^{-1}\exp(-H_B/T)$ where
$H_B$ is the boost Hamiltonian of the accelerated reference frame. In standard quantum field theory the
infinite number of infrared degrees of freedom near the horizon leads to
a formally infinite entanglement entropy. However, if we regulate our theory in the ultraviolet with a fundamental cutoff
length, the entropy becomes finite and is proportional to the horizon area.

 We also measure energy flux which defines the heat flow from the same accelerated reference frame.
This will give different results depending on the acceleration of the observer and
the acceleration becomes infinite as the observer's worldline approaches the horizon.
Thus the energy flux and temperature diverge. However, the ratio of the two is kept finite.
It is in this limit which we shall consider the thermodynamics of the system.

Now consider a small neighborhood of $\cal{P}$
that is essentially flat with the usual associate Poincar\'e symmetries. There is an
approximate Killing field $\chi^a$ which generates boosts orthogonal to $\cal{P}$ and vanishing at $\cal{P}$.
A boosted reference frame with acceleration $a$, possesses an Unruh temperature $T=\frac{a\hbar}{2\pi}$.
The heat flow can be determined by the boost-energy current of matter $T_{ab}\chi^a$ where $T_{ab}$ is the matter
energy momentum tensor.
Thus we find the heat flow past $\cal{P}$ to be
\begin{equation}
\delta Q=\int_{\cal H} d\Sigma^b \, T_{ab}\chi^a.
\end{equation}
Here we are integrating over a pencil of generators of the inside past horizon $\cal{H}$ of $\cal{P}$.
Now let $k^a$ be the tangent vector to the horizon generators for an affine parameter $\lambda$ that vanishes
 at $\cal{P}$ and is negative in the past. This implies $\chi^a=-a\lambda k^a$ and $d\Sigma^b=k^bd\lambda d^2A$. 
Thus we may rewrite our expression for the heat flux as
\begin{equation}
\delta Q=-a\int_{\cal H} d\lambda d^2A \, \lambda T_{ab}k^a k^b.
\end{equation}
Recalling the result for the temperature we find the entropy change to be
\begin{equation}
\frac{\delta Q}{T} = (2\pi/\hbar)\int d\lambda d^2A \, T_{ab}k^a k^b (-\lambda). \label{jak}
\end{equation}
The entropy change is also given by the change in the area of the horizon
\begin{equation}
\delta S=\alpha\, \delta A.
\end{equation}
Now recall the Raychaudhuri equation (see also App. C),
\begin{equation}
\frac{d\theta}{d\lambda} =-\frac{1}{2}\theta^2-\sigma_{ab}\sigma^{ab}-R_{ab}k^a k^b.
\end{equation}
Also recall that the expansion parameter, $\theta$ and the shear, $\sigma$ vanish at $p$.
This condition yields
\begin{equation}
\theta= -\lambda R_{ab}k^a k^b + O(\lambda^2).
\end{equation}
This implies that to lowest order in $\lambda$ the entropy change is given by
\begin{equation}
\delta S = \alpha \int_{\cal{H}} d\lambda d^2A \, R_{ab}k^a k^b
(-\lambda). \label{bak}
\end{equation}
If we now require that Clausius relation hold for each local Rindler horizon we
find that the integrands of the (\ref{jak}) and (\ref{bak}) are equivalent for all null vectors $k^a$.
Equating coefficients of $\lambda$ yields
\begin{equation}
R_{ab}  +\Phi g_{ab} = (2\pi/\hbar\alpha)T_{ab} \label{hi},
\end{equation}
where $\Phi$ is unknown function. To determine $\Phi$ we invoke the requirement of local matter energy conservation.
Now taking the divergence of both sides of (\ref{hi}) and applying the contracted Bianchi identity $R_{ab}{}^{;a}=\frac{1}{2} R_{,b}$
we find
\begin{equation}
\Phi=-\frac{1}{2}R - \Lambda
\end{equation}
where $\Lambda$ is an undetermined constant. Inserting this into our previous expression yields
\begin{equation}
R_{ab}-\frac{1}{2} Rg_{ab}-\Lambda g_{ab}=8\pi G_N T_{ab}
\end{equation}
where $G_N=(4\hbar\alpha)^{-1}$ is Newton's constant and $\Lambda$ is the cosmological constant. Note that
this implies that the universal entropy density is $\alpha=(4G_N\hbar)^{-1}$ which is in full agreement with the standard expression
for Bekenstein--Hawking black hole entropy, and we are consistent with standard general relativistic
results.

Now we wish to consider allowing for higher order curvature terms which we expect
from the effective field theory point of view \cite{guess}. One way to accomplish this is
to allow the entropy density to be dependent on the Ricci scalar
\begin{equation}
\alpha f(R)=1+O(R).
\end{equation}
Then the entropy change becomes
\begin{equation}
\delta S = \alpha \int_{\cal{H}} d\lambda d^2A \, (\theta f+\dot{f})
\end{equation}
where differentiation of $f$ is with respect to $\lambda$. However,
now if the Clausius relation is to hold $\theta$ must be non-vanishing.
Thus the area of the horizon is dynamic and this would seem to indicate
the equilibrium condition at $p$ no longer holds.
However, the rate at which the area changes is exponentially vanishing with respect to the Killing time.
So we may consider the system as approaching equilibrium near $p$. However,
the Clausius relation does not strictly hold. Instead we have $dS>\frac{\delta Q}{T}$. This
implies the entropy balance relation,
\begin{equation}
dS=\delta Q/T+d_iS
\end{equation}
where $d_iS$ is a contribution of to the entropy created internally due to
the system being out of equilibrium. In addition it was shown in \cite{ted2} that in order to
maintain local conservation of energy the additional entropy term must
be of the form
\begin{equation}
d_iS=\int_{\cal{H}}d\lambda d^2A \, \sigma 
\end{equation}
where $\sigma=-\frac{3}{2}\alpha\Phi\theta^2\lambda$. Now following a similar method used in the equilibrium case and
defining $\cal{L}$ by $f=d{\cal{L}}/dR$ we find the following equation of state,
\begin{equation}
fR_{ab}- f_{;ab} +(\Box f-\frac{1}{2}{\cal L})g_{ab}=
(2\pi/\hbar\alpha) T_{ab}.
\end{equation}
This is the equation of motion given by the Lagrangian $(\hbar\alpha/4\pi){\cal{L}}(R)$ for which the black hole entropy
density is $\alpha f(R)$ \cite{ted3}. Thus the thermodynamic equation of state is consistent with the Lagrangian field equation as in the case of pure
General Relativity. The inclusion of higher order curvature terms is indicative of the incorporation of quantum mechanical effects into General Relativity \cite{guess}. When
considering more traditional approaches to quantum gravity comparisons to non-equilibrium thermodynamics may be fruitful. We will make
use of this parallel in subsequent discussions.

\subsection{Gravitational Statistical Mechanics}
Thermodynamics is a coarse-grained description of a microscopic theory of physics.
Consider, for example, the molecules of air in a room.
The gas in the room is characterized by thermodynamic variables, temperature, pressure, and the chemical potentials of the different types of molecules.
To describe the physics it is not necessary to know the precise configuration of the molecules in the room.
An extensive thermodynamic state function, the entropy, captures the observer's ignorance regarding the microscopic details, and properties of the gas are determined by the behavior of this state function.
The entropy enumerates the possible configurations of the molecules in the room.
Likewise, the entropy of a black hole $S_{BH} = A/4G_N\hbar$ is expected to enumerate the microphysical states of the black hole.
There must be $e^{S_{BH}}$ such microstates with the charges of the black hole spacetime.\footnote{
Arguably, as horizon area is associated with entropy, geometries associated to these microstates cannot have horizons, for if they did, we would have $e^{S_{BH}}$ microstates each with their own entropy $e^{S_{BH}}$.
This furthermore already suggests that black hole entropy in classical gravity is a consequence of the thermodynamic limit, {\em viz.}\ an averaging or coarse-graining over the microstates.}
Identifying the microscopic states has proved challenging because in General Relativity the geometry of the spacetime, at least in four dimensions, is uniquely specified by conserved charges, mass, angular momentum, and electric charge.\footnote{
The discovery of black objects with non-spherical topologies --- objects like the {\em black ring} \cite{br} and the {\em black Saturn} \cite{bs} --- in higher dimensions indicates a violation of the uniqueness theorems.
To distinguish these states from black holes it becomes necessary to specify as well higher moments such as multipole charges, which are in general not conserved.}
General Relativity is expected to supply an effective description of the physics of spacetime that breaks down at some fundamental scale, say the Planck length.
The semiclassical limit of this fundamental theory --- {\em i.e.}, the long-wavelength description of gravity --- cannot probe physics at the Planck scale.
The individual microstates may therefore be expected to be distinguished from each other at Planck distances.

A few remarks about entropy in quantum gravity may be in order.
The entropy $S_{BH}$ scales as the area of the black hole horizon and not the volume of the spacetime encompassed by the horizon.
This is unlike the scaling of degrees of freedom of quantum mechanical systems decoupled from gravity for which entropy is extensive with the volume.
Quantum gravity has a {\em holographic} character in that degrees of freedom are codimension one \cite{holog}.
Any microscopic theory underlying gravitational thermodynamics must explain this unexpected scaling behavior.

Suppose we consider pure states in quantum mechanics.
Such states can collapse semiclassically to form black hole.
If the degrees of freedom that differentiate one microstate from another are confined to a Planck sized region around the singularity which the horizon shields, this leads to the so-called {\em information paradox}.
Recall that the vacuum in a quantum field theory is dynamical.
Consider the pair production of particle modes near the black hole horizon.
Hawking showed that when one particle in a pair crosses the horizon and the other escapes to infinity, the black hole's mass is reduced \cite{hawk2}.
The Hawking radiation is insensitive to the Planck scale structure of spacetime near the singularity.
Rather, it is determined by the thermodynamic properties of the system.
When the black hole has evaporated, the asymptotic observer at infinity is incapable of deducing which pure state formed the black hole.
If this information is in principle lost, this is a violation of the unitarity of quantum mechanics!
This cannot happen.

A resolution to these issues may lie in fundamental misconceptions regarding the nature of black hole spacetimes.
String theory has made progress in understanding the microscopic origin of black hole entropy.
In particular, for certain supersymmetric black holes, the Bekenstein--Hawking entropy is reproduced by an enumeration of degenerate vacua of D-brane configurations \cite{asen, sv}.
This is not, however, enough to extrapolate the identity of the microstates in the strong coupling regime where the semiclassical description of the spacetime as a black hole solution to General Relativity applies.
More recently, Mathur and his collaborators have argued that the characteristic features of black hole spacetimes in General Relativity, namely the existence of horizons and singularities, are associated to a thermodynamic coarse-graining, or averaging, over regular geometries \cite{mathur}.
There is no information paradox because the geometry of spacetime is smooth.\footnote{
The information paradox may as well be resolved if almost all states have a typical character, and almost no probes are capable of examining the deviations from typicality \cite{bdjs}.}
What is spectacular is that in cases where the microstates are each associated to geometries, the spacetimes start to differ from each other at the scale of the horizon of the semiclassical black hole.
Mathur's ideas rely on the existence of powerful string dualities, and while this identification of microstates may be compelling within these settings, it is difficult to see how to apply them to more general asymptopia or to cosmological horizons.
This would entail a radical reassessment of causal structure in General Relativity.
We do not pursue this program here.

\subsection{Observables and Measurement in QFT and GR}

Now we wish to discuss the nature
of observables and measurement in quantum theory and General Relativity.
This will aid us in attempting to uncover the general structure of
a background independent quantum theory of gravitation.

As Wigner pointed out, measurements are by nature different in quantum field theory and the General Theory of Relativity \cite{wig}.
The Special Theory of Relativity and quantum mechanics, as well as their conflation, are formulated in terms of particle trajectories or wave functions or fields that are functions (functionals) of positions or momenta.
Such coordinates are auxiliary constructs in the General Theory of Relativity.
Consistent with diffeomorphism invariance, we can assign almost any coordinates to label events in classical gravity and, by extension, in quantum gravity.
The coordinates are not in themselves meaningful.
Moreover, gravity is non-local, as perhaps best manifested in the concept of gravitational energy \cite{mtw}.
No local gauge invariant observables can exist \cite{einstein}.
The decoupling of scales familiar to local quantum field theories is simply not possible.
The ultraviolet physics mixes inextricably with dynamics in the infrared.

Measurements in a theory of gravitation are founded upon the relational properties of spacetime events.
Timelike separation of events is measured by clocks, whereas spacelike separation is determined more indirectly.\footnote{
There are subtleties regarding the interplay between a suitably microscopic clock and a macroscopic apparatus that records the measurement \cite{swig}.}
Within a quantum theory, events cannot themselves be localized to arbitrary precision.
Only for high-energies does it even make sense to speak of a local region in spacetime where an interaction takes place.
This is a simple consequence of the energy-time uncertainty relation.

The measurements that a bulk observer makes within a quantum theory of gravity are necessarily restricted, however.
Experiments are performed in finite times and at finite scales.
The interactions accessed in the laboratory also take place in regions of low spacetime curvature.
In local quantum field theory, we operate successfully under the conceit that the light cone is rigid.
There is an approximate notion of an $S$-matrix that applies to in- and out-states with respect to the vacuum in flat space.
Computing scattering amplitudes in string theory proceeds through analytic continuation of Lorentzian spacetimes into Euclidean spaces with fixed asymptopia.
On cosmological scales, this is of course a cheat.
Light cones tilt.
The causal structure, in particular, is not static.
We do not in general know the asymptotic behavior of the metric at late times.
The only data available about spacetime are events in an observer's past light cone.
Each observer has a different past light cone consistent with the histories of all the other observers, and the future causal structure is partially inferred from these data.

A manifold is constructed out of an atlas of local coordinate charts.
A sufficiently small neighborhood about any point is flat.
To solve Einstein's equations, the vacuum energy of empty Minkowski space vanishes exactly.
Globally, the ratio of the vacuum energy density to the expectation of Planck scale physics is extremely close to zero but does not identically vanish.
We wish to regard the measured, small cosmological constant as the consequence of patching together the physics of locally flat spaces consistent with
the existence of canonical gravitational quanta.
Instead of working with the spacetime manifold, we employ a larger geometric structure whose tangent spaces are the canonical Hilbert spaces of
a consistent quantum mechanics of gravitons.
The equivalence principle we employ relies on the universality and consistency of quantum mechanics at each point.
In every small, local neighborhood of this larger structure, the notion of quantum mechanical measurement is identical.
In particular, local physics in the laboratory is decoupled from the global physics. Nevertheless, as we will emphasize in what follows, there is
a non-trivial non-decoupling of local and global physics when one discusses the quantum origin of vacuum energy. %

\section{Background 3: Matrix theory}

\subsection{M-theory}

Quantizing General Relativity the way classical field theories are quantized leads to well-known ultraviolet divergences.
To renormalize the theory, an infinite number of counterterms are necessary.
It is possible that this signals an inappropriate use of the perturbative expansion and that the quantum theory of gravitation is in fact finite when treated exactly.
We do not, however, know of the existence of an ultraviolet fixed point of the renormalization group (RG) that would render the theory finite in this way.
The leading candidate theory of quantum gravity consistent with what we
know about the Standard Model of particle physics is {\em string theory}.\footnote{
For background on string theory, we refer the reader to the canonical textbooks \cite{gsw, polc}.}
The exchange of closed strings provides a mechanism for obtaining a quantum theory of gravity that is ultraviolet finite and consistent with Poincar\'e invariance and general covariance.
Moreover, string theory is a theory without free parameters.
The string moves self-consistently in the background spacetime that it generates.
The Polyakov action for the string (constant dilaton, antisymmetric tensor) is
\be
S = -\frac{1}{4\pi\alpha'}\int\dee^2\sigma\ \sqrt{-\gamma}\, \gamma^{ab} \pa_a X^\mu \pa_b X^\nu G_{\mu\nu},
\ee
where $X^\mu(\tau,\sigma)$ describes the embedding of the string in target space, $\gamma_{ab}$ is the worldsheet metric, and $G_{\mu\nu}$ is the spacetime metric.
The renormalization group equation for the non-linear sigma-model on the string worldsheet implies that
\be
\beta_{\mu\nu} = \frac{\dee G_{\mu\nu}}{\dee\log\Lambda} = \alpha' R_{\mu\nu} + O(\alpha'^2) = 0.
\ee
Thus the Einstein vacuum field equation for gravity is
encoded in the quantum structure of the conformal worldsheet theory.
The algebraic structure of the quantum worldsheet theory is
nicely captured by the operator product expansion (OPE)
\be
{\cal O}_\mu(u) {\cal O}_\nu(v) = \lambda_{\mu\nu}^\rho \frac{{\cal O}_\rho}{(u-v)^{\#}} + \ldots~
\ee
specified by the stress-energy tensor $T^{\mu\nu}$.
The $\beta$-function equation may equivalently be couched as a definition of the stress-energy tensor for the worldsheet theory:
\be
\beta^{\mu\nu} = T^{\mu\nu} = \frac{\delta\Gamma_{\rm eff}}{\delta G_{\mu\nu}},
\ee
where
\be
e^{-\Gamma_{\rm eff}} = \exp\left(\int\dee^2\sigma\ G_{\mu\nu} T^{\mu\nu}\right).
\ee
Thus the worldsheet quantum theory knows about the spacetime physics.
This remarkable connection can be extended in the context of stringy non-perturbative
physics from a spacetime point of view, via the holographic renormalization group as discussed in App. C.

Anomaly cancellation requires that the supersymmetric string (superstring) propagates in a spacetime with a critical dimension $D=9+1$.
There are five known perturbative string theories in ten dimensions.
All of these string theories are in turn different perturbative limits of M-theory \cite{hull, wdual}.

To date, little is known about M-theory itself beyond the following salient facts.
\begin{itemize}
\item The low-energy limit of M-theory is $\CN=1$ supergravity in eleven dimensions.
\item M-theory compactified on a small circle recovers type IIA string theory in ten dimensions.
The relation between the radius of the M-theory circle $R_{11}$ and the string coupling $g_s$ and string scale $\alpha'$ is
\be
R_{11} = g_s \sqrt{\alpha'}.
\ee
The eleven-dimensional Planck length $\ell_{\rm Pl} = g_s^{1/3}\sqrt{\alpha'}$.
The string coupling $g_s$ is determined by the expectation value of the dilaton field, $g_s = e^{\vev{\phi}}$.
The description in terms of type IIA string theory is therefore appropriate at weak string coupling.
At strong coupling the theory grows an eleventh dimension.
\item M-theory compactified on an interval bounded by Ho\v{r}ava--Witten domain walls gives heterotic $E_8\times E_8$ string theory.
\item The BPS spectrum of M-theory contains the M$2$-brane and its magnetic dual the M$5$-brane.
The AdS/CFT correspondence indicates that there is an exact duality between the six-dimensional worldvolume $\CN=(2,0)$ gauge theory on a stack of $N$ M$5$-branes with M-theory on $\AdS{7}\times S^4$ with $N$ units of four-form flux on $S^4$ and radius of curvature $R_{\AdS{7}} = 2R_{S^4} = 2(\pi N)^{1/3}\ell_{\rm Pl}$ and another exact duality between the three-dimensional $\CN=8$ gauge theory on a stack of $N$ M$2$-branes with M-theory on $\AdS{4}\times S^7$ with $N$ units of flux dual to the four-form on $S^7$ and radius of curvature $2R_{\AdS{4}} = R_{S^7} = (32\pi^2 N)^{1/6}\ell_{\rm Pl}$.
\end{itemize}

\begin{figure}[h]
\begin{center}
\epsfysize=1.75in
\epsfxsize=5in
\epsffile{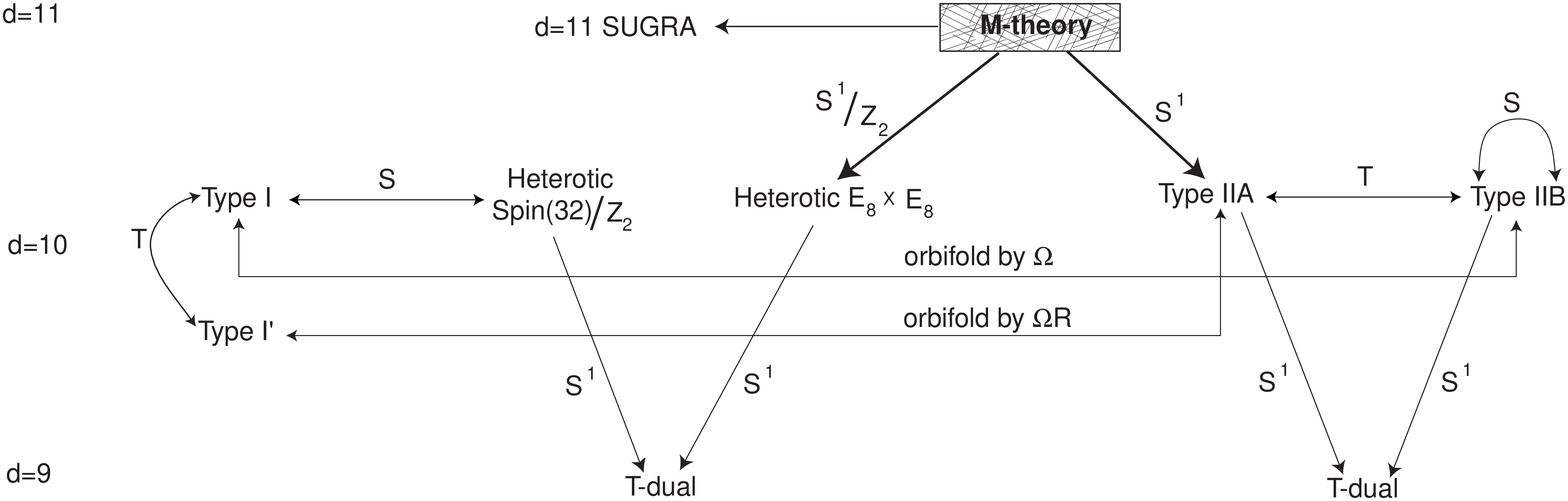}
\caption{M-theory and her children.}
\label{fig:Mth}
\end{center}
\end{figure}

The remaining perturbative string theories are implicated in the description of type IIA string theory as a weak coupling limit of M-theory.
There is a web of dualities that connects the various perturbative string theories together and relates them to M-theory.
Aspects of this web of dualities are diagrammed in Figure~\ref{fig:Mth}.
Using T-duality, which is a duality that leaves the coupling constant invariant up to a radius dependent rescaling, and S-duality, a duality under which the coupling changes non-trivially, as well as orbifold maps that involve the gauging of a discrete worldsheet symmetry, we can map each of the perturbative string theories to any of the others.

Our capacity to explore M-theory is frustrated by the fact that the theory is strongly coupled and there is no good perturbation theory to apply.
M-theory nevertheless admits a microscopic description as the supersymmetric quantum mechanics of matrix degrees of freedom.
We are led to this conclusion through two separate lines of investigation:
from the quantization of the supermembrane and from the Hamiltonian mechanics of $N$ D$0$-branes in type IIA string theory considered in light-cone frame \cite{matrix}.
Matrix theory thus formulated produces a non-perturbative definition of M-theory within a fixed background.
There are several outstanding reviews devoted to this subject \cite{srev,drev,brev,wrev}.
In our brief synopsis, we shall draw in particular from \cite{wrev}.

\subsection{From Membranes to Matrices}

As in the case of quantizing the string, we must have a quantization procedure for membranes.
Quantization of the supermembrane, however, presents a formidable challenge.
As we shall see, it is in attempting to overcome these difficulties that we are led to Matrix theory.

We first consider the classical bosonic membrane in flat spacetime.
Just as a particle traces out a worldline as it propagates in spacetime and the string traces out a worldsheet, the membrane traces out a worldvolume.
We shall parameterize coordinates on the worldvolume with the variables $\{\tau,\sigma_1,\sigma_2\}$.
The motion of the membrane in target space is given by $X^{\mu}(\tau,\sigma_1,\sigma_2)$.

The action that describes the membrane's motion is the membrane equivalent of the Nambu--Goto action for the string:
\begin{equation}
S=-T\int \dee^3\sigma\ \sqrt{-g}.
\end{equation}
Here we are using the signature $(-,+,+)$.
The membrane tension is defined as $T={1}/{(2\pi)^2\ell_{\rm Pl}^3}$, and $g_{\alpha\beta}$ is the pull-back of the Minkowski metric on the worldvolume.
We may rewrite this as the membrane equivalent of the Polyakov action,
\begin{equation}
S=-\frac{T}{2}\int \dee^3\sigma\ \sqrt{-\gamma}\, (\gamma^{\alpha\beta}\partial_\alpha X^{\mu}\partial_\beta X_{\mu}-1).
\label{pol}
\end{equation}
Here $\gamma_{\alpha\beta}$ is the metric on the worldvolume.
Now we are confronted with serious discrepancies between the analysis of the bosonic membrane and that of the bosonic string.
First, we take note of the constant term in (\ref{pol}).
It is absent in the Polyakov action for the bosonic string.
Its presence in the action of the membrane is indicative of the lack of conformal invariance.
The action does possess three diffeomorphism symmetries.
However, unlike the worldsheet metric of the bosonic string which contained three components, the worldvolume metric contains six components.
Due to the lack of conformal invariance and the increased number of components in $\gamma_{\alpha\beta}$, we are only able to partially gauge fix the metric.
We shall do this as follows,
\begin{equation}
\gamma_{00}=-\frac{4}{N^2}\textrm{det}\,g_{ab}, \quad
\gamma_{0b}=0,
\end{equation}
where $N$ is a constant that will be fixed when we come to the matrix regularization of the membrane.
Also note that for constant $\tau$ the Poisson bracket is defined
\begin{equation}
\{f,g\}\equiv\epsilon^{ab}\partial_a f \partial_b g.
\end{equation}
Now we may rewrite the action in terms of the gauge fixed metric and the Poisson bracket \cite{wrev},
\begin{equation}
S=\frac{N T}{4}\int \dee^3\sigma\Big(\dot{X}^{\mu}\dot{X}_{\mu}-\frac{2}{N^2}\{X^{\mu},X^{\nu}\}\{X_{\mu},X_{\nu}\}\Big).
\end{equation}
This action yields the following equations of motion
\begin{equation}
\ddot{X}^{\mu}=\frac{4}{N^2}\{\{X^{\mu},X^{\nu}\},X_{\nu}\},
\end{equation}
with the constraint equations
\begin{equation}
\dot{X}^{\mu}\dot{X}_{\mu}=-\frac{2}{N^2}\{X^{\mu},X^{\nu}\}\{X_{\mu},X_{\nu}\},
\end{equation}
and
\begin{equation}
\dot{X}^{\mu}\partial_aX_{\mu}=0\quad\Rightarrow\quad\{\dot{X}^{\mu},X_{\mu}\}=0.
\end{equation}
This theory has manifest covariance but because of the non-linear nature of the equations of motion and the constraint equations it is difficult to quantize.

Quantization of the bosonic string is simplified by going to light-cone gauge.
In that spirit let us consider the following light-cone coordinates,
\begin{equation}
X^{\pm}=\frac{1}{\sqrt{2}}(X^0\pm X^{D-1}),
\end{equation}
and proceed to a light-cone gauge,
\begin{equation}
X^+=\tau.
\end{equation}
This allows us to solve the constraint equations
\begin{equation}
\dot{X}^-=\frac{1}{2}\dot{X}^i\dot{X}^i+\frac{1}{N^2}\{X^i,X^j\}\{X^i,X^j\}, \quad \partial_aX^-=\dot{X}^i\partial_aX^-.
\end{equation}
(The indices $i,j$ run from $1$ to $D-2$.)
If we now turn to the Hamiltonian formalism while remaining in the light-cone gauge, we find
\begin{equation}
H=\frac{N T}{4}\int \dee^2\sigma \Big( \dot{X}^i\dot{X}^i+\frac{2}{N^2}\{X^i,X^j\}\{X^i,X^j\}\Big),
\label{mh}
\end{equation}
with only remaining constraint on the transverse degrees of freedom being,
\begin{equation}
\{\dot{X}^i,X^i\}=0.
\end{equation}
Even with these simplifications quantization remains difficult because of the non-linearity of the equations of motion.

In order to quantize this theory we must turn to the matrix regularization scheme originated in \cite{hop} and \cite{gold}.
In this regularization, we map functions on the membrane's surface to $N\times N$ matrices.
We will assume the membrane surface to be the sphere $S^2$.
Thus at any fixed time we may describe the membrane surface as a unit two-sphere with an associated $SO(3)$-invariant symplectic form.
We may describe a function on the sphere using functions of the Cartesian coordinates which we denote $q_1,q_2,q_3$ with the obvious constraint that $\sum_i q_i^2=1$.
The symplectic structure is encoded in the Poisson bracket $\{q_a,q_b\}=\epsilon_{abc}q_c$.
The Lie algebras of $SU(2)$ and $SO(3)$ are identical.
The algebra of $SU(2)$ may be expressed in terms of the Lie bracket
$[\mathbf{J}_a,\mathbf{J}_b]=i\epsilon_{abc}\mathbf{J}_c$.
Thus we are led to relate the coordinate functions on the membrane to matrices that are an $N$-dimensional representation of the generators of $SU(2)$
\begin{equation}
q_a\rightarrow\frac{2}{N}\mathbf{J}_a.
\end{equation}
All functions on the sphere can be expanded in terms of spherical harmonics
\begin{equation}
f(q_1,q_2,q_3)=\sum_{l,m}a_{lm}Y_{lm}(q_1,q_2,q_3),
\end{equation}
which are themselves functions of the coordinates on the sphere
\begin{equation}
Y_{lm}(q_1,q_2,q_3)=\sum b^{(lm)}_{a_1\dots a_l} q_{a}\dots q_{l}.
\end{equation}
Note that that because we are constrained to the unit sphere the coefficients $b^{(lm)}_{a_1\dots a_l}$ are symmetric and traceless.
Thus we may extend our correspondence to spherical harmonics for $l<N$,
\begin{equation}
Y_{lm}(q_1,q_2,q_3)\rightarrow\mathbf{Y}_{lm}=\bigg(\frac{2}{N}\bigg)^{l}\sum b^{(lm)}_{a_1\dots a_l} \mathbf{J}_{a}\dots \mathbf{J}_{l}.
\end{equation}
We may only allow for spherical harmonics with $l<N$ because the higher order monomials generated by $\mathbf{J}_a$ are not linearly independent matrices.
This naturally leads to a matrix correspondence for an arbitrary function,
\begin{equation}
f(q_1,q_2,q_3)\rightarrow \mathbf{F}=\sum_{l<N,m}a_{lm}\mathbf{Y}_{lm}.
\end{equation}
Likewise we define a correspondence between the Poisson bracket and the Lie bracket
\begin{equation}
\{f,g\}\rightarrow -\frac{iN}{2}[\mathbf{F},\mathbf{G}].
\end{equation}
There is also a relation between an integral of a function over the membrane and the trace of the corresponding matrix,
\begin{equation}
\frac{1}{4\pi}\int \dee^2\sigma\, f\rightarrow\frac{1}{N} \textrm{Tr}\,\mathbf{F}.
\end{equation}
Applying this prescription to (\ref{mh}) yields the following regularized Hamiltonian
\begin{equation}
H=\frac{1}{2\pi \ell^3_{\rm Pl}}\textrm{Tr}\Big(\frac{1}{2}\dot{\mathbf{X}}^i\dot{\mathbf{X}}^i- \frac{1}{4}[\dot{\mathbf{X}}^i,\dot{\mathbf{X}}^j][\dot{\mathbf{X}}^i,\dot{\mathbf{X}}^j]\Big).
\end{equation}
 From this we may derive the equations of motion
\begin{equation}
\ddot{\mathbf{X}}^i+[[\mathbf{X}^i,\mathbf{X}^j],\mathbf{X}^j]=0
\label{gruga}
\end{equation}
and the constraint
\begin{equation}
[\dot{\mathbf{X}}^i,\mathbf{X}^j]=0.
\end{equation}
Even after instituting this procedure, we are left with a classical theory but quantization is now straightforward.

The result is a quantum theory whose fundamental degrees of freedom are $N\times N$ matrices with $U(N)$ symmetry.
Here we have assumed the membrane to have the topology of the two-sphere.
However, this regularization procedure can be generalized to membranes of arbitrary topology \cite{fair, bord}.

Now we turn to the case of a supersymmetric membrane.
There is unfortunately no known method to formulate the supermembrane with manifest worldvolume supersymmetry.
That is to say that there is no analogue of the NSR formalism that is applied to superstrings.
We shall instead follow the Green--Schwarz formalism with $\kappa$-symmetry present in the worldvolume.

Briefly, there are eleven bosonic degrees of freedom corresponding to the embedding of the membrane.
The reparameterization invariance of the worldvolume gauges away three of these so that there are eight bosonic degrees of freedom at the end.
The fermions start out as thirty-two component spinors.
The mass-shell condition and the $\kappa$-symmetry each halve the available degrees of freedom.
The bosonic and fermionic degrees of freedom are then organized as an $\CN=8$ multiplet of the three-dimensional worldvolume theory.

This procedure yields the following Hamiltonian \cite{wrev},
\begin{equation}
H=\frac{N T}{4}\int \dee^2\sigma \Big( \dot{X}^i\dot{X}^i+\frac{2}{N^2}\{X^i,X^j\}\{X^i,X^j\} -\frac{2}{N}\theta^T\Gamma_i\{X^i,\theta\}\Big). \label{gogu}
\end{equation}
Here $\theta$ is a sixteen component Majorana spinor of $Spin(9)$, and $\Gamma_i$ are the $\Gamma$-matrices associated to the Clifford algebra.
As before we are working in light-cone gauge, and we have gauge fixed the $\kappa$-symmetry as follows: $\Gamma^+\theta=0$.
As in the bosonic case we implement the matrix regularization which yields,
\begin{equation}
H=\frac{1}{2\pi \ell^3_{\rm Pl}}\textrm{Tr}\Big(\frac{1}{2}\dot{\mathbf{X}}^i\dot{\mathbf{X}}^i-
\frac{1}{4}[\dot{\mathbf{X}}^i,\dot{\mathbf{X}}^j][\dot{\mathbf{X}}^i,\dot{\mathbf{X}}^j]+
\frac{1}{2}\theta^T\Gamma_i[\mathbf{X}^i,\theta]\Big).
\label{bing}
\end{equation}
However, as is the case with quantization of the superstring in the Green--Schwarz formalism, Lorentz invariance is lost.
Attempts at covariant membrane quantization were addressed in the early work of \cite{tzeminic} making use of Nambu brackets \cite{nambu} in place of Poisson brackets,
\begin{equation}
\{x,y,z\}=\epsilon^{\alpha\beta\gamma}(\partial_{\alpha}x)(\partial_{\beta}y)(\partial_{\gamma}z).
\end{equation}

Unfortunately the membrane theory we have described in this section suffers from an apparent pathological instability.
Consider a bosonic membrane with constant tension.
Because its energy is proportional to its area, it can develop long, very thin, quills emanating from its surface because these are energetically inexpensive.
This renders the membrane highly delocalized.
On the quantum level this instability leads to a continuous spectrum.
However, as we shall see, this difficulty is resolved because the Matrix theory is a manifestly second quantized theory.

\subsection{The BFSS Conjecture and Discrete Light Cone Quantization}

Now we will explore another path to Matrix theory that has entirely different roots than membrane quantization.
To begin we consider M-theory compactified on a circle of radius $R_{11}$.
This compactification provides the theory with an infrared cutoff and ensures that momentum is quantized according to the Kaluza--Klein condition. %
We separate the momentum in the compactified dimension, $p^{11}={N}/{R_{11}}$, from the momenta $p^i$ in the other ten dimensions to write the dispersion relation as
\begin{equation}
E^2=\Big(\frac{N}{R_{11}}\Big)^2+p^2+m^2.
\end{equation}
We work in the large-$N$ limit, and
the infinite momentum frame which corresponds to taking $p^{11}$ to infinity.
We shall at the end decompactify by sending $R_{11}$ to infinity as well.
Expanding our previous result
\begin{equation}
E=\frac{N}{R_{11}}+\frac{1}{2}\frac{R_{11}}{N}(p^2+m^2)+\mathcal{O}\Big(\frac{R_{11}}{N}\Big)^2.
\end{equation}
In the infinite momentum frame, the partons are the D$0$-branes \cite{matrix}.
D-branes are non-perturbative solitons (defects) upon which open strings end;
their masses are proportional to $1/g_s$, and thus these are precisely the states that become massless as we send $g_s\to\infty$ or equivalently decompactify the M-theory circle.
In the infinite momentum frame the dynamics are non-relativistic, and thus we may describe physics using supersymmetric quantum mechanics \cite{wmatrix, kogut}.

Let us explore the dynamics in this limit in more detail.
We wish to find the low-energy effective Lagrangian for a system of $N$ D$0$-branes.
To do this we begin with the Lagrangian for ten-dimensional super-Yang--Mills theory in the low-energy limit and dimensionally reduce it to $0+1$ dimensions,
\begin{equation}
\mathcal{L}=\frac{1}{2g_s\ell_s}\textrm{Tr}\Big[\dot{\mathbf{X}}^a\dot{\mathbf{X}}^a+
\frac{1}{2}[\mathbf{X}^a,\mathbf{X}^b]^2+\theta^T(i\dot{\theta}-\Gamma_a[\mathbf{X}^a,\theta])\Big].
\end{equation}
Here we have gauge fixed $A_0=0$.
The Yang--Mills coupling is expressed in terms of the string coupling and the string length as
\be
g_{YM}^2 \propto g_s \ell_s^{-3}.
\ee
We choose to work in units where $2\pi\ell_s = 1$, such that $R_{11} = g_s \ell_s = 2\pi \ell_{\rm Pl}^3$.
Now we will rescale $\mathbf{X}/g_s^{\frac{1}{3}}\rightarrow\mathbf{X}$ and put $\ell_{\rm Pl}=1$.
This yields the Hamiltonian
\begin{equation}
H=\frac{R_{11}}{2}\textrm{Tr}\Big(\dot{\mathbf{X}}^i\dot{\mathbf{X}}^i-
\frac{1}{2}[\dot{\mathbf{X}}^i,\dot{\mathbf{X}}^j][\dot{\mathbf{X}}^i,\dot{\mathbf{X}}^j]+
\theta^T\gamma_i[\mathbf{X}^i,\theta]\Big).
\end{equation}
There are several striking things to notice about this Hamiltonian.
First, it is equivalent to the Hamiltonian arrived at by quantization of the supermembrane, (\ref{bing}).
Remarkably, D$0$-branes describe the M$2$-brane. %
Next, it is also composed of eleven-dimensional quantities although we began with a ten-dimensional super-Yang--Mills theory.
We are led to a conjecture due to Banks, Fischler, Shenker, and Susskind \cite{matrix}.
\begin{BFSS}
The infinite momentum limit of M-theory is equivalent to the $N\rightarrow\infty$ limit of $N$ coincident D$0$-branes, given by $U(N)$ super-Yang--Mills theory.
\end{BFSS}

Although the conjecture was made in the $N\to\infty$ limit it was later suggested that for finite-$N$, Matrix theory is equivalent to the discrete light-cone quantized (DLCQ) sector of M-theory.
There are several different points of view on this subject that come in the form of different limiting procedures.
Here we will review the limit due to Seiberg \cite{sei} and Sen \cite{sen}, but other limiting procedures were suggested in \cite{bala, deal}.
The various limits were nicely reviewed by Polchinski in \cite{poll}.

Following \cite{sei}, we begin with M-theory compactified on a {\em lightlike} circle of radius $R$ with momentum $p^-={N}/{R}$ with the following identifications:
\begin{equation}
\binom{x}{t}\sim\binom{x}{t}+\binom{\frac{R}{\sqrt{2}}}{-\frac{R}{\sqrt{2}}}.
\label{lit}
\end{equation}
This is a particular limit of a compactification on a spacelike circle with the identification
\begin{equation}
\binom{x}{t}\sim\binom{x}{t}+\binom{\sqrt{\frac{R^2}{2}+R^2_s}}{-\frac{R}{\sqrt{2}}}\approx
\binom{\frac{R}{\sqrt{2}}+\frac{R_s^2}{\sqrt{2}R}}{-\frac{R}{\sqrt{2}}}~
\label{sec}
\end{equation}
with $R_s\ll R$.
We may rescale the value of $R$ as well as the light-cone energy $p^+$ by way of a longitudinal boost.
Consider the following boost parameter
\begin{equation}
\beta=\frac{R}{\sqrt{R^2+2R_s^2}}\approx 1-\frac{R_s^2}{R^2}.
\end{equation}
If we boost (\ref{sec}) in this way we obtain greatly simplified identifications
\begin{equation}
\binom{x}{t}\sim\binom{x}{t}+\binom{R_s}{0}.
\label{simp}
\end{equation}
Now in analogy with the $N\rightarrow\infty$ case we consider a compactification on (\ref{simp}).
This yields the following string scale and coupling \cite{sei},
\begin{equation}
g_s = \left(\frac{R_s}{\ell_{\rm Pl}}\right)^{\frac{3}{2}}, \quad \alpha' = \ell_s^2 = \frac{\ell_{\rm Pl}^3}{R_s}.
\end{equation}
As expected, in the limit $R_s\rightarrow 0$ the string coupling vanishes, meaning that higher genus contributions may be dropped.
The string tension also vanishes in this limit as $\alpha'\to\infty$.
However, the boost rescales $p^+$ to be of the order of $R_s/\ell_{\rm Pl}^2$.
In order to analyze more clearly the modes with this energy we must rescale the parameters of the theory.
We will replace the M-theory with Planck scale $M_{\rm Pl} = \ell_{\rm Pl}^{-1}$ compactified on a lightlike circle of radius $R$ with
$\widetilde{\textrm{M}}$-theory with Planck scale $\widetilde{M}_{\rm Pl}$ compactified on a spacelike circle of radius $R_s$.
The transverse geometry of $\widetilde{\textrm{M}}$-theory replaces that of the original M-theory.
The relationship between the parameters of the two theories is found by taking
$R_s\rightarrow 0$, $\widetilde{M}_{\rm Pl}\rightarrow\infty$, and holding $p^+\sim R_s\widetilde{M}_{\rm Pl}^2$ fixed.
This yields the relation,
\begin{equation}
R_s\widetilde{M}_{\rm Pl}^2=R_sM_{\rm Pl}^2.
\end{equation}
Also, because the boost does not affect the transverse direction, we may equate
\begin{equation}
M_{\rm Pl}R_i=\widetilde{M}_{\rm Pl}\widetilde{R}_i.
\end{equation}
Thus we find the following rescaled string tension and Planck scale
\begin{eqnarray}
\widetilde{g}_s=(R_s\widetilde{M}_{\rm Pl})^{\frac{3}{2}}=R_s^{\frac{3}{4}}(RM_{\rm Pl}^2)^{\frac{3}{4}}
\nonumber\\
\widetilde{M}_s^2=R_s\widetilde{M}_{\rm Pl}^3=R_s^{-\frac{1}{2}}(RM_{\rm Pl}^2)^{\frac{3}{2}}.
\end{eqnarray}
The rescaled theory in the $R_s\rightarrow 0$ limit has a vanishing string coupling and a large string tension.
We have mapped our original M-theory with momentum $p^-={N}/{R}$ in the compactified direction to a new $\widetilde{\textrm{M}}$-theory with momentum $p={N}/{R_s}$ in the compactified direction.
We have already established that M-theory with this type compactification is dual to type IIA string theory, and its dynamics are determined by D$0$-branes when the radius of compactification is small.

\subsection{Matrix Theory and Second Quantization}

It is important to note that because the dynamics of Matrix theory are governed by matrices the theory is manifestly second quantized.
To see this explicitly, consider the equations of motion of a bosonic Matrix theory (\ref{gogu}) with equations of motion (\ref{gruga}).
For block diagonal matrices
\begin{displaymath}
\mathbf{X}^i =
\left( \begin{array}{ccc}
\mathbf{A}^i & 0  \\
0 & \mathbf{B}^i   \\
\end{array} \right),
\end{displaymath}
the equations of motion of each block are separable:
\begin{equation}
\ddot{\mathbf{A}}^i+[[\mathbf{A}^i,\mathbf{A}^j],\mathbf{A}^j]=0, \quad
\ddot{\mathbf{B}}^i+[[\mathbf{B}^i,\mathbf{B}^j],\mathbf{B}^j]=0.
\end{equation}
Remarkably, we may consider each block as representing a different object with a different center of mass.
We could extend this type of analysis to include an arbitrary number of objects.
Matrix theory therefore describes the dynamics of multiple independent objects and is by its very nature second quantized in target space.
Moreover, this resolves the issues of the apparent instability noted above.
 From a classical perspective we are faced with the problem of all of the energetically inexpensive quills on the surface of the membrane.
However, in the case of a multiple membrane configuration these quills become tubes connecting the various membranes.
The tubes in question have a small radius and have a negligible effect on the dynamics of each independent membrane.
On the quantum level the D$0$-branes represent multiple graviton bound states with a single unit of momentum.
Thus we expect to find the continuous spectrum that was previously considered so problematic.

\subsection{Symmetries of Matrix Theory}

Notice that we started with a theory in eleven-dimensions, namely M-theory.
Going to light-cone frame required that we isolate $p^+$ as the generator of time translations.
The subgroup of the Lorentz group that acts invariantly upon the light-cone frame is the Galilean group, which is to say that $p^- \simeq p^{11} = N/R_{11}$ is identified with a non-relativistic mass and $p^i$ with a non-relativistic momentum.
The $p^i$ transform under the Galilean group as follows:
\be
p^i \mapsto p^i + p^- v^i.
\ee
The other Lorentz generators are a longitudinal boost, which oppositely rescales $p^\pm$, and the rotations of the null plane.

The Hamiltonian for Matrix theory is expressed in terms of $N\times N$ matrices ${\mathbf X}^i$, $i=1,\ldots,9$ that describe the transverse space to the light-cone.
These matrices ${\mathbf X}^i$ are invariant under $U(N)$ (crucially this is not $SU(N)\times U(1)$).
Commuting matrices may be simultaneously diagonalized.
Their eigenvalues denote the classical positions of D$0$-branes.
The D$0$-branes, however, interact via open strings.
These interactions are determined by off-diagonal entries in the matrices.
Such quantum fluctuations lead to a failure of matrix commutativity.
Since the matrices are a description of target space, the spacetime geometry becomes non-commutative once interactions are turned on.
Because D$0$-branes are themselves graviton bound states, the gravitational interaction and thus the geometry of spacetime are contained in the open string dynamics, {\em viz.}\ the fluctuations of the matrix degrees of freedom.
Moreover, although it is not as transparent as in AdS/CFT, the Matrix theory description of physics is holographic and exhibits the phenomenon of UV/IR mixing.
The theory also has the curious property that the density of states grows as we compactify on higher dimensional tori.

Bulk spacetime is as well an emergent feature of the Matrix theory construction.
The transverse space arises from the parameterization of the moduli space of the Yang--Mills theory on the $N$ D$0$-branes.
The longitudinal direction is canonically conjugate to $p^-$, which is specified by the compactification scale (which acts an infrared cutoff on the theory) and the rank of the gauge group.

This is a very appealing picture in which we have developed a non-perturbative, non-local formulation of M-theory in a particular limit and understood both the origins and the dynamics of the microscopic degrees of freedom.
Indeed, calculations of scattering amplitudes in Matrix theory correctly reproduce expected results in eleven-dimensional supergravity and vindicate this paradigm \cite{matrix}.
One drawback in the formulation is that the construction is not manifestly covariant as we work {\em ab initio} in an infinite momentum frame (or equivalently in light-cone gauge).
As well, in restricting to degrees of freedom with positive longitudinal momentum, we have forsaken background independence.
Adding states with zero longitudinal momentum corresponds to shifting the background, for example by adding five-branes that wrap the light-cone directions \cite{berkdoug}.
In these cases, the prescription for obtaining a Hamiltonian changes because there are new interactions invoked (for instance, D$4$-brane dynamics from the point of view of the type IIA string theory).
To date, Matrix theory provides a holographic description of physics only in asymptotically flat eleven-dimensional space.
We insist upon Minkowski asymptopia because it is in this setting that the particle description at infinity exactly applies, and we have an $S$-matrix theory at long wavelengths.
Clearly, such a description is unsatisfactory for cosmological spacetimes.
Indeed, although a number of attempts have been made in this direction \cite{verl}, even for symmetric spacetimes with a non-zero cosmological constant, there is no explicit Matrix theory formulation that captures the bulk physics.

\subsection{A Recapitulation}

Matrix theory summarizes very nicely what we know about non-perturbative physics
of string theory in asymptotically flat spaces. Because it is a theory
in which the configuration space involves (non-commuting) degrees of freedom that
can be related to spatial coordinates, it is of special interest given the discussion
about the purely quantum theoretic viewpoint on the measurements of time intervals
and spatial distances, as reviewed in Sec. 2.
This is why we will take Matrix theory as a candidate canonical quantum theory suitable for
a generalized geometric formulation to be discussed in the following section.

Finally we conclude this section by remarking that the explicit solution of Matrix theory does
not exist at present. Given the recent progress in understanding the
the spectrum of non-supersymmetric (bosonic) 2+1 Yang--Mills theory \cite{lmy} using the method of
wave-functionals, it is tempting to
think that the spectrum of Matrix theory can be understood along similar lines.
This would be important in view of the generalization presented below: the usual Hilbert space
of Matrix theory would provide a ``coordinate basis'' for understanding the general background
independent formulation of Matrix theory, based on the background independent formulation of quantum theory
to which we now turn.

\section{M-theory $=$ Background Independent Matrix Theory}

In this section we put all the physics motivations and all the
necessary background material together
to put forward a quantum background independent version of Matrix theory. This can be taken
as a proposal for a definition of M-theory, or a suggested answer
to the question ``What is string theory?''
The crucial physical element is a quantum version of the equivalence principle which suggests
a gauging of the unitary geometric structure of canonical quantum theory reviewed in previous sections.
The only extra element needed apart from this principle is the choice of a quantum theory
to be gauged, for which we take Matrix theory, as an example of a quantum theory
which encodes the quantum physics of asymptotically Minkowski space.
\subsection{Background Independent Matrix Theory}
We begin by recalling the given Riemannian structure of quantum mechanics and 
the observed connection between the Fubini--Study and spatial metrics. 
It behooves us to inquire if a more general Riemannian structure of 
space can be induced from a more flexible state space than $\CP{n}$.
We have previously noted that the Fisher--Fubini--Study metric of the Gaussian coherent state recapitulates the metric in configuration space.
Instead of $\psi_l(x)$, let us, for example, consider $\psi_{k(l)}(x)$.
The corresponding expression for the spatial metric results from the overlap of two Gaussians
$
\psi_{k(l)}(x) \sim \exp(- \frac{(x-k(l))^2}{\delta k(l)^2})
$
which in turn follows from
$
\int dx\ g_{\psi_l, \psi_{l+dl}} \psi_l^* \psi_{l+dl}
$
where the ``quantum metric'' reads
$
g_{\psi_l, \psi_{l+dl}} \equiv \frac{ \psi_k^* \psi_{k+dk}}{ \psi_l^* \psi_{l+dl}}.
$
Clearly the transformation that takes $\psi_l \to \psi_{k(l)}$ is {\em not} in general unitary. Thus,
if we insist on the desired relation between the quantum metric and an arbitrary metric on the classical configuration space, then the kinematics of the quantum theory must be altered.
Moreover, if the induced classical configuration space is to be the actual space of spacetime, only a special quantum system will do.
We are thus induced to make the state manifold suitably flexible by doing General Relativity on it.
The resultant metric on the Hilbert space is generally curved with its distance function modified, an extended Born rule, and hence a new meaning is assigned to probabilities.
By insisting on diffeomorphism invariance in the state space and on preserving the desirable complex projective properties of Cartan's rank one symmetric spaces such as $\CP{n}$, we
apparently arrive at the ensuing coset state space $\Diff(n, \BC)/\Diff(n-1,\BC) \times \Diff(1,\BC)$ with $n\to\infty$ as the minimal phase space candidate for a background independent quantum mechanics.
By background independence, we mean that in the configuration space, no {\em a priori} choice of asymptotics in made.

In summary, the axioms of standard geometric quantum mechanics are enlarged as follows.
\begin{enumerate}
\item The state space $\CP{\infty}$ is extended to $\Diff(\infty,\BC)/\Diff(\infty-1,\BC) \times \Diff(1,\BC)$, deriving from the generalized inner product
\begin{equation}
dS^2 = \sum h_{ab}[(dq_a)^2 + (dp_a)^2] \equiv h_{ab}\, dX^a dX^b,
\end{equation}
where $h_{ab}$ is Hermitian.
The ``Born rule'' now reads
\begin{equation}
{1 \over 2} \sum_{a,b} h_{ab} [ (p^a p^b) + (q^a q^b) ] =1.
\end{equation}
These equations provide the metric relation on and the geometrical shape(s) of the new state space, and implicitly defines $\hbar$.
The probabilistic interpretation lies in the definition of geodesic length on the new space of quantum states (events).
The relation $\hbar\, ds = 2 \Delta E\, dt$ gives meaning to the ``evolution parameter'' $t$!
Notably different metrics imply different ``evolution parameters'' with $t$ relational and akin to the ``multifinger time'' of General Relativity \cite{mtw}.
Given the $X$ space, we can introduce a natural $\Diff(1,\BC)$ map, $X \to f(X)$.
The $\Diff(1,\BC)$ identification of the points on the submanifold determined by the ``Born rule'' defines the generalized projective Hilbert manifold.

\item The observables are functions of the natural distance on the quantum phase space $h_{ab} X^{a} X^{b}$, $O = O(h_{ab} X^{a} X^{b})$.
They reduce to the usual ones when the Riemannian structure is canonical.
More explicitly
\begin{equation}
O = \sum_{a,b} o_{a} h_{ab} X^a X^b
\end{equation}
where the ``eigenvalue'' $o_a$ is given as (see \cite{weinberg})
\begin{equation}
{ d O \over d X^a}= o_a \omega_{ab} X^b.
\end{equation}
Here the symplectic form $\omega^{ab}$ as well as $O$ depend on the invariant combination $h_{ab} X^{a} X^{b}$.

\item The temporal evolution equation reads
\begin{equation}
{d u^a \over d\tau} + \Gamma^{a}_{bc} u^b u^c = \frac{1}{2 \Delta E}{\rm Tr}(H {\cal{F}}^{a}_{b}) u^b
\end{equation}
where now $\tau$ is given through the metric $\hbar\, d\tau = 2 \Delta E\, dt$, as in the original work of Aharonov and Anandan \cite{anandan}.
Note that ultimately we can generalize the line element so that the energy uncertainty is
measured in terms of a fundamental energy scale, the Planck energy ($E_{\rm Pl}$), so that
$\hbar\, d\tau = 2 E_{\rm Pl}\, dt$. Note that
$\Gamma^{a}_{bc}$ is the affine connection associated with this general metric ${\cal{G}}_{ab}$ and ${\cal{F}}_{ab}$ is a general curvature two-form in $\Diff(\infty-1,\BC) \times \Diff(1,\BC)$.
(Here we have adopted a stylized notation to indicate the specific generalization being considered.)
\end{enumerate}

Next we reformulate geometric quantum mechanics in the above background independent setting.
Due to the $\Diff(\infty,\BC)$ symmetry, ``coordinates'' $z^a$ ({\em i.e.}, quantum states themselves) make no sense physically, only quantum events do, which is the quantum counterpart of the corresponding statement on the meaning of spacetime events in General Relativity.
Probability is generalized and given by the notion of diffeomorphism invariant distance in the space of quantum configurations.
The dynamical equation is a geodesic equation on this space.
Time, the evolution parameter in the generalized Schr\"{o}dinger equation, is not global and is given in terms of the invariant distance.
Our basic starting point of a background independent quantum mechanics is to notice that the evolution equation (the generalized Schr\"{o}dinger equation) as a geodesic equation, can be derived from an Einstein-like equation with the energy-momentum tensor determined by the holonomic non-abelian field strength ${\cal{F}}_{ab}$ of the $\Diff(\infty-1,\BC) \times \Diff(1,\BC)$ type and the interpretation of the Hamiltonian as a ``charge''.
Such an extrapolation is logical since $\CP{n}$ is an Einstein space; its metric obeying Einstein's equation with a positive cosmological constant given by $\hbar$:
${\cal{R}}_{ab} - \frac{1}{2} {\cal{G}}_{ab} {\cal{R}}  - \lambda {\cal{G}}_{ab}=0$.
The Ricci curvature of $\CP{n}$ is
${\cal{R}}_{ab} \equiv \frac{n+1}{\hbar} {\cal{G}}_{ab} = \frac{1}{2} c (n+1) {\cal{G}}_{ab}$,
where $c$ is the constant holomorphic sectional curvature of $\CP{n}$ given by $c=\frac{2}{\hbar}$.

The geodesic equation follows from the conservation of the energy-momentum tensor
$
\nabla_a {\cal{T}}^{ab} =0
$
with
${\cal{T}}_{ab} = {\rm Tr}({\cal{F}}^{ac}{\cal{G}}_{cd}{\cal{F}}^{cb} -\frac{1}{4} {\cal{G}}_{ab} {\cal{F}}_{cd}{\cal{F}}^{cd}
+ \frac{1}{2\Delta E}H u_a u_b)
$
by way of the usual argument in General Relativity (see {\em e.g.}, \cite{mtw}, chapter 20).
With quantum gravity in mind, we set $\Delta E$ to the Planck energy $E_{\rm Pl}$, the proper deformation parameter.
When $E_{\rm Pl} \to \infty$ we recover the usual flat metric on the Hilbert space or the Fubini--Study metric on the projective Hilbert space.
Since both the metrical and symplectic data are also contained in $H$, we have here the advertised non-linear ``bootstrap'' between the space of quantum events and the dynamics.
The diffeomorphism invariance of the new phase space suggests the following dynamical scheme for the background independent quantum mechanics:
\begin{equation}
\label{BIQM1}
{\cal{R}}_{ab} - \frac{1}{2} {\cal{G}}_{ab} {\cal{R}}  - \lambda {\cal{G}}_{ab}= {\cal{T}}_{ab}
\end{equation}
with ${\cal{T}}_{ab}$ given as above (as determined by ${\cal{F}}_{ab}$ and the Hamiltonian (``charge'') $H$).
Furthermore
\begin{equation}
\label{BIQM2}
\nabla_a {\cal{F}}^{ab} = \frac{1}{2\Delta E} H u^b.
\end{equation}
The last two equations imply via the Bianchi identity a conserved energy-momentum tensor, $\nabla_a {\cal{T}}^{ab} =0$.
The latter, taken together with the conserved ``current'' $j^b \equiv \frac{1}{2\Delta E} H u^b$, {\em i.e.}, $\nabla_a j^a =0$, implies the generalized geodesic Schr\"{o}dinger equation.
So (\ref{BIQM1}) and (\ref{BIQM2}), being a closed system of equations for the metric and symplectic form on the space of events, define our background independent quantum mechanics.
We emphasize once again that in the limit $E_{\rm Pl} \to \infty$  we recover the usual structure of linear quantum mechanics.
Moreover this limit does not affect the geodesic equation
${d u^a \over d\tau} + \Gamma^{a}_{bc} u^b u^c = \frac{1}{2 \Delta E}{\rm Tr}(H {\cal{F}}^{a}_{b}) u^b$
due to the relation
$\hbar\, d\tau = 2 \Delta E\, dt$.
As such our formulation offers a tantalizing non-linear linkage between the metric and symplectic data embodied in $H$ and the quantum metric and symplectic data.
The space of quantum events is {\em dynamical} paralleling the dynamical r\^{o}le of spacetime in General Relativity, as opposed to the rigid, absolute state space of standard quantum mechanics.
This is then, in our view, the price of quantum background independence.
To draw more concrete consequences of this kinematics made dynamical, we next specify a quantum system with its $H$.
The configuration space of the quantum metric defines a new form of
``superspace'' (as in canonical General Relativity \cite{super}) and the dynamics on it presumably select a particular background.
Note that this formulation of general background independent
geometric quantum theory essentially repeats the lesson of General Relativity, as a gauged
theory of the Lorentz group, so we indeed end up with a gauged quantum theory, where
now the unitary group of canonical quantum theory is being gauged.
Thus problems with violations of unitarity are avoided in this
very general formulation of background independent quantum theory.

We now demand that the configuration space metric be the actual physical {\it spatial} metric.
The suitable quantum system must then have a very special configuration space and should describe a quantum theory of gravity.
Specifically, we seek a canonical quantum mechanics of a non-perturbative form of quantum gravity in a fixed background, with a well defined perturbative limit and a configuration space being the actual space.
The only example we know fulfilling these criteria is Matrix theory \cite{matrix}.
(The latter is also ``holographic'' \cite{holog}, in the sense of mean-field theory.\footnote{
The relationship between holography, unitarity, and diffeomorphism invariance was explored in \cite{rob}.})
As with other roads to quantum gravity, Matrix theory which leaves quantum mechanics intact suffers from the problem of background dependence \cite{seiberg, mback}.\footnote{
We should mention here that different arguments for revising quantum mechanics in the framework of quantum gravity have been advanced for example in \cite{thooft, penrose, holobdbm}.}

In implementing our scheme, we assume that the metric on the transverse space is encoded in the metric on the quantum state space.
Then we take the Matrix theory Hamiltonian in an arbitrary background and insert it into the defining equations of the above background independent quantum mechanics.
The evolution of our system then reads
$
{d u^a \over d\tau} + \Gamma^{a}_{bc} u^b u^c = \frac{1}{2 E_{\rm Pl}} H_M {\cal{F}}^a_b u^b
$
where $H_M$ is the Matrix Hamiltonian ($i,j$ denote the transverse space indices ($i=1,...,9$), and $R_{11}$ is the extent of the longitudinal eleventh direction):
\begin{equation}
H_M = R_{11} {\rm Tr}(\frac{1}{2} P^i P^j G_{ij}(Y) + \frac{1}{4}[Y^i, Y^l][Y^k, Y^j] G_{ij}(Y) G_{lk}(Y)) + {\rm fermions}.
\end{equation}
Here $P^i$ is the conjugate momentum to $Y^i$ ($N\times N$ Hermitian matrices) given a symplectic form $\omega$.
(We adopt the symmetric ordering of matrices, see \cite{mback}.)
{\it Given this expression for $H_M$, the general equations (\ref{BIQM1}) and (\ref{BIQM2}) then define a background independent Matrix theory.}
Note that in (\ref{BIQM1}) and (\ref{BIQM2}) $a,b$ denote the indices on the quantum space of states, whose span is determined by the dimension of the Hilbert space of Matrix theory, given in terms of $N$.

The time coordinate of background independent Matrix theory is manifestly not global, but is defined by the invariant distance on the space of quantum events.
The light-front (light-cone) $SO(9)$ symmetry is only ``local'' (in the sense of the generalized quantum phase space).
Supersymmetry is generally broken since generically the background will not admit globally defined supercharges.
Only ``locally'' (again in the sense of the generalized quantum space) may we talk about the correspondence between the moduli space of the Matrix theory supersymmetric quantum mechanics and the transverse space \cite{matrix}.

The longitudinal coordinate, and longitudinal momentum specified in terms of the ratio $N/R_{11}$ \cite{matrix} can be made dynamical in our proposal.
The rank $N$ of the matrices implicitly defines the size of the Hilbert space, which is seemingly fixed (the dimension of the index space is fixed).
On the other hand, one of the fundamental features of Matrix theory is that of being automatically second quantized;
it encodes the Fock space $\{n_k \}$ in terms of block diagonal $n_k \times n_k$ matrices \cite{matrix}.
Taking a cue from this defining feature, we promote the points on the quantum phase space into Hermitian matrices.
This is the final ingredient in our proposal.
In practice, the $u^a$s appear as Hermitian matrices in the defining equations (\ref{BIQM1}) and (\ref{BIQM2}).
So the rank of matrix-valued non-commuting transverse coordinates $Y^i$ ($N$) is made dynamical by turning the ``coordinates'' $z^a$ of our background independent quantum phase space into non-commutative objects.
The asymptotic causal structure (and thus a {\it covariant background independent structure}) only emerges in the Matrix theory limit \cite{matrix}, $N\to \infty$, $R \to \infty$ while keeping $N/R$ fixed.
The above defining dynamical equations (\ref{BIQM1}) and (\ref{BIQM2}) can also be cast in the context of Connes' non-commutative geometry \cite{connes}.

The gist of our proposal lies in the non-linear interconnection between the metric ($G_{ij}$) and symplectic data ($\Omega_{ij}$) contained in the Hamiltonian $H$ and the quantum metric (${\cal{G}}_{ab}$) and symplectic data ($\omega_{ab}$, or equivalently, ${\cal{F}}_{ab}$).
This non-linear connection may well explain how
(a) different degrees of freedom are associated to different backgrounds, and
(b) how the observed four-dimensional spacetime background dynamically emerges in Matrix theory,
the pre-geometry being the dynamical stochastic geometry of the space of events.
Furthermore we can't but ponder the fascinating possibility that the very form of the Matrix theory
Hamiltonian $H_M$ is already encoded in the non-trivial topological structure of the space of quantum events.
This may be so if the latter manifold is non-simply connected and is non-commutative.\footnote{This in complete analogy with the concept of ``charge without charge'' of the Einstein--Maxwell system of equations in vacuum, as discussed by Misner and Wheeler \cite{mw}.}

\subsection{Global Structure and ${\rm Gr}(\BC^{n})$}

We may recast standard quantum mechanics in the language of complex geometry by way of only two compatible postulates.
The latter show that, just as thermodynamics, Special Relativity, and General Relativity, quantum mechanics, in spite of its appearance, belongs in Einstein's categorization to ``theories of principles'' \cite{stachel}.
These two postulates to be stated below form a physically more intuitive rendition of the mathematical axioms of Landsman \cite{landsman}.
They make manifest the very rigid structure of the underlying state space (the space of quantum events), the complex projective space $\CP{n}$.
As such they also underscore the relational \cite{qrel} and information theoretic nature of quantum theory \cite{wootters}.
Most importantly, this perspective points to a possible extension of quantum mechanics along the line discussed in \cite{tzeminic}, one relevant to a background independent formulation of quantum gravity.
Such a generalization is achieved, in analogy to what is done with the spacetime structure in General Relativity, in a two-fold way:
firstly, by relaxing the integrable complex structure of the space of events, and
secondly, by making this very space of events (that is, its metric and symplectic and therefore its almost complex structure), the arena of quantum dynamics, into a dynamical entity in its own right.
One of the byproducts of such an extended quantum theory is the notion of an intrinsic, probabilistic local time.
This quantum time is rooted in the strictly almost K\"ahler geometry of a dynamically evolving, diffeomorphism invariant state space of events.
In physical terms, this non-integrable almost complex structure implies a relaxation of the absolute global time of quantum mechanics to an intrinsic, relative, local time.
This novel feature is, in our view, the key missing conceptual ingredient in the usual approaches to the background independent formulation of quantum gravity.
The main technical thrust of the present paper is contained in a set of very recent mathematical results of Haller and Vizman \cite{vizman} concerning the category of infinite dimensional almost K\"ahler manifolds which, in our view, naturally replaces the category of complex projective spaces of standard quantum mechanics.
These results enable us to significantly sharpen the geometric formulation of our previous more heuristic proposal \cite{tzeminic}.

First we recall that, among their many available formulations, the axioms of standard quantum mechanics \cite{auletta} can take a very elegant, simple $C^*$-algebraic form \cite{landsman}.
The Landsman formulation offers an {\em unified} view of both quantum and classical mechanics thereby suggesting its structural closeness to geometric quantum mechanics \cite{geomqm}.
We recall that in the latter setting, a quantum system is described by an infinite classical Hamiltonian system, albeit one with very specific K\"ahler constraints.
Here, we seize on this formal closeness by providing the physical, geometric counterparts of these Landsman axioms.
Paraphrasing \cite{landsman}, the first axiom states that the space of pure states is a Poisson space with a transition probability.
More precisely the definition of the Poisson bracket is exactly that of geometric quantum mechanics \cite{geomqm}.
Then, as detailed in Landsman's book \cite{landsman}, the first axiom says that the essential physical information is carried by a well defined symplectic ({\em i.e.}, a non-degenerate symplectic two-form) and metrical structures on the space of states.
The second axiom further specifies the transition probability to be that of standard quantum mechanics, namely the metric information of the Cayley--Fubini--Study type \cite{geomqm}, the natural, unique metric on $\CP{n}$.
(The third axiom deals with superselection sectors, which, for simplicity we do not concern ourselves with here.).
It suffices to say that Landsman's axioms can be shown \cite{landsman} to imply the usual geometric structure of quantum mechanics, in particular the uniqueness of $\CP{n}$ as the space of pure states.

Moreover the Landsman axioms as translated above can be understood in the following physically more intuitive manner.
To do so, we first recall Bohr's dictum that ``(quantum) physical phenomena are observed {\it relative} to different experimental setups'' \cite{jammer}.
This statement closely parallels the r\^{o}le that inertial reference frames play in relativity theory.
More accurately, as paraphrased by Jammer \cite{jammer}, this viewpoint reads:
``just as the choice of a different frame of reference in relativity affects the result of a particular measurement,
so also in quantum mechanics the choice of a different experimental setup has its effect on measurements, for it determines what is measurable.''
Thus while the observer does choose what to observe by way of a particular experimental setup, he or she cannot influence quantitatively the measured value of a particular observable.
Thus in analogy with the postulates of Special Relativity and in the place of Landsman's axioms, we propose the following two quantum postulates:

\noindent
(I.) The laws of physics are invariant under the choice of the experimental set up.
Mathematically, we thus prescribe that, as in classical mechanics, there is a well-defined symplectic structure which stands for the classical kinematical features of the measurement process.

To expand on this Postulate I, we should first note that, in a broader setting, it actually allows for a general Poisson structure.
However, by confining for simplicity, to a theory with no selection rules, we thus restrict ourselves to a symplectic structure.
Now the classical symplectic structure is an inherent property that comes with the measurement device whose readings are then statistically analyzed in the sense of statistical inference theory.
That a measurement device always comes together with a symplectic structure can be seen as follows.
Take a system on which we perform physical measurements.
It is described by a certain Hamiltonian (or Lagrangian) so that the classical dynamics can be well defined.
Consider a coupling of this system to another one, a measurement device, so that both the interaction Hamiltonian and the Hamiltonian of the measurement device are in principle known.
(This is the classic setup considered for example in the literature on decoherence \cite{decoh}.)
The measurement process is then in principle described by the interaction part of the total Hamiltonian.
Knowing the Hamiltonian assumes knowledge of a well-defined symplectic structure.
Thus the existence of a symplectic structure is an intrinsic property that comes with a measuring setup.
So the first postulate asserts the existence of a natural classical closed symplectic two-form $\Omega$,
$
d\Omega=0,
$ as well as the associated canonical Hamiltonian flow ({\em i.e.}, dynamical equations of motion).
Namely, the state space is an even dimensional symplectic Poisson manifold.
This is the mathematical rendition of Postulate I.

Next we make a principle out of another dictum of Bohr and his school on the existence of {\em primary} probabilities in Nature:

\noindent
(II.) Every quantum observation (reading of a given measurement device) or quantum event, is irreducibly statistical in nature.
These events, being distinguishable by measurements, form points of a statistical (informational) metric space.
There is then a natural, unique (maximally symmetric) statistical distance function on this space of quantum events, the famous Fisher distance \cite{wootters} of statistical inference theory \cite{arima}.

More precisely, from the seminal work of Wootters \cite{wootters}, a natural statistical distance on the space of quantum events is uniquely determined by the size of statistical fluctuations occurring in measurements performed to tell one event from another.
This distance between two statistical events is given in terms of the number of distinguishable events, thus forming a space with the associated Riemannian metric
$ds^2 \equiv  \sum_i \frac{dp_i^2}{p_i} = \sum dX_i^2$, where $p_i \equiv X_i^2 $ denote individual probabilities.
This distance in the probability space is nothing but the celebrated Fisher distance of information theory and can be rewritten as \cite{wootters}
\begin{equation}
ds_{12} = {\cos}^{-1}(\sum_i \sqrt{p_{1i}} \sqrt{p_{2i}}).
\end{equation}
This is the mathematical content of our Postulate II.

Our principles (I) and (II) as stated above clearly display quantum theory as what might be called a Special Theory of Quantum Relativity.
It is then only natural to take the next logical step, to go beyond this and formulate a General Theory of Quantum Relativity as well.
This extension is accomplished by allowing both the metric and symplectic form on the space of quantum events to be no longer rigid but fully dynamical entities.
In this process, just as in the case of spacetime in General Relativity, the space of quantum events becomes dynamical and only individual quantum events make sense
observationally.

Specifically, we do so by relaxing our Postulate II to allow for {\it any} statistical (information) metric all the while insisting on the compatibility of this metric with the symplectic structure underlying our Postulate I.
Physics is therefore required to be diffeomorphism invariant in the sense of information geometry \cite{arima} such that the information geometric and symplectic structures remain compatible, requiring only a {\em strictly} ({\em i.e.}, non-integrable) almost complex structure $J$.
Once we relax Postulate II, so that any information metric is allowed, the relativity of canonical quantum mechanical experiments (such as the double-slit experiment) becomes possible and would provide an experimental test of our proposal.

Our extended framework readily implies that the wave functions labeling the event space, while still unobservable, are no longer relevant.
They are in fact as meaningless as coordinates in General Relativity.
There are no longer issues related to reductions of wave packets and associated measurement problems.
At the basic level of our scheme, there are only dynamical correlations of quantum events.
 From the previous analysis and in the spirit of constructing an {\em ab initio} quantum theory %
of matter and gravity, we can enumerate the main structural features one may want in such a scheme for the space of quantum events:
\begin{itemize}
\item it must have a symplectic structure;
\item it must be strictly almost K\"ahler;
\item it must be the base space of a $U(1)$ bundle; and
\item it must be diffeomorphism invariant.
\end{itemize}
We recall that the state space $\CP{\infty}$ is a linear Grassmannian manifold, $\CP{n}$ being the space of complex lines in $\BC^{n+1}$ passing through the origin.
We seek a coset of $\Diff(\BC^{n+1})$ such that locally looks like $\CP{n}$ and allows for a compatibility of the metric and symplectic structures, expressed in the existence of a (generally non-integrable) almost complex structure.

The following non-linear Grassmannian
\begin{equation}
{\rm Gr}(\BC^{n+1}) = \Diff(\BC^{n+1})/\Diff(\BC^{n+1},\BC^n \times \{0\}),
\end{equation}
with $n = \infty$ satisfies the above requirements,
thus sharpening the geometrical information of the na\"{\i}ve proposal
(concerning the coset state space $\Diff(n, \BC)/\Diff(n-1,\BC) \times \Diff(1,\BC)$)
made in the beginning of this section!

Indeed this infinite (even for finite $n$) dimensional space ${\rm Gr}(\BC^{n+1})$ is modeled on a Frechet space.
Very recently, its study was initiated by Haller and Vizman \cite{vizman}.
Firstly it is a non-linear analog of a complex Grassmannian since it is the space of (real) co-dimension two submanifolds, namely a hyperplane $\BC^n \times [0]$ passing through the origin in $\BC^{n+1}$.
Its holonomy group $\Diff(\BC^{n+1}, \BC^n\times \{0\})$ is the group of diffeomorphisms preserving the hyperplane $\BC^n \times \{0\}$ in $\BC^{n+1}$.
Just as $\CP{n}$ is a coadjoint orbit of $U(n+1)$, ${\rm Gr}(\BC^{n+1})$ is a coadjoint orbit of the group of volume preserving diffeomorphisms of $\BC^{n+1}$.
As such it is a symplectic manifold with a canonical Kirillov--Kostant--Souriau symplectic two form $\Omega$ which is closed ($d\Omega=0$) but not exact.
Indeed the latter two-form integrated over the submanifold is non-zero;
its de Rham cohomology class is integral.
This means that there is a principal one-sphere, a $U(1)$ or line bundle over ${\rm Gr}(\BC^{n+1})$ with curvature $\Omega$.
This is the counterpart of the $U(1)$-bundle of $S^{2n+1}$ over $\CP{n}$ of quantum mechanics.
It is also known that there is an almost complex structure given by a $90^\circ$ rotation in the two-dimensional normal bundle to the submanifold.
While $\CP{n}$ has an integrable almost complex structure and is therefore a complex manifold, in fact a K\"ahler manifold, this is {\em not} the case with ${\rm Gr}(\BC^{n+1})$.
Its almost complex structure $J$ is by a theorem of Lempert \cite{lempert} strictly {\em not} integrable in spite of its formally vanishing Nijenhius tensor.
While the vanishing of the latter implies integrability in the finite dimensional case, one can no longer draw such a conclusion in the infinite dimensional Frechet space setting.
However what we do have in ${\rm Gr}(\BC^{n+1})$ is a strictly ({\em i.e.}, non-K\"ahler) almost K\"ahler manifold \cite{almost} since there is by way of the almost complex structure $J$ a compatibility between the closed symplectic two-form $\Omega$ and the Riemannian metric $g$ which {\em locally} is given by $g^{-1}\Omega = J$.\footnote{
It would be very interesting to understand how unique is the structure of ${\rm Gr}(\BC^{n+1})$.}

Next, just as in standard geometric quantum mechanics, the probabilistic interpretation lies in the definition of geodesic length on the new space of quantum states (events) as we have emphasized before \cite{tzeminic, geomqm}.
Notably since ${\rm Gr}(\BC^{n+1})$ is only strictly almost complex, {\em i.e.}, its $J$ is only locally complex.
This fact translates into the existence of only local time and local metric on the space of quantum events.
These local dynamical equations are precisely the ones one of us has proposed in a previous paper \cite{tzeminic}.
The fact that the space of quantum events should be ${\rm Gr}(\BC^{n+1})$ sharpens the global geometric structure of our proposal.
As in General Relativity it will be crucial to understand the global features of various solutions to the above dynamical equations.

Finally, we have argued above, that the form of $H$ (the Matrix theory Hamiltonian in an arbitrary background), viewed as a ``charge''  may be determined in a quantum theory of gravity by being encoded in the non-trivial topology of the space of quantum events.
This may well be the case here with our non linear Grassmannian which is non-simply connected \cite{vizman}.
However definite answers to this and many other more concrete questions must wait until greater details 
are known on the topology and differential geometry ({\em e.g.}, invariants, curvatures, geodesics) of ${\rm Gr}(\BC^{n+1})$.
In the meantime we hope to have laid down here the conceptual and mathematical foundations of what may
be called a General Theory of Quantum Relativity in which its fundamental kinematical structure follows
from its dynamical structure.

\subsection{A Comprehensive View}
\label{sec:acv}

We now consider the issue of quantum gravity in a broader context.
Our goal in doing this is two-fold.
We wish to demonstrate that there is a natural and consistent way of relating different views of spacetime.
Also taking a more global viewpoint can provide greater insight into the nature of the difficulties quantum gravity presents.
As we shall see the proposal in this review offers viable and perhaps compelling answers to these open questions.

\begin{figure}[h]
\begin{center}
\epsfysize=3.5in
\epsfxsize=3.5in
\epsffile{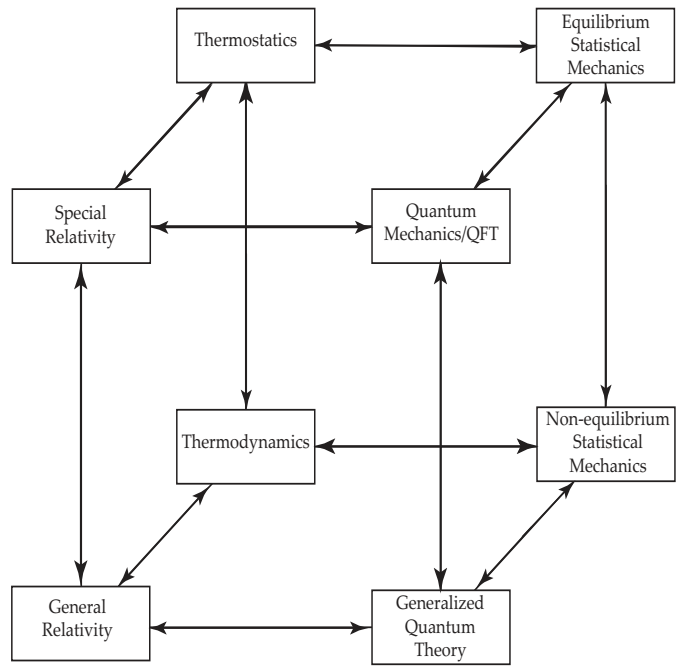}
\caption{Toward a generalized quantum theory.}
\label{fig:cube}
\end{center}
\end{figure}

A schematic overview of how we view spacetime is contained in Figure~\ref{fig:cube}.
The corners of the bottom face of the cube provide generalizations of the top face.
Special Relativity provides a unifying framework of spacetime by applying the principle of relativity to inertial frames.
General Relativity renders this dynamical by invoking the principle of equivalence.
Spacetime in Special Relativity is globally Minkowski space whereas General Relativity, which invokes the mathematics of differential geometry, only requires flat space locally.
Thus transitioning from Special Relativity to General Relativity entails gauging the Lorentz group $SO(3,1)$ to the general diffeomorphism group $\Diff(\cal{M})$.
Just as we generalize Special Relativity to General Relativity via the gauge principle, so too do we generalize quantum mechanics by treating it geometrically.
We may transition from classical mechanics to quantum mechanics by defining a quantum phase space.
Classical phase space carries the standard symplectic structure while quantum phase space possesses a Riemannian metric as well as a symplectic structure.
The progression from canonical quantum mechanics to generalized quantum mechanics parallels the transition from Special to General Relativity.
As we have stressed, the generalization of quantum mechanics is accomplished by means of a quantum version of the equivalence principle.
We have established that the space of quantum events in standard quantum mechanics is $\CP{n}$.
In transitioning to generalized quantum mechanics $\CP{n}$ mirrors the r\^{o}le of Minkowski space.
As previously discussed ${\rm Gr}(\BC^{n+1})$ is locally $\CP{n}$ in the $n\rightarrow\infty$ limit, and thus we choose this as our generalized space of quantum events.
Moreover, we may gauge the unitary group $U(n+1)$ in analogy with the gauging of the Lorentz group in Special Relativity in order to properly generalize quantum mechanics.

We have seen that the Einstein equation emerges as an equation of state upon application of the laws of black hole thermodynamics and the Raychaudhuri (focusing) equation.
Thermostatics describes the thermodynamics of systems in exact equilibrium.
It is essentially a theory of only the first law (conservation of energy and the conversion of one form of energy to another).
The dynamics is introduced when we invoke the second law as well.
Thus thermostatics is the analogue of Special Relativity, while thermodynamics is the analogue of General Relativity.

The connection between quantum mechanics and equilibrium statistical mechanics has long been understood
as they are both fundamentally statistical theories.
There is a direct correspondence between quantum field theory and equilibrium statistical mechanics with a Gibbsian measure.
This can be seen most clearly by considering the generating functional of correlation functions in quantum field theory and the partition function in equilibrium  statistical mechanics.
We begin with the generating functional of correlation functions,
\begin{equation}
Z[J]= \int {\cal{D}} \phi \, \exp \big[ {i} \int  d^4x\,({\cal{L}}+J\phi) \big].
\end{equation}
Now we Euclideanize it through a Wick rotation $t\rightarrow -ix^0$.
This yields an expression equivalent to the partition function describing the equilibrium statistical mechanics of a macroscopic system,
\begin{equation}
Z[J]=\int {\cal{D}} \phi \, \exp\big[-\int d^4x_E\, ({\cal{L}}_E+J\phi) \big].
\end{equation}
Here the subscript indicates the transition to Euclidean space, and $J$, which is a source in the setting of quantum field theory and is an external field in the context of statistical mechanics.
(For further discussion, see, for example, \cite{pesky}.)

Now we consider the transition to non-equilibrium statistical mechanics.
As indicated in \cite{ted2} inclusion of quantum gravitational corrections necessitates the use of non-equilibrium thermodynamics.
Correspondingly, this has a profound implication from the quantum mechanical perspective.
The gauging of the unitary group means that in general we do not have path integrals, and the aforementioned analogy between quantum field theory and equilibrium statistical mechanics with a Gibbsian measure implies that the generalized quantum theory should be of non-equilibrium type.
There is no real meaning assigned to the states (wave functions), but there is nevertheless a general dynamical statistical geometry of quantum theory.

Thus each of the view points on spacetime can be related in a fundamental way to the others. In addition certain aspects of the internal structure of each perspective are similar.
Each of the formalisms contain a highly constrained theory which may be generalized such that the constrained theory holds only locally.
Thus we are left with a more cohesive picture of how the various views of spacetime explored in this review are related, and we may draw on the relationships between the various formalisms in order to come to a more complete understanding of each one individually.
The gauge principle explains how the ``static'' situations that arise along the top face of the cube to the ``dynamic'' condition that underlies this physics.
In going from the top face to the bottom face, the conception of spacetime is dramatically altered.

\begin{figure}[h]
\begin{center}
\epsfysize=4in
\epsfxsize=4in
\epsffile{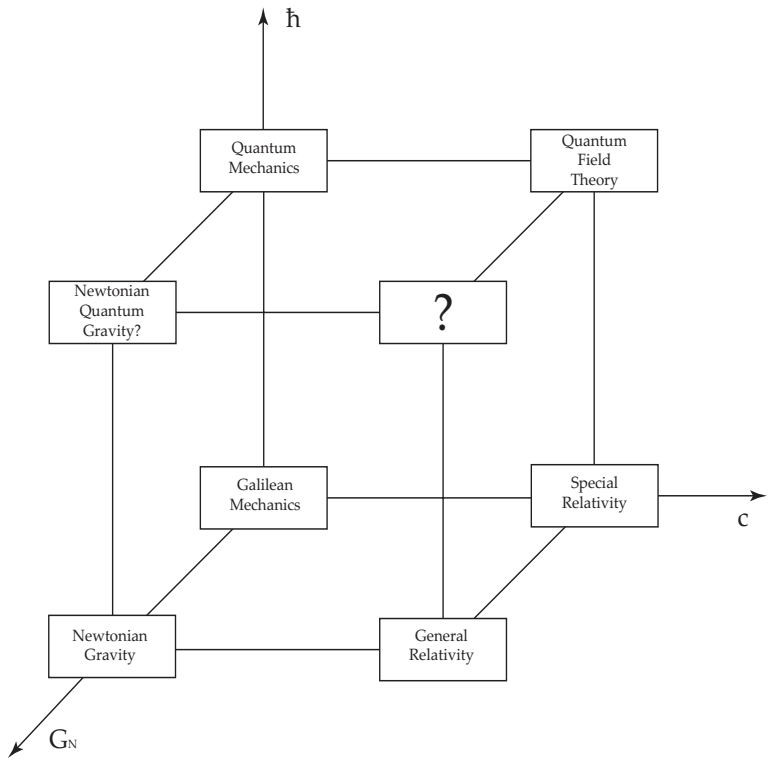}
\caption{What is string theory?}
\label{fig:cube2}
\end{center}
\end{figure}

We claim an identification between string theory as the fundamental theory of quantum gravity and the generalized quantum mechanics necessary to reconcile the dynamical nature of spacetime geometry with the low-energy description of Nature as a quantum field theory.
The path to string theory is defined in Figure~\ref{fig:cube2}.

We recall that the Planck length that defines the fundamental physical scale in Nature is $\ell_{\rm Pl} = \sqrt{\hbar G_N/c^3}$.
We can imagine turning on the parameters $G_N$, $1/c$, and $\hbar$ individually.
Finite values of these deformation parameters give rise to Newtonian gravity, the Special Theory of Relativity, and quantum mechanics, respectively.
With both $c$ and $\hbar$ present, we have a synthesis of quantum mechanics and Special Relativity, namely quantum field theory.
Likewise, by adding to Special Relativity the principle of equivalence, that there is no way to distinguish locally between gravity and acceleration, we have a theory with $c$ and $G_N$ turned on, which is couched in the language of differential geometry; this is of course the General Theory of Relativity.
Presumably, there is a theory with the speed of light infinite that has only $G_N$ and $\hbar$ turned on.
This is a non-relativistic, or Newtonian, theory of quantum gravity.
The theory of quantum gravity with finite values of $G_N$, $c$, and $\hbar$ should be
identified with
non-perturbative string theory, defined by the
proposal discussed in this review.

\begin{figure}[h]
\begin{center}
\epsfysize=4in
\epsfxsize=4in
\epsffile{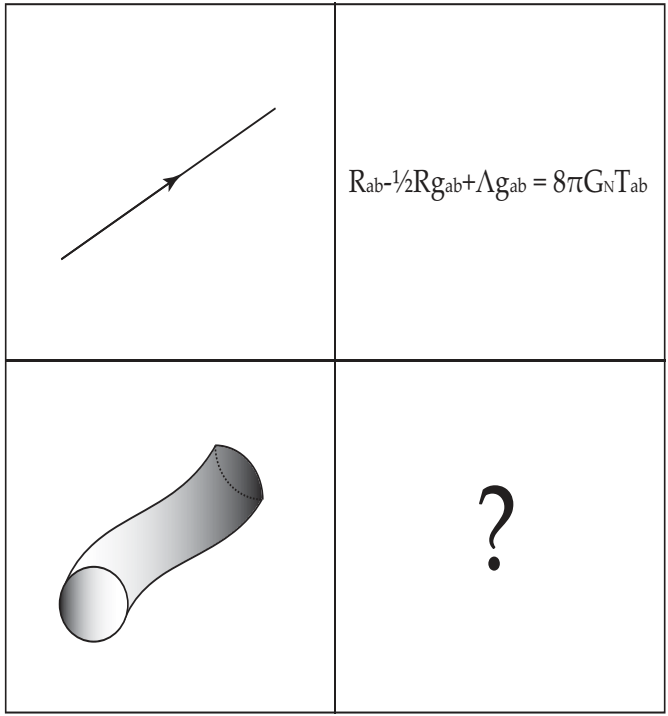}
\caption{Equations of motion.}
\label{fig:ms}
\end{center}
\end{figure}

Though familiar four-dimensional quantum field theories are low-energy theories associated to compactifications of string theory and the Einstein equations emerge as a characterization of the background on which the string consistently propagates, there is no string equation of motion in the sense that it exists for General Relativity and quantum mechanics.
As Figure~\ref{fig:ms} illustrates, string theory is by nature a different type of theory than the General Theory of Relativity,
even though we have argued that it should be understood as a quantum background independent Matrix theory,
constructed by emulating the structure of General Relativity.

\begin{figure}[h]
\begin{center}
\epsfysize=3.5in
\epsfxsize=3.5in
\epsffile{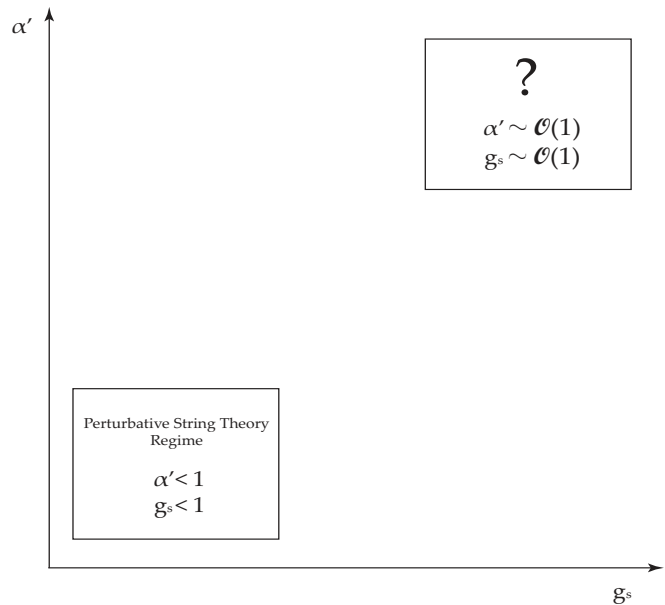}
\caption{Beyond perturbation theory.}
\label{fig:fig3}
\end{center}
\end{figure}

String theory is as well only understood in a perturbative regime, where the string coupling $g_s$ is small and the string scale $1/\sqrt{\alpha'}$, which characterizes the mass of the tower of string excitations, is large.
In this review we have provided
hints at what the non-perturbative theory (M-theory) is, but
we have given no detailed understanding of the strong coupling regime where $g_s \gg 1$ and $\alpha' \gg 1$.
Figure~\ref{fig:fig3} highlights our technical ignorance of these matters.

In this review we have argued that the formulation of string theory consistent with general, {\em a priori} unspecified spacetime asymptopia and dynamical causal structure demands that we generalize the canonical framework of quantum mechanics, which is absolutely rigid.
The hallmarks of string theory --- holography, the UV/IR correspondence, the absence of free parameters --- must be incorporated into the generalized theory of quantum mechanics.
This we accomplish by insisting that the physics at any point is Matrix theory, which supplies a non-perturbative definition of M-theory (string theory).
The dynamics emerge in patching together the physics at local neighborhoods about points in spacetime.
The kinematical structure of the theory is determined by its dynamics, and space, time, and matter appear as emergent concepts.

As we have observed, the various views of spacetime outlined in this review crystallize into a cohesive picture.
This provides a way to more clearly understand the problems inherent in developing a theory of quantum gravity.
We believe in light of this analysis perhaps a novel approach to this issue is not simply reasonable but required.
We believe that we have presented such an approach which offers natural solutions to each of the questions raised.
In the next section we ask whether our rather abstract proposal has any observational implications.

\section{Implications: The Problem of Dark Energy}

The most important physical implication of the background independent Matrix theory is
in one of its fundamental motivations mentioned in the introduction of this
review --- the problem of vacuum energy, {\em i.e.}, the cosmological constant problem.
Recently we have presented a new viewpoint on the cosmological constant problem based upon
the background independent Matrix quantum theory. This argument also has ramifications
for the possibly observable spectrum of dark energy, both of which we review in this section.

\subsection{The Cosmological Constant}

The new viewpoint on the cosmological constant problem runs as follows:
\begin{enumerate}
\item First, from the quantum diffeomorphism invariance the expectation value of the vacuum energy is zero.
This is to be compared with the red-shift formula in General Relativity which follows from the diffeomorphism invariance of the theory.
Starting from $E=h\nu$, the red-shift due to a mass $M$ gives
\be
E_{\rm corrected} = h\nu(1 - M/R)
\ee
as the photon climbs out of a potential well of characteristic size $R$.
For a closed universe $E_{\rm corrected} = 0$.
This is a statement of the familiar equivalence principle.
We argue that diffeomorphism invariance in the space of quantum configurations of the system leads to a red-shift of the zero-point energy.
This quantum diffeomorphism invariance, captured by a quantum equivalence principle, means that $E_{\rm vacuum} = \sum\frac12 \hbar\omega$ is ``red-shifted'' as
\be
E_{\rm vacuum} = \sum \frac12 \hbar\omega (1 - M_{\rm Pl}/R).
\ee
For a closed universe $E_{\rm vacuum} = 0$.

\item In general our extended background independent geometric quantum theory is non-linear (because the metric is dynamical, not fixed, unlike the usual canonical quantum theory) and non-local (as Matrix theory uses non-commuting matrices).
It is difficult to compute in the framework of a non-linear, non-local, probabilistic theory.
To a first approximation we expand around the standard Fisher--Fubini--Study metric of complex projective spaces.
Non-linear corrections to the Schr\"odinger equation are written as a geodesic equation in the configuration space.
We may interpret this non-linear Schr\"odinger equation from the point of view of third quantization and view it as a non-linear Wheeler--de-Witt equation.
Vacuum energy is a dynamical variable from the context of ordinary quantization.
The relevant coupling constant that becomes third quantized is $\Lambda$ or the vacuum energy density in the canonical quantum theory limit.

\item The vacuum energy density $\Lambda$ is dynamical and fluctuates around zero (because this value is fixed by quantum diffeomorphism invariance).
We use the large volume approximation of the non-linear Wheeler--de-Witt equation with $\Lambda$ non-zero;
$\Lambda$ and the volume of spacetime are here conjugate quantities and realize an uncertainty relation:
\be
\Delta\Lambda\, \Delta V \sim \hbar.
\label{unc}
\ee
Here, $\Lambda$ is an ``energy'', while the observed volume of spacetime is ``time.''
This point is elaborated upon in more detail in App. C.
(For work in a similar spirit, see \cite{by, pad, vol}.)
The notion of conjugation is well-defined, but approximate in our scheme, as implied by the expansion about the static Fubini--Study metric.

\item While it is true that the uncertainty relation $\Delta\Lambda\,\Delta V\sim \hbar$ is consistent with the observed vacuum energy of our Universe, there is a problem with this approximate conjugate relation: what fixes the volume?
The smallness of the measured cosmological constant relies on the largeness of the observed spacetime.
(This is also a problem with unimodular gravity and related approaches \cite{by}, in which there is no {\em a priori} explanation for why the Universe is big.)
We motivate the largeness of observed $V$ through a gravitational see-saw as follows.
The scale of the vacuum energy is set by the balancing of the scale of cosmological supersymmetry breaking with the Planck scale.
The UV/IR correspondence inherent to this argument depends crucially on the spacetime uncertainty relations of Matrix theory \cite{ur}.
In perturbative string theory, modular invariance on the worldsheet translates in target space to the spacetime uncertainty relation:
\be
\Delta T \, \Delta X_{\rm tr} \sim \ell_s^2 \sim \alpha'.
\ee
Here, $T$ is a timelike direction, and $X_{\rm tr}$ is a spacelike direction transverse to the light-cone.
In Matrix theory this becomes a cubic relation
\be
\Delta T \, \Delta X_{\rm tr} \, \Delta X_{\rm long} \sim \ell_{\rm Pl}^3,
\ee
where $X_{\rm long} $ is the longitudinal direction.
After using one of the defining relations of the gauged quantum theory,
\be
\hbar\, \Delta s \sim M_{\rm Pl}\, \Delta T,
\label{eq:deltas}
\ee
and an estimate that the line element on the space of probabilities scales as
$ds \sim e^{-S_{\rm eff}}$,
where $S_{\rm eff}$ denotes a hard-to-compute-from-first-principles low-energy (Euclidean) effective action for the matter degrees of freedom propagating in an emergent (fixed) spacetime background,
we get a gravitational see-saw formula
\begin{equation}
\Delta X_{\rm tr} \, \Delta X_{\rm long} \sim e^{S_{\rm eff}}\, \ell_{\rm Pl}^2.
\end{equation}
The product of the ultraviolet cutoff (the maximal uncertainty in the transverse coordinate) and the infrared cutoff (the maximal uncertainty in the longitudinal coordinate) is thus exponentially suppressed compared to the Planck scale.
The mid-energy scale should be naturally related to a supersymmetry breaking scale, supersymmetry being broken ``cosmologically.''

\item Even though, the breaking of supersymmetry is crucial for the stability of local regions of the global spacetime manifold, in Minkowski space the cosmological constant vanishes identically.
Locally, physics is described by Matrix theory, which is a supersymmetric theory of quantum mechanics.
The fluctuations in $\Lambda$ which account for the measured vacuum energy arise as a consequence of the tension between local and global physics (UV and IR).
This is a statement about the failure of decoupling in quantum gravity.
Effective field theory, which is extraordinarily successful in its domain of validity, relies on the separation of scales, which we do not have.

We expect that the fluctuation about the zero value is biased towards the positive sign by the cosmological breaking of supersymmetry.
It is therefore our generic expectation that the vacuum energy ought to scale as $m_{\rm susy}^8/M_{\rm Pl}^4$, which is consistent with the cosmology of the present de Sitter epoch.\footnote{
See also \cite{tb}.}
The considerations presented here are, however, thermodynamic in nature.
As well, a more refined statistical analysis is necessary in order for us to explore the fluctuations about $\Lambda=0$ and their possible observation.

The coincidence problem ---
why $\Omega_\Lambda \approx \Omega_{\rm matter}$ today ---
is considerably more subtle.
Weinberg's classic argument based on the Bayesian distribution of the cosmological constant and observer bias \cite{anthropics, msw} may perhaps be replaced by a bias towards a certain set of observables in the proposed background independent quantum theory of gravity.
These observables would be relevant for describing the low-energy physics in which the supersymmetry breaking scale is related to the cosmological scale by the gravitational see-saw.

\end{enumerate}

\subsection{Fine Structure of Dark Energy}

As mentioned in the introduction, there exists an illuminating analogy to be drawn
between the problem of vacuum energy with the
problem of black-body radiation (and the related problem of specific heats) in pre-quantum physics. This deep analogy enables us to
think about a possible observable fine structure of dark energy.

First we review the analogy between black-body radiation puzzle and the vacuum energy puzzle.
In the case of the black-body there is a $\frac12 k_B T$ contribution to the energy for each independent degree of freedom:
\be
dE = \sum_n \left( \frac{1}{2} k_B T \right),
\ee
where $n$ is an abstract index that labels the degrees of freedoms.
This should be compared to the cosmological constant which counts degrees of freedom in the vacuum.
Heuristically, we sum the zero-point energies of harmonic oscillators and write
\be
E_{\rm vac} = \sum_{\vec{k}} \left( \frac{1}{2}\hbar \omega_{\vec{k}} \right),
\ee
where, unlike the fixed temperature $T$, $\omega_{\vec{k}} = \sqrt{|\vec{k}|^2 + m^2}$.
The divergence of the blackbody $dE$ is the ultraviolet catastrophe that the Planck distribution remedies.
Quantum mechanics resolves the over counting.
In asking why the vacuum energy is so small, we seek to learn how quantum gravity resolves the over counting of the degrees of freedom in the ultraviolet.\footnote{
Similarly, in the infrared, the proper formulation of quantum theory of gravity should resolve the stability problem
(``Why doesn't the Universe have a Planckian size?''),
once again in analogy with the resolution of the problem of atomic stability offered by quantum mechanics.}

This analogy between blackbody and the vacuum energy problems extends even further:
\begin{itemize}
\item The total radiation density of a blackbody at a temperature $T$ is given by the Stefan--Boltzmann law:
\be
u(T) = \sigma T^4.
\ee
This is to be compared with the quartic divergence of the vacuum energy,
\be
E_{\rm vac} \sim E_0^4,
\ee
$E_0$ being the characteristic energy cut-off, for bosons, or fermions separately, up to a sign difference.
We disregard, for the moment, the cancellation that happens in supersymmetric theories which leads to a quadratic divergence.

\item From adiabaticity, we obtain the Wien displacement law:
\be
\omega R = {\rm constant}, \qquad
\frac{\omega}{T} = {\rm constant},
\ee
where $R$ is the size of the blackbody cavity and $\omega$ the angular frequency.
This is to be compared with the uncertainty relation \eref{unc}, which tells us that $\Delta \Lambda\, \Delta V \sim \hbar$.

More precisely, fluctuations in the volume of spacetime are fixed by statistical fluctuations in the number of degrees of freedom of the gauged quantum mechanics.
In Matrix theory, the eigenvalues of the matrices denote the positions of D$0$-branes which give rise to coherent states in gravity.
Off-diagonal terms in Matrix theory break the permutation symmetry and render the D$0$-branes distinguishable.
Therefore, to enumerate the degrees of freedom, we employ the statistics of distinguishable particles (which will be of central importance in what follows).
The fluctuation is given by a Poisson distribution, which is typical for coherent states.
The fluctuation of relevance for us is in the number of Planck sized cells that fill up the configuration space (the space in which quantum events transpire), that is to say in four-dimensional spacetime:
\be
\CN_{\rm cells} \sim \frac{V}{\ell_{\rm Pl}^4} \Longrightarrow
\Delta \CN_{\rm cells} \sim \sqrt{\CN_{\rm cells}} \Longrightarrow
\Delta V \sim \sqrt{V}\ \ell_{\rm Pl}^{2},
\ee
and thus
\be
\Delta \Lambda\, \sqrt{V}\, G_N\sim 1,
\ee
where $V$ is the observed spacetime volume and $G_N$ is the four-dimensional Newton constant.\footnote{
In $D$ spacetime dimensions, \eref{unc} informs us that
$$
\Delta\Lambda \sim \frac{\hbar}{\Delta V}
\sim \frac{\hbar^{(D-4)/2(D-2)}}{\sqrt{V}\ G_D^{D/2(D-2)}}.
$$}

\end{itemize}

The Stefan--Boltzmann law and Wien's law are implicated in the derivation of the Planck distribution for blackbody radiation.
If the analogy holds, what does this say for vacuum energy?
A natural question to ask here is whether there is a universal energy distribution for dark energy.
If so, what is its nature and what are the observational consequences?
Here we will start with an assumption that there is such a distribution, which is natural from the
point of view of the new physics advocated in the previous section.
Given our proposal for a background independent quantum theory of gravity, we
investigate the nature of such a distribution and consider its observational consequences.

We should note that an important consequence of this analogy is that one should compare the temperature of the cosmic microwave background radiation (CMBR) we see now, $T_{\gamma} = 2.7\ {\rm K}$, to the cosmological constant we observe now!
The spectral distribution of dark energy should then be a function of energy for the fixed present value of the cosmological constant, corresponding to the energy scale of $10^{-3}\ {\rm eV}$, in analogy with the CMBR spectral distribution.
The question of why this scale is so low (why the Universe is so big), the proposed answer to which has been outlined above, is thus analogous to the question why the background CMBR temperature is so close to the absolute zero.

According to the proposal discussed in this review M-theory is background independent Matrix theory.
The infinite momentum limit of M-theory is equivalent to the $N\to\infty$ limit of coincident D$0$-branes given by $U(N)$ super-Yang--Mills gauge theory \cite{matrix}.
In particular, Matrix theory gravitons are bound states of D$0$-branes and the gravitational interaction, and thus the geometry of spacetime, is contained in the open string dynamics, {\em viz.}\ the quantum fluctuations of matrix degrees of freedom.
D$0$-branes obey $U(\infty)$ statistics.
Infinite statistics \cite{green,st,inf,mi} can be obtained from the $q=0$ deformation of the Heisenberg algebra
\be
a_i a_j^\dagger - q a_j^\dagger a_i = \delta_{ij}, \qquad a_i|0\rangle = 0.
\ee
(The cases $q=\pm 1$ correspond to Bose and Fermi statistics; $q=0$ is the so called Cuntz algebra \cite{cuntz} corresponding to infinite statistics.)
In particular, the inner product of two $N$-particle states is
\be
\langle0| a_{i_N} \cdots a_{i_1} a_{j_1}^\dagger \cdots a_{j_N}^\dagger |0\rangle = \delta_{i_1 j_1}\cdots \delta_{i_N j_N}.
\ee
Thus any two states obtained from acting with the same creation and annihilation operators in a different order are mutually orthogonal.
The partition function is
\be
Z = \sum_{\rm states} e^{-\beta H}.
\ee
The D$0$-branes are distinguishable. Thus there is no Gibbs factor.
Therefore, we can argue that
{\em the spectral distribution of dark energy that follows from infinite statistics is the familiar Wien distribution}.
\be
\label{eq:wien}
\rho_{\rm DE} (E, E_0) = A E^3 e^{- B\frac{ E}{E_0}}
\ee
which implies that
\be
\rho_{\rm vac} = \int_0^{E_0} dE\ \rho_{\rm DE}(E,E_0) \sim \frac{6 A}{B^4}\ E_0^4,
\ee
with $A,B$ universal constants, and $E_0 \sim 10^{-3}\ {\rm eV}$, which corresponds to the observed cosmological constant.
The integrated energy density is proportional to $E_0^4$, as it must be.
This Wien-like spectral distribution for dark energy is thus the central prediction of a detailed analogy between the blackbody radiation and dark energy.
This in turn is rooted in our new viewpoint on the cosmological constant problem as summarized in the introduction to this section.
The constants $A$ and $B$ are in principle computable in the framework of the background independent Matrix theory, but that computation is forbidding at the moment.
We will therefore only concentrate on global features of this viewpoint on the fine structure of dark energy.
Also, the precise dispersion relation of the dark energy
quanta
(ultimately determined by the degrees of freedom of Matrix theory within the framework of the generalized quantum theory that we have proposed)
is not relevant for the general statistical discussion of possible observational signatures presented below.

Vacuum energy ({\em i.e.}, $\sum_{\vec{k}} \frac12 \hbar\,\omega_{\vec{k}}$) has negative effective pressure.
The Wien and Planck distribution share a common prefactor, which is the reason why we argue that at low energies our proposal is consistent with the positive cosmological constant, the dark energy being modeled as vacuum energy.
 From the effective Lagrangian point of view, the positive cosmological constant accounts for the accelerated expansion.
At short distances, we have a radically different situation.
The pressure in this scenario is positive and set by the scale of $E_0$.
The proposed dark energy quanta that are physically responsible for such an effective view of the cosmological constant have a strange statistics fixed by symmetry requirements, and which has certain parameters that should be bounded by observation.\footnote{
A useful comparison is the following.
For photons in the CMBR there exists a vacuum contribution and then the usual Planck distribution.
Ours is a completely analogous claim:
we have the vacuum part and the distribution of the quanta which constitute the vacuum.
The only difference here is that the quanta are unusual and the distribution is unusual due to the infinite statistics invoked.}

To summarize,
in accordance with our view of the cosmological constant problem, we think of dark energy as vacuum energy.
Just as in the case of a photon gas, the Wien distribution for vacuum energy exhibits both a classical and a quantum nature.
In Matrix theory the degrees of freedom, in the infinite momentum frame, are non-relativistic and distinguishable D$0$-branes whose dynamics are obtained from a matrix quantum mechanics.
The UV/IR correspondence at the heart of Matrix theory (and holographic theories in general) encodes the essential dualism of the cosmological constant problem:
vacuum degrees of freedom determine the large-scale structure of spacetime.

Direct observation of the Wien distribution for dark energy from calorimetry, {\em i.e.}, the analogue of measurements of the CMBR, is probably impossible,
given the gravitational nature of Matrix theory degrees of freedom.
We mention some more practical tests that one might be able to make of our proposal.

\begin{itemize}
\item Recently, a possibility for a direct observation of dark energy in the laboratory has been discussed in the literature \cite{beck,copeland}.
The idea is simple and fascinating.
One simply relies on identifying dark energy as the quantum noise of the vacuum, as governed by the fluctuation-dissipation theorem.
For example, by assuming that vacuum fluctuations are electromagnetic in nature, the zero point energy density is given by the phase space factor of the Planck distribution
(the same as the one discussed above in the case of the Wien distribution).
The integrated expression, which formally diverges, if cut-off by the observed value of dark energy, $E_0$, would correspond to the cut-off frequency
\be
\nu_{\rm DE} \sim 1.7 \times 10^{12}\ {\rm Hz}.
\ee
The present experimental bound \cite{beck,copeland} is around $\nu_{\rm max}\sim 6 \times 10^{12}\ {\rm Hz}$.

If our proposal is correct, and the dark energy is endowed with its own spectral distribution of the Wien type, then there is a window around the $\nu_{\rm DE}$ determined by the fluctuations $\delta E_0$ of dark energy around $E_0$.
The right way to look at this is that the present maximum frequency sets a bound on the possible fluctuation $\delta E_0$.
The theoretical value of this fluctuation is tied to the precise value of the parameters in the Wien distribution, which are determined by the underlying new physics.

The fluctuation in the dark energy distribution (\ref{eq:wien}) is
\be
\frac{\delta\rho_{\rm DE}}{\rho_{\rm DE}} = \frac{B E}{E_0^2}\ \delta E_0.
\ee
We have as well
\be
\delta E^2 = \vev{E^2} - \vev{E}^2 = \frac{4 E_0^2}{B^2},
\ee
where
\be
\vev{E^a} = \frac{\int_0^\infty dE\ E^a\ \rho_{\rm DE}(E,E_0)}{\int_0^\infty dE\ \rho_{\rm DE}(E,E_0)}.
\ee

The observed vacuum energy is given as
\be
\int_{0}^{\nu_{\rm DE}} d\nu\ \rho_{\nu} = \frac{\pi h}{c^3} \nu_{\rm  DE}^4.
\ee
Now, we identify $\delta E$ with the fluctuation of the vacuum energy around $E_0$.
The energy density corresponding to the maximum observed frequency should bound the fluctuation of $E_0$.
This implies
\be
\delta E = \delta E_0 = \frac{2 E_0}{B} \leq E_0\left(1-\frac{\nu_{\rm max}}{\nu_{\rm DE}}\right).
\ee
Inserting the current observational bound, $\nu_{\rm max}$ and the values for $E_0$ and $\nu_{\rm DE}$ noted above, yields the following bound on the vacuum energy fluctuation
\be
\delta E_0 \lesssim 6.47\times 10^{-4}\ {\rm eV},
\label{qnoise}
\ee
which in turn implies
\be
B\gtrsim 3.1\,.
\ee

\item The Greisen--Zatsepin--Kuzmin (GZK) bound provides a theoretical upper limit on the energy of cosmic rays from distant sources \cite{gzk}.
In the usual GZK setup a CMBR photon is scattered off a proton producing positively charged or neutral pions (plus a neutron or a proton), thus degrading the incoming proton's energy.
The rough estimate of the energy cutoff is the threshold when the final products are both at rest.
Neglecting the split between proton and neutron masses one gets from simple kinematics
\be
E_{\rm threshold} \sim \frac{(m_p + m_{\pi})^2 - m_p^2}{4 E_{\gamma}} \sim 5\times 10^{19}\ {\rm eV}.
\ee
Note, $E_{\gamma} \sim 6.4 \times 10^{-4}\ {\rm eV}$, from the temperature of $T_{\gamma} = 2.7\ {\rm K}$,
and there are on average in one ${\rm cm}^3$ $400$ CMBR photons.
This depletion occurs on distances of $O(10)\ {\rm Mpc}$. Recently, the GZK cutoff was observed by the
Pierre Auger Observatory \cite{auger} which found a suppression in the cosmic ray spectrum above $10^{19.6}\ {\rm eV}$
at six sigma confidence.

We now consider the interaction of high energy cosmic rays with the proposed
dark energy distribution for which there should be an analogous GZK effect.
Although the coupling for the interaction responsible for this effect would be quite
small, over cosmological distances
the effect could be observable.
In our case the modification of the corresponding GZK formula, comes from a simple replacement 
of $E_{\gamma}$ by $E_0 + \delta E$, which implies
\begin{equation}
E_{\rm threshold}\simeq\frac{1}{4 E_{0}}\bigg[(m_p+m_{\pi})^2-m_p^2 - \frac{\delta E}{E_0} \Big( (m_p + m_{\pi})^2 - m_p^2 \Big)\bigg].
\label{gzk}
\end{equation}
If the fluctuation in the dark energy distribution is too great the analogous  
GZK cutoff considered here would fall below that of the standard cutoff 
and would be observed as an unexplained suppression in the cosmic ray spectrum.
No such suppression has been detected. Thus we may use the observed cosmic ray 
spectrum to further constrain the fluctuation in the dark energy distribution. 
Taking as our lower bound the observed standard GZK cutoff and making use
of (\ref{gzk}) we find 
\be
\delta E \lesssim 4.37\times 10^{-4}\ {\rm eV}.
\ee
This is a similar but more stringent bound than the one provided by quantum noise measurements, 
(\ref{qnoise}). It is worth noting that these two bounds were derived from
unrelated physical phenomena but are of the same order of magnitude. This suggests a
level of consistency in the proposal for dark energy quanta presented above.

\end{itemize}

\section{Outlook}

In this review we have collected various aspects of our recent proposal for a background independent formulation of a holographic theory of quantum gravity.
We have included the necessary background material on geometry of canonical quantum theory, holography and spacetime thermodynamics, Matrix theory, as well as our specific abstract proposal for a dynamical theory of geometric quantum theory, as applied to Matrix theory.
We have placed particular emphasis on the conceptual problem of time and a closely related and phenomenologically relevant problem of vacuum energy in quantum gravity.
We have also summarized our recent discussion of observational implications of the new viewpoint on the problem of vacuum energy.

Obviously we have only explored the surface.
There are many technical developments needed for full exploration of the consequences of our proposal.
There are many open questions.
Here we collect a few perhaps more obvious open issues.
Much of this constitutes present research that will appear in future publications.

\begin{itemize}
\item The gauging of the unitary group of quantum mechanics led us to the infinite dimensional Grassmannian ${\rm Gr}(\BC^{n+1}) = \Diff(\BC^{n+1})/\Diff(\BC^{n+1},\BC^n \times \{0\})$, with $n = \infty$.
This is the configuration space of the generalized quantum theory.
We do not, however, know the topology and geometry of this space.

\item We wish to know how to recover in detail the gravitational physics of asymptotically flat spacetimes.
This is the background of much of ordinary low-energy physics, and it is crucial for some of the outstanding problems in General Relativity and quantum field theory in curved spacetime.
In particular, in asymptotically flat spaces, the $S$-matrix encodes $n$-point interactions.
This is a correspondence limit of the gauged quantum theory that we have proposed.
We would like to make this explicit.

\item The sharpest non-perturbative formulations of string theory is in asymptotically AdS backgrounds, as provided by AdS/CFT duality with its manifold ramifications.
How do we compare to the AdS/CFT intuition and results?
Here we can offer some general comments:
Based on general symmetry considerations, the gauged Matrix theory is compatible with other non-perturbative formulations of string theory in curved backgrounds.
The dual CFT, in the case of $16$ supercharges, is reducible (upon dimensional reduction to $0+1$ dimensions) to Matrix theory.
The Matrix theory must capture the physics of local flat regions of the AdS space, which according to the principle of equivalence, are physically independent from the AdS asymptotics.
Thus there is no conflict between gauged Matrix theory and AdS/CFT from the global point of view, but the gauged Matrix theory offers a more general formulation of string theory, because it can in principle be applied to other backgrounds, such as time-dependent cosmological situations, in which case it is not necessarily physically meaningful to adhere to the notions of fixed asymptopia, or global holographic screens, or $S$-matrix observables, or dual CFTs.

As we mentioned in the introduction a useful analogy to draw here is to compare the Bohr hydrogen atom of old quantum theory to AdS/CFT ("static" holography) and ask about holography, {\em viz.}\ the structure of quantum gravity, in more general (cosmological) backgrounds.
String theory anywhere except AdS, especially on time-dependent backgrounds, or backgrounds
in which knowledge of future causal and asymptotic structure is not an {\em a priori} given, is analogous to
the helium atom for which the techniques of the Bohr atom are insufficient.

Obviously, we will need to reconsider the contents of the black hole information paradox from our point of view and compare to the perspective offered by the gauge theory/gravity duality.
In particular, it may well be the case that horizons and singularities are artifacts of the semiclassical (or thermodynamic) limit.
The emergence of the causal structure of these spacetimes is therefore mysterious, and something we should explore.

\item Another related avenue of investigation is the application of this formalism to the problem of cosmological spacelike or null singularities ({\em e.g.}, the Big Bang/Big Crunch).
As previously noted resolving the problem of an initial cosmological singularity is closely tied to the problem of time in quantum theory.
Because the approach presented here offers a viable solution to this issue, progress towards an understanding of the initial cosmological singularity should be possible.
Also, the question of cosmological singularities has been studied using the holographic nature of Matrix theory, albeit in a fixed background \cite{mbb}.
Thus a background independent formulation of Matrix theory may be a promising route toward understanding this problem.

\item A gauge theory dual to flat spacetimes can be considered by taking the infinite radius of curvature limit of AdS spacetimes.
Consider, for example, the outstanding challenge of formulating a dual to four-dimensional Minkowski space.
To do this, we consider the infinite radius limit of $\AdS{4}\times S^7$.
Recall that on a $p$-brane the gauge theory coupling is specified as
\be
g_{YM}^2 = (2\pi)^{p-2} g_s (\alpha')^{(p-3)/2} = {\rm constant}
\ee
as $\alpha'\to 0$.
The infinite radius limit we wish to take sends
\be
\frac{R_{\AdS{4}}}{(\alpha')^{1/2}} = \frac12 \frac{R_{S^7}}{(\alpha')^{1/2}} = g_s^{1/3}\left( \frac{\pi^2}{2} N \right)^{1/6}\longrightarrow \infty.
\ee
In this limit $\AdS{4}\times S^7$ becomes $M_4\times [\BR^7 \bigcup \{{\rm pt.}\}]$.
The supergravity approximation, valid for energies smaller than $R_{\AdS{4}}^{-1}$, appears to break down completely in the $R_{\AdS{4}}\to \infty$ limit.
The $S^7$, also of infinite size, is threaded by $N$ units of flux, where we have sent $N\to \infty$.
The isometry group of $S^7$ is $SO(8)$.
Common to the Penrose diagrams of an infinite radius $\AdS{4}$ and four-dimensional Minkowski space is $i_0$.
This is where the dual gauge theory should live.

The gauge theory dual is $\CN=8$ super-Yang--Mills in three dimensions.
This theory has eight scalars in the adjoint representation and eight fermions.
The R-symmetry is $SO(8)$.
This theory itself is not conformal, but flows under the renormalization group to a three-dimensional CFT.
We can write the $\CN=8$ super-Yang--Mills theory as a $(2+1)$-dimensional Matrix model.
When we reduce this theory to $(0+1)$ dimensions, we obtain an $\CN=16$ super-Yang--Mills theory with nine scalars and 16 fermions.
This Matrix theory should teach about physics in Minkowski space.
We can then in principle patch together the Matrix theory on different Minkowski spaces to construct a gauged quantum mechanics applicable to general backgrounds with unspecified future asymptopia.

\item Can we say something precise about asymptotically de Sitter backgrounds and more general cosmological backgrounds?
What is, in particular, the r\^{o}le of ``quasi-local'' holography we have tried to emphasize in the background independent formulation of quantum theory of gravity discussed in this review, in the context of cosmological backgrounds?
Here we wish to offer a comment about the tension between local and global notions of holography which is one of the crucial elements of gauge theory/gravity duality:
It is well known that global holography provides heuristic support for a cosmological constant far smaller than the exaggerated expectations of effective field theory.
According to holography, the degrees of freedom of gravity in $D$ spacetime dimensions are captured by equivalent non-gravitational physics in $D-1$ dimensions \cite{holog}.
For example, the relation between holography and the cosmological constant was explored in \cite{tb, ckn, hm}.

To be precise \cite{thomas}, suppose there are $D$ spacetime dimensions, each with a characteristic scale $R$.
The holographic bound demands that the entropy (the number of degrees of freedom $\CN_{\rm dof}$) scales as the area.
This is a $(D-2)$-dimensional surface, so
$
\CN_{\rm dof} \leq \frac{R^{D-2}}{4 G_D \hbar}.
$
 From the uncertainty principle, the energy of each independent degree of freedom scales as
$
E \sim \frac{\hbar}{R}.
$
All the degrees of freedom contribute equally to the vacuum energy density:
$
\Lambda \sim \CN_{\rm dof} \frac{E}{R^{D-1}}
\sim \frac{1}{R^2 G_D}.
$
The dependence on $\hbar$ has cancelled in the last expression, so this vacuum energy density should survive the semiclassical ($\hbar\to 0$) limit.
Here, the cosmological constant $\Lambda$ is a prescribed, fixed number.
It is determined by the size $R$ of the regularized ``box.'' Note that this is somewhat na\"{\i}ve, because the characteristic sizes
of spatial and temporal directions do not have to be the same.

We notice that the fluctuation $\Delta\Lambda$ that we promoted above matches this scaling, but only when $D=4$.
A Poisson fluctuation in the holographic degrees of freedom $\CN_{\rm dof}$ will not recover the holographic scaling of the cosmological constant.
The fluctuation we have considered is in the volume of the quantum mechanical configuration space rather than in a codimension one structure.
The reason that this is consistent is that we have applied the principle of equivalence at each point in spacetime.
In the scheme that we have proposed, holography enters in the choice of the quantum theory compatible with having Minkowski space as a local solution, namely through Matrix theory.
Holography is ``local,'' in the sense of the equivalence principle.
Therefore there does seem to be some tension with global holography, which might be a useful concept only for certain physics questions.

\item In a related vein, because the future causal structure of spacetime is unknown, a global $S$-matrix description of our Universe is unavailable.
 From the viewpoint of the generalized quantum theory, a wave functional approach \`a la Wheeler--de-Witt \cite{wdw} seems poorly formulated.
In the general quantum theory we can have dynamical statistical correlations between the past and today.
The observables of quantum gravity are these dynamical correlations in the configuration space of the quantum theory.
A functional approach to quantum gravity is to consider a canonical theory of quantum mechanics (Matrix theory) at every point in spacetime, where spacetime is here regarded as a semiclassical geometry that arises from identifying the configuration space with the physical space in the $\hbar\to 0$ limit.
This is an application of the correspondence principle that ensures that at long wavelengths we recover General Relativity.

\item In gauging quantum mechanics to lift the ten-dimensional (or eleven-dimensional) vacuum, we obtain a vacuum energy that corresponds to the cosmological scale of supersymmetry breaking.
The second cosmological constant problem nevertheless persists:
why is $\Omega_\Lambda \approx \Omega_{\rm matter}$ {\em today}?
This could be an accident of living in the present epoch \cite{anthropics}.
We would, in view of the main point of this paper, prefer to view the cosmic concordance problem through a different dynamical lens, one might want to term ``the Universe as an attractor.''

The classic reference \cite{msw} considers linear perturbation theory in a Friedmann--Robertson--Walker (FRW) background for a certain density of matter and a certain vacuum energy and then deduces an {\em a priori} probability for the vacuum energy so that
(1) gravitational bound states appear at large scales;
(2) the fundamental constants are held fixed; and
(3) the probability distribution is independent of a bare vacuum energy, which permits the use of Bayesian statistics.

It is in holding the constants fixed that considerations based on observer bias, {\em i.e.}, anthropic selection, takes place.
We may be able to apply a similar reasoning with generalized quantum mechanics, but without resorting to anthropic selection.
In the framework of gauged quantum mechanics, non-Gibbsian quantum probability distributions are dynamically possible, for example as perturbations of the usual path integral, around the Fisher metric.
Anthropic reasoning is evaded because we have an $S_{\rm eff}$ that can in principle be obtained directly and exactly from Matrix theory.
Thus a dynamical resolution of the coincidence problem might be possible
in this more general ``non-equilibrium'' quantum theoretic approach.

\item Finally, in a more phenomenological vein, our discussion of the vacuum energy seems to imply that dark energy has a fine structure embodied in a very particular energy distribution of a Wien type.
This distribution is compatible with the statistics of the underlying quantum gravitational degrees of freedom we have argued are relevant for a new viewpoint on the cosmological constant problem.
This new point of view offers other possible theoretical perspectives.
For example, in view of some intriguing phenomenological scaling relations found in studies of dark matter \cite{Mil, Kap}, which are apparently sensitive to the vacuum parameters, such as the cosmological constant, it is natural to ask whether within our discussion one can get both dark energy and dark matter in one go.
In Matrix theory, the open string degrees of freedom (without which we would not have infinite statistics) could thus be responsible for dark energy, and the D$0$-brane quanta attached to the open strings could provide natural seeds of large-scale structure, {\em i.e.}, dark matter, especially when treated as non-relativistic degrees of freedom fixed to a background.
This would also imply that infinite statistics is relevant for dark matter as well!
It is intriguing that in the formal studies of infinite statistics one finds non-local expressions for the canonical fermion and boson operators in terms of Cuntz algebra ({\em i.e.}, infinite statistics) operators.
Could this mean that the standard model matter is just a collective excitation around the dark matter condensate?
Such a collective ``condensed matter'' view of the emergence of the Standard Model would be radically different from the usual compactification based approaches to particle physics phenomenology in the framework of string theory.

\end{itemize}

\section*{Acknowledgments}
We would like to give special thanks to Chia Tze for his integral contributions to the foundational work on which this review is based.
We also thank many colleagues for their comments, and in particular Tatsu Takeuchi and Alexandr Yelnikov.
Most recently we have enjoyed conversations with
Vijay Balasubramanian, Jan de Boer, Sumit Das, Petr Ho\v{r}ava, Dan Kabat, Nemanja Kaloper, Rob Leigh, Tommy Levi,
and other participants in the recent Sowers Theoretical Physics Workshop ``What is String Theory?''\ at Virginia Tech.
MK gratefully recognizes Jon Cook, Bing Feng, Chrysostomos Dimitrios Kalousios, Can Koz\c{c}az, Zack Lewis, Manavendra Mahato, Andy O'Bannon, Bora Orcal, Mark Pitt, Greg Stock, and Roger Wendell for informative discussions and thoughtful insight and would like to especially acknowledge the shared wisdom of Louise Marie Olsofka.
MK also thanks the Theoretical Advanced Study Institute in Elementary Particle Physics held at the University of Colorado at Boulder for hospitality during the completion of this work.
VJ was supported by PPARC.
DM was supported in part by the U.S.\ Department of Energy under contract DE-FG05-92ER40677.
The Amsterdam summer workshop in string theory, the Aspen Center for Physics, and the KITP, Santa Barbara are gratefully acknowledged for their stimulating atmospheres.

\section{Appendices}

 \subsection{Appendix A: Weinberg's Non-linear Quantum Mechanics}
We will now review the proposed non-linear generalization of quantum mechanics
due to Weinberg \cite{weinberg}. This generalization is not an attempt to create
the most general framework in which to formulate quantum mechanics. It is instead an attempt to
generalize quantum mechanics in a way which motivates
experimental tests of the linearity of the theory. Indeed it is
surprising how few direct tests of quantum mechanics have been performed over the years.
Every test of a particular quantum field theory ({\em i.e.}, QED) is in a sense a test of
quantum mechanics, but a high precision test of quantum mechanics that is
independent of any particular theory is of obvious interest. Weinberg's non-linear generalization was
conceived with this goal in mind. We will summarize the formalism of Weinberg
as originally devised. We shall also summarize the results of experiments inspired
by his work. We will also address some interesting questions that have arisen regarding
this formalism and the Einstein--Podolsky--Rosen (EPR) paradox \cite{epr}.
Finally, as previously noted Weinberg's generalization can be
found within the geometric quantum mechanical framework. We will detail how
one can arrive at Weinberg's formalism by choosing a particular
generalization of the dynamical structure in the geometric framework.
Note though, that this generalization is very restricted, compared to the
general geometric formulation of quantum theory discussed in the main body of the review.

\subsubsection{Basic Formalism}

One direct approach to endowing quantum mechanics with non-linear structure is
to simply add non-linear terms to the Schr\"{o}dinger equation. However,
it is difficult to do this in such a way as to yield physically reasonable results.
Weinberg's formalism is conservative in the sense that it focuses on the elements
that seem to be a requirement of non-linearity.

We begin with the wave function. As in standard quantum mechanics, the states $\psi$ and
$Z\psi$ are identified where $Z$ is an arbitrary complex number. For
clarity in this appendix we will use natural units and consider $\psi$ to be a function of the
discrete variable, $k$.
We may now define observables. In standard quantum mechanics observables
are represented by a Hermitian matrix, $A_{ij}$ or equivalently the bilinear function,
$\psi_{i}^{*}A_{ij}\psi_j$. Let us generalize this to a {\em non-bilinear} function,
$a(\psi,\psi^{*})$. In order to maintain the identification of $\psi$ with
$Z\psi$ we require that the observables be homogeneous of degree one both
in $\psi$ and $\psi^*$
\begin{equation}
\psi_{k}\frac{\partial a}{\partial \psi_k}=\psi_{k}^*\frac{\partial a}{\partial \psi_k^*}=a.
\end{equation}
We may define the sum of observable functions as follows,
\begin{equation}
(a+b)(\psi,\psi^*)=a(\psi,\psi^*)+b(\psi,\psi^*).
\end{equation}
However, we must be careful in defining the product of observable functions.
In order to do this we must generalize the matrix multiplication of standard quantum mechanics
\begin{equation}
a*b=\frac{\partial a}{\partial \psi_k}\frac{\partial b}{\partial \psi_k^*}.
\end{equation}
We make note of a particular function, the norm
\begin{equation}
d=\psi^*_k\psi_k,
\end{equation}
which is the unit element of the previously defined product.
Most of the differences between classical physics and standard
quantum mechanics can be traced to the fact that the product of observables
becomes non-commutative. Now note that the product of observables in this
formalism is neither commutative nor associative.
In analogy with classical physics and commutativity, the lack of associativity is
the source of most of the discrepancies between
Weinberg's formalism and standard quantum mechanics.

Now we address the issue of symmetries in this formalism.
Initially this may seem problematic because of the lack of
associativity. However, we shall see that Lie algebras still
play a vital r\^{o}le.
We begin by noting that a linear transformation in quantum mechanics
may be expressed as
\begin{equation}
\delta\psi_k=-i\epsilon  A_{kl}\psi_l.
\end{equation}
In order to generalize this we consider the change in the wave function
with regard to an infinitesimal function, $\epsilon a(\psi,\psi^*)$
\begin{equation}
\epsilon \delta_a\psi_k\equiv-i\epsilon\frac{a}{\psi_k}.
\end{equation}
This implies the change in a function $b$ with respect to $\epsilon a$
is given by
\begin{eqnarray}
\delta_a b=
-i\epsilon\bigg[\frac{\partial b}{\partial \psi_{k}}\frac{\partial a}{\partial \psi^*_{k}}
-\frac{\partial b}{\partial \psi^*_{k}}\frac{\partial a}{\partial \psi_{k}}\bigg],\nonumber
\end{eqnarray}
which may be rewritten as
\begin{equation}
\delta_a b=i[a,b]\equiv i(a*b-b*a).
\end{equation}
The commutator is antisymmetric,
and the Jacobi identity is satisfied.
Thus as in standard quantum mechanics we are able to
make use of Lie algebras of the symmetry transformations.
Thus we require,
\begin{equation}
[\xi_i,\xi_j]=iC_{ijk}\xi_k.
\end{equation}
Here $C_{ijk}$ is the structure constant of a given Lie algebra.
Now we consider, in particular, the symmetry of time translation
generated by the Hamiltonian function $h(\psi,\psi^*)$. We define the time dependence
of the wave function as,
\begin{equation}
\psi_k(t+\epsilon)=\psi_k(t)+\delta_k\epsilon \psi_k(t)
\end{equation}
which yields the time dependent non-linear Schr\"{o}dinger equation,
\begin{equation}
\frac{d\psi_k}{dt}=-i\frac{\partial h}{\partial \psi_k^*}.
\end{equation}
As in standard quantum mechanics we may establish a direct correspondence
between the Poisson bracket and the commutator. We use this correspondence
to define the time dependence of any function $a(\psi,\psi^*)$ where $\psi$ is a
function of $t$
\begin{equation}
\frac{da}{dt}=-i[a,h].
\end{equation}
Note that we can immediately establish two conserved functions.
The Hamiltonian which of course commutes with itself and the norm
which commutes with all observables by its definition as the unit element.
However, note that because of the lack of associativity, a product of
conserved functions is not necessarily a conserved function.

This leads to an important distinction between standard
quantum mechanics and the non-linear generalization considered
here. The wave functions in standard quantum mechanics display
quasiperiodic behavior. Now consider the class of
non-linear Hamiltonian that is $integrable$. The wave functions associated
with theses systems are guaranteed to be quasiperiodic on the $n$-torus.
In order for the Hamiltonian to be integrable there must be no
independent quantities, $a_n(\psi,\psi^*)$, that commute with themselves as well as
$h(\psi,\psi^*)$. As previously noted in Weinberg's formalism there are only two
guaranteed commuting observable $d$ and $h$. Thus in this formalism systems are not
generically integrable for $n>2$ components. Thus the quasiperiodic
behavior of the wave function may be replaced by chaotic behavior.
However, consider Hamiltonians of the form
\begin{equation}
h=h_0+h_i.
\end{equation}
Here $h_0$ is integrable and $h_i$ is not.
If $h_i$ is small compared to $h_0$ the averaged equations of motion will be the same as
if the Hamiltonian function had been integrable. This implies that if the deviation
from standard quantum mechanics is small the time dependent Schr\"{o}dinger equation
will be integrable.

We would also like to address the issue of combining two isolated systems
in this formalism. Consider two isolated states $\psi_I$ and $\psi_{II}$ and their
combined wave function $\psi_{III}$. Also consider their associated Hamiltonians
$h_I$ and $h_{II}$.
In standard quantum mechanics both Hamiltonians would be bilinear. Thus
there sum $h_{III}$ would be bilinear with a matrix coefficient being given
by the direct sum of $h_I$ and $h_{II}$. If we generalize this sum for non-bilinear functions
we arrive at the following expression for the new Hamiltonian,
\begin{equation}
h_{III}(\psi_{III},\psi^*_{III})=\sum_nh_I(\psi_{I}^{(n)},\psi_{I}^{(n)*})+\sum_mh_{II}(\psi_{II}^{(m)},\psi_{II}^{(m)*}).
\label{wwd}
\end{equation}
Note that any additive observable for such combined system would be constructed in a similar way and
would be of this form.
\subsubsection{Eigenvalues and Expectation Values}

Next we will address the issue of eigenvalues in this formalism.
In standard quantum mechanics the wave function will give a definitive
value, $\lambda$ for an observable represented by a matrix if it is an eigenvector
of that matrix. Thus $\lambda$ is an eigenvalue. However, in accordance with our
new definition
of observables we will generalize this notion. Thus
we consider
observables represented by a non-bilinear function, $a(\psi,\psi^*)$. This implies
the wave function will give a definitive value only if
\begin{equation}
\frac{\partial a}{\partial \psi_k^*}=\lambda\psi_k
\label{wwa}
\end{equation}
and
\begin{equation}
\frac{\partial a}{\partial \psi_k}=\lambda\psi_k^*.
\end{equation}
As in standard quantum mechanics we may
use the variational principle to define eigenvectors.
Thus
we may define the eigenvectors of the observable
function $a(\psi,\psi^*)$ as the stationary points of the
following equation,
\begin{equation}
\Lambda(\psi,\psi^*)=\frac{a(\psi,\psi^*)}{d(\psi,\psi^*)}.
\label{wwc}
\end{equation}
The eigenvalues are then the values of $\Lambda$ at those stationary points.
These can be obtained by means of differentiating (\ref{wwc}) with
respect to $\psi$ and $\psi^*$. This yields the following equations,
\begin{eqnarray}
\frac{\partial \Lambda}{\partial\psi_k^*}=
\frac{1}{d}\frac{\partial a}{\partial\psi_k^*}
-\frac{a}{d^2}\psi_k
\nonumber\\
\frac{\partial \Lambda}{\partial\psi_k}=
\frac{1}{d}\frac{\partial a}{\partial\psi_k}
-\frac{a}{d^2}\psi_k^*.
\end{eqnarray}
Thus if both of these equations vanish $\psi$ is an eigenvector of $a(\psi,\psi^*)$
with an eigenvalue of $a/d$. Now note that because $\psi$ is identified with $Z\psi$,
$\Lambda$ is invariant under such a transformation.
\begin{equation}
\Lambda(Z\psi,Z\psi^*)=\Lambda(\psi,\psi^*).
\end{equation}
Thus an $n+1$ component wave function $\psi_k$ is defined on the
projective space $\CP{n}$. It is important to note that if we assume
a small departure from linear quantum mechanics we may observe a shift
in a given eigenvalue. This shift is similar to the first Born approximation
in standard quantum mechanics.

We must also try to make sense of the expectation value in this
formalism. In standard quantum mechanics we may define the expectation
value of an observable $A$ in state $\psi_k$ as
\begin{equation}
\langle \hat{A} \rangle_{\psi}=\frac{\psi_k^*A_{kl}\psi_l}{\psi_k^*\psi_l}.
\end{equation}
In analogy we generalize the expectation value of an
observable function $a$ of state $\psi$ to
\begin{equation}
\langle a \rangle_{\psi}=\frac{a(\psi,\psi^*)}{d(\psi,\psi^*)}.
\end{equation}
Note that this is identical to (\ref{wwc}).

In standard quantum mechanics we may use the expectation value
to define the probability distribution for values of any observable
in a given state. This is because a given observable commutes with itself
to all orders. Thus it may be measured simultaneously to all orders.
However this is not the case in Weinberg's formalism. As before this
discrepancy can be traced back to the lack of associativity among observables.
Thus there is no unique way to determine a probability distribution from
the expectation value of an observable function to arbitrary order.

There are, however, some observables in this formalism that do commute with
themselves to all orders. These observables have an associated symmetry
principle that requires that they be bilinear functions.
Observables such as momentum and angular momentum
belong to this class. We can use these observables to determine the probability
distribution for an arbitrary observable. This can be accomplished by allowing
the observable in question to interact with one of the observables which is
represented by a bilinear function.

\subsubsection{Experimental Tests of Linearity}
It was Weinberg's purpose to discover a precise
way of directly testing the linearity of quantum
mechanics. In order to see how this
may be accomplished we consider a small non-linear perturbation added
to a given observable. This will shift the eigenvalue associated with that
observable while still leaving the system
integrable. More specifically consider a particle with
a given spin in a weak uniform magnetic field.
It is possible to coherently observe the oscillating particle
over long periods of time and thus to measure its characteristic
frequencies extremely precisely. This would allow
a shift in the frequencies due to non-linear perturbations to be observed.

Subsequent to the introduction of Weinberg's formalism several
such experiments were performed \cite{chu,maj,pitt}.
These experiments used spin-$\frac{3}{2}$ nuclei.
For such nuclei the shift in the characteristic frequencies is given
in action angle variables by
\begin{equation}
\delta\omega_n=\frac{\partial \langle h_i \rangle}{\partial J_n},
\end{equation}
where $h_i$ is a small non-bilinear term added to the unperturbed
Hamiltonian which is averaged to ensure integrability \cite{weinberg}.
 From the measurement of characteristic frequencies
we may calculate the size of corrections to the energy eigenvalues, $\epsilon$
or place an
upper bound on it. This parameter is indicative of the level
of non-linearity present in the system. The best current upper
bound is $|\epsilon|<6.2\times 10^{-21}$ eV given
by Chupp and Hoare \cite{chu} using freely precessing
$^{21}$Ne.

\subsubsection{The EPR Paradox}

It was noted by Polchinski in \cite{pol} that there were potential
difficulties
with Weinberg's non-linear quantum mechanics involving the EPR paradox.
It was shown for a given non-linear
generalization
of quantum mechanics, if there is to be no EPR type communication among isolated systems,
the observables of a given system (I) must be defined as follows,
\begin{equation}
a_I^{(P)}(\psi,\psi^*)=h_I\bigg(\sum_n\psi^{(n)}_i\psi^{(n)*}_j\bigg).
\label{wwe}
\end{equation}
However, if we recall (\ref{wwd}) we see that we
would express an observable for an isolated system as
\begin{equation}
a_I^{(W)}(\psi,\psi^*)=\sum_nh_I(\psi^{(n)}_i,\psi^{(n)*}_j).
\end{equation}
Thus there must be EPR type communication which is permitted
in Weinberg's formalism. Even if observables were of the form of
(\ref{wwe})
Polchinski went on to show there is still the potential for a special
type of
communication between the branches of the wave function. This type of communication
is a realization of
an Everett phone, which might be of interest in thinking about
the multiverse and the string landscape. It was conjectured that
these forms of superluminal communication may be isolated to Weinberg's formalism. This, however,
was challenged by Mielnik in \cite{miel} where it was claimed that this was generic
attribute of non-linear models.
Once again, we want to emphasize that this particular generalization is
very restrictive and is
very different from the background independent, or gauged, quantum theory we
have discussed in the main text of the review.

\subsubsection{Connection to Geometric Quantum Mechanics}

It was alluded to earlier that Weinberg's formalism may be arrive at
by considering a particular type of generalization of quantum mechanics
in the geometric framework. Now that we have explored both
formalisms we seek to understand the connection between the two. This
connection
was lucidly explained in \cite{as}. We begin by
considering a generalized class of Hamiltonians $C_H$. This will
consist of densely defined functions on the projective Hilbert space
$\cal{P}$.
They must be functions which are smooth on their domain of definition
and whose associated Hamiltonian vector fields set up a flow on
$\cal{P}$.

Now recall our previous consideration of generalized dynamics. We lifted
the dynamical flows from $\cal{P}$ to the full Hilbert space $\cal{H}$ on
the
constraining surface $S$. Thus
we will also lift the class of Hamiltonians under consideration and denote
this
by $C_H^{'}$. As suggested by our previous analysis we may extend
the elements of
$C_H^{'}$
off of $S$ in whichever way we like, but we will choose the extension
suggested
in (\ref{ext}). This extension may be reformulated as
\begin{equation}
A_{ext}(\Psi)=\parallel\Psi\parallel^{2}A(\Psi/\parallel\Psi\parallel).
\label{exta}
\end{equation}
We may use this equation to extend all of the elements of $C_H^{'}$ to all
of the Hilbert space defined by $\cal{H}^{'}=\cal{H}-\{$0$\}$. Note that (\ref{exta})
implies the elements of $C_H$ and the smooth
gauge invariant functions
on $\cal{H}^{'}$ which are homogeneous of degree two are in a one-to-one correspondence.
If we view $\cal{H}$ as the vector space over complex numbers, this is
equivalent
to homogeneity of degree one in both $\psi$ and $\psi^*$. This is exactly the class
of Hamiltonians
which were considered in Weinberg's formalism. Note that the homogeneity
requirement
restricts the freedom to extend functions on $\cal{H}^{'}$. Also, as previously stated, because
$\psi$ is equivalent to $Z\psi$, Weinberg's space of physical states is $\CP{n}$.
This is the
quantum phase space $\cal{P}$ considered in the geometric quantum framework.
It is this essential fact that permits the connection between
the two formalisms.

\subsection{Appendix B: Foundations of Quantum Theory Redux}

Here we collect a streamlined version of the fundamental
structure of quantum theory from the geometric axiomatic point of view.
This material is already present in the main body of the review, but
here we assemble it all in one place, as a relatively simple way to rationalize
quantum theory from first principles, as well as offer a natural route to its
generalization, as discussed in this review.

We reason
as follows: Assume that individual observable events are
statistical and statistically distinguishable (Axiom I). (This of course is a huge
conceptual jump in comparison to classical physics, but it is absolutely
crucial for the structure of QM. This leap is centrally connected with the
seemingly counter-intuitive view on the concept of ``physical states'' or
``physical reality'' in canonical
quantum theory.)
On the space of probability distributions there is a
natural metric, called Fisher metric, which provides a geometric measure of statistical
distinguishability \cite{wootters}
\begin{equation}
ds^2=\sum_i \frac{dp_i^2}{p_i}, \quad \sum_i p_i =1, \quad p_i \geq 0.
\end{equation}
(This distance can be reasoned out as follows: to estimate probabilities
$p_i$ from frequencies $f_i$, given $N$ samples, when $N$ is large,
use the central limit theorem which says that the probability for the
frequencies is given by the
Gaussian distribution
\be
\exp(-\frac{N}{2}\frac{(p_i-f_i)^2}{p_i}).
\ee
Thus a probability
distribution $p^1_i$ can be distinguished from a given probability
distribution $p^2_i$ provided the
Gaussian $\exp(-\frac{N}{2}\frac{(p^{(1)}_i-p_i)^2}{p_i})$ is small. Hence it follows that
the quadratic form $\frac{(p^{(1)}_i-p_i)^2}{p_i}$, or its infinitesimal
form $\sum_i \frac{dp_i^2}{p_i}$, the Fisher distance, is a natural measure of
distinguishability.)
Now, change variables $p_i = x_i^2$, to make $p_i$ manifestly non-negative.
The Fisher distance is then
\be
ds^2 = \sum_i dx_i^2, \quad \sum_i x_i^2 =1
\ee
Therefore
the Fisher distance in the probability space is nothing but the shortest distance along this unit sphere \cite{wootters}
$
ds_{12} = \cos^{-1}(\sum_i \sqrt{p_{1i}} \sqrt{p_{2i}}).
$

Next, we demand that on this metrical space of probabilities one can define a canonical
Hamiltonian flow (Axiom II).
This is where the correspondence principle with the
canonical classical theory resides.
Note that the two postulates do seem to be incompatible and the
resolution of incompatibility is quantum theory.
There is here a similarity with the axiomatic approach to
Special Theory of Relativity, except that in the present case the
we have a radical departure with regards to what is meant by ``physical reality''.

To be able to implement Axiom II, the space of $x_i$ has to
be even dimensional (hence, $\sum x_i^2 =1$ defines an odd-dimensional sphere). Then the Hamiltonian flow is given (locally) as
$
\frac{df(x_i)}{dt} = \omega_{ij}\frac{\partial h(x_i)}{\partial x_i}\frac{\partial f(x_i)}{\partial x_j} \equiv \{h,f\},
$
where $\omega$ is a closed non-degenerate two-form.
The compatibility of the symplectic form $\omega$ and the metric $g$
allows for the introduction of an almost complex structure (in matrix notation)
\be
J \equiv \omega g^{-1}
\ee
Now, it follows that
\be
J^2 =-1
\ee
because the compatibility between the metric and symplectic structures demands
\be
\omega_{ij} g^{jk} \omega_{kl} = g_{il}.
\ee
Given this constant complex structure introduce complex coordinates on this even dimensional space $\psi_a$ (and their conjugates $\psi_a^{*}$), so that
\be
\sum_i x_i^2 \equiv \sum_a \psi_a^{*} \psi_a =1,
\ee
and thus $p_a = \psi_a^{*} \psi_a$.
This statistical distance is invariant under
\be
\psi \to e^{J \alpha} \psi,
\ee $J$ being the above {\it integrable} almost
complex structure.
Thus $\psi$ can be identified with $e^{J \alpha} \psi$.
It is a fact that an odd dimensional sphere can be viewed as a $U(1)$ fibration of a complex
projective space $\CP{n}$ which is a coset space
$\frac{U(n+1)}{U(n) \times U(1)}$. A $\CP{n}$ is a homogeneous, isotropic and
simply connected K\"ahler manifold with a constant holomorphic sectional
curvature.
The canonical metric on $\CP{n}$ is the Fubini--Study metric (which is nothing but the above
statistical Fisher metric up to a multiplicative constant,
the Planck constant $\hbar$), which reads, in the canonical Dirac notation
and the derived Born rule, $p_a = \psi_a^{*} \psi_a$):
$
ds_{12}^2=4(\cos^{-1}|\langle \psi_1|\psi_2\rangle|)^2 = 4(1 -
|\langle \psi_1|\psi_2 \rangle|^2)\equiv 4(\langle d\psi|d\psi\rangle
- \langle d\psi|\psi\rangle\langle \psi|d\psi\rangle).
$
Thus $\CP{n}$ is the underlying manifold of statistical events on which we
have a well defined Hamiltonian flow and as such provides a kinematical background on which a Hamiltonian
dynamics is defined.
The only Hamiltonian flow compatible with the isometries of $\CP{n}$
(which are the unitaries $U(n+1)$) is given by a quadratic function of
$x_i$ or, alternatively, a quadratic
form in the pair $q_a \equiv Re(\psi)$ and $p_a \equiv Im(\psi)$), $h = {1 \over 2} \sum_a [ (p^a)^2 + (q_a)^2 ] \omega_a$, or in the usual notation, ${h = \langle \hat{H}\rangle}$, $\omega_a$ being the eigenvalues
of $\hat{H}$.
The Hamiltonian equation for the $\psi$ and its conjugate becomes therefore
the linear evolution equation (Schrodinger equation),
$
{d p_a \over dt} = \{h, p_a \}, \quad {d q^a \over dt} = \{h, q^a\}
$, that is $J \frac{ d|\psi\rangle}{dt} = H |\psi\rangle$.
(In this formulation both the Heisenberg and the Schrodinger picture have a
canonical Hamiltonian form!)
Any observable, consistent with the isometries of the underlying space of statistical events, is given as a quadratic function in the $q_a, p_a$. These are just the
usual expectation values of linear operators.

Note that this compact approach is rather close to other axiomatic
systems invented to make quantum theory more palatable,
in particular, such as the one
advocated by Hardy \cite{hardy} and Aharonov \cite{aharonov}.
Also, the geometric formulation of quantum theory based on above axioms
naturally lends itself to the generalization discussed in the main body of the text,
which in turn sheds new light on the fundamental structure of canonical quantum theory,
in a way very much analogous to the relationship between the General and Special Theories of
Relativity.

\subsection{Appendix C: $\Lambda$ vs.\ V and UV vs.\ IR}

Here we make some comments about the canonical Wheeler--de-Witt (WdW) equation
(which should be a limit of some more general non-linear and non-local
Wheeler--de-Witt-like equation of the background independent Matrix theory)
with the cosmological constant, and the relation between
the cosmological constant and the spacetime volume. Furthermore, we
review the relation between the Wheeler--de-Witt equation and
the holographic renormalization group, as a hallmark of the
UV/IR correspondence.
\subsubsection{WdW and $\Lambda$ vs.\ V }
We start from the usual Wheeler--de-Witt equation
\be
H\Psi_\Lambda = 0
\ee
on a spacetime with cosmological constant $\Lambda$.
Let's explore what this means.

We can write the spacetime metric in a local neighborhood in the ADM form:
\be
ds^2 = g_{\mu\nu}\, dx^\mu dx^\nu
= -N^2 dt^2 + h_{ij} (dx^i - N^i dt) (dx^j - N^j dt).
\ee
We find that the extrinsic curvature $K_{ij}$ is
\be
K_{ij} = -\frac{1}{2N} \left( \pa_t h_{ij} + \nabla_i N_j + \nabla_j N_i \right),
\ee
which can obviously be rewritten as the evolution equation
\be
\pa_t h_{ij} = -2N K_{ij} - \nabla_i N_j - \nabla_j N_i.
\ee
We also have the Hamiltonian and momentum constraints
\bea
&& H = R^{(3)} + K^2 - K_{ij} K^{ij} - 2\Lambda = 0, \\
&& M_i = \nabla_j K^j_i - \nabla_i K = 0,
\eea
and a second evolution equation
\be
\pa_t K_{ij} = N R^{(3)}_{ij} + N K K_{ij} - 2 N K_{ik} K^k_j - \nabla_i \nabla_j N -
\nabla_i N^k K_{kj} - \nabla_j N^k K_{ki} - N^k \nabla_k K_{ij} - N \Lambda h_{ij},
\ee
where $K = h^{ij} K_{ij}$ and $R^{(3)}_{ij}$ and $R^{(3)}$ are the Ricci and scalar
curvatures of the spatial metric $h_{ij}$.
The Schr\"odinger equation then is
\be
\frac12 \left( R^{(3)} + K^2 - K_{ij} K^{ij} \right) \Psi_\Lambda = \Lambda \Psi_\Lambda.
\ee

It is convenient to rewrite this in a slightly different form.
Following Brown and York and Unruh \cite{by2}, define
\be
G_{ijkl} = \frac{1}{2\sqrt{h}} (h_{ik} h_{jl} + h_{il} h_{jk} - h_{ij} h_{kl}).
\ee
Define the conjugate momentum to the spatial metric $h_{ij}$ as
\be
\pi_h^{ij} := -i\hbar\frac{1}{\sqrt{h}} \frac{\delta}{\delta h_{ij}}.
\ee
We will restore powers of $\hbar$ and put $\kappa = 8\pi G_N = \hbar M_{Pl}^{-2}$.
Dimensional analysis tells us that $\pi_h^{ij}$ has units $M L^{-2}$.
The functional Schr\"odinger equation is the Wheeler--de-Witt equation
\be
\left( -2\kappa\hbar^2 \frac{1}{\sqrt{h}} G_{ijkl} \frac{\delta}{\delta h_{ij}} \frac{\delta}{\delta h_{kl}}
       -\frac{1}{2\kappa} R^{(3)} + \frac{1}{\kappa} \Lambda \right) \Psi_\Lambda[h] = 0.
\ee
Quantization maps
\be
\frac{\Lambda}{\kappa} \mapsto -i\hbar \frac{\delta}{\delta \tau} =: \pi_\tau.
\ee
(The dimensions of $\tau$ and $\pi_\tau$ are $L$ and $M L^{-3}$, respectively.)
In this sense $\tau$ and $\Lambda/\kappa = M_{Pl}^2 \Lambda/\hbar$ are conjugate variables.
The solution to the time-dependent Wheeler--de-Witt equation
\be
\left( 2\kappa \sqrt{h}\, G_{ijkl} \pi_h^{ij} \pi_h^{kl}
       -\frac{1}{2\kappa} R^{(3)} - i\hbar \frac{\delta}{\delta\tau} \right) \Psi_\Lambda[h, \tau] = 0
\ee
is a rotation of the $\tau$-independent solution:
\be
\Psi_\Lambda[h,\tau] = \exp\left\{ \frac{i}{\hbar} \int d^3x\, \tau\frac{\Lambda}{\kappa} \right\} \Psi_\Lambda[h],
\ee
and the wave function defines the probability measure for the spatial three-geometry defined by $h_{ij}$
to be found with spacetime volume $V$ in the region of superspace with volume element $d\mu[h]$:
\be
dP = |\Psi_\Lambda[h,\tau]|^2 d\mu[h].
\ee

The spacetime volume is determined by $\tau =: \tau^0$.
We have
\be
{\rm Vol} = \int d^4x\, \sqrt{-g} = \int d^4x\, \pa_\mu \tau^\mu.
\ee
Defining the spatial average
\be
V_\Sigma = \int_\Sigma d^3x\, \tau,
\ee
the spacetime volume is nothing but the difference of $V$ on the initial and final hypersurface:
\be
{\rm Vol} = \int d^4x\, \pa_\mu \tau^\mu = \int_{\Sigma_f} d^3x\, \tau - \int_{\Sigma_i} d^3x\, \tau = V_f - V_i.
\ee
Note that $\tau$ is a timelike direction.
We can only employ this prescription when we have a spatial slicing, \ie on a local neighborhood.

Now, let's explore the effect of fluctuations in $V$.
A small fluctuation can be defined by a shift in $\tau$ for fixed $h_{ij}$:
\be
\tau \mapsto \tau' = \tau + \epsilon.
\ee
This means that
\be
\pi_{\tau'} = -i\hbar \frac{\delta}{\delta(\tau + \epsilon)}
= -i\hbar \frac{1}{1+\frac{d\epsilon}{d\tau}} \frac{\delta}{\delta\tau}
= \left( 1 - \frac{d\epsilon}{d\tau} + \ldots \right) \pi_\tau.
\ee
We have chosen $\epsilon(\tau, x^i)$ such that
\be
\delta V = \left( \int_{\Sigma_f} d^3x\, \tau' - \int_{\Sigma_i} d^3x\, \tau' \right) -
           \left( \int_{\Sigma_f} d^3x\, \tau - \int_{\Sigma_i} d^3x\, \tau \right)
         = \left( \int_{\Sigma_f} d^3x\, \epsilon - \int_{\Sigma_i} d^3x\, \epsilon \right).
\ee
We do not know whether the choice of $\epsilon$ is unique (up to boundary terms), but some such $\epsilon$ must exist.
We will work to leading order in $\epsilon$.
We are interested in what the shift in $\tau$ does to $dP$.
If the configuration space (the space of quantum events) is space, then the fluctuations in the probability measure
$dP$ induced by the fluctuation $\delta V$ should also describe fluctuations in the space of quantum events.
The fluctuations in the space of quantum events are the fluctuations of the almost complex structure on the
infinite dimensional Grassmannian that are compatible with metric and symplectic structure.
The dynamics of these fluctuations must be described by the Einstein--Yang--Mills equation on the space of quantum events,
but we cannot be more precise without knowing what the metric and symplectic structures on this space are.

Let's do a quick calculation.
We have the analogue of the usual energy-time uncertainty relation for $\Lambda/\kappa$ and $\tau$:
\be
\left( \frac{\Delta\Lambda}{\kappa} \right) \left( \int d^3x\, \Delta\tau \right)
= \left( \frac{\Delta\Lambda}{\kappa} \right) \delta V \simeq \hbar.
\ee
This implies that
\be
\Lambda' = \Lambda + \frac{\hbar\kappa}{\delta V}.
\ee
Now,
\bea
\Psi_{\Lambda'}[h,\tau'] &=& \exp\left\{ \frac{i}{\hbar} \int d^3x\, \tau'\frac{\Lambda'}{\kappa} \right\} \Psi_\Lambda[h]
= \exp\left\{ \frac{i}{\hbar} \int d^3x\, \left[ (\tau + \epsilon) \left( \frac{\Lambda}{\kappa} + \frac{\hbar}{\delta V} \right) \right] \right\} 
\Psi_\Lambda[h] \nn \\
&=& \exp\left\{ \frac{i}{\hbar} \int d^3x\, \left[ \epsilon \frac{\Lambda}{\kappa} + \frac{\hbar(\tau + \epsilon)}{\delta V} \right] \right\} \Psi_\Lambda[h,\tau]
= \exp\left\{\frac{i}{\hbar} \frac{\Lambda}{\kappa} \delta V + i \frac{V'}{\delta V} \right\} \Psi_\Lambda[h,\tau]. \nn \\
\eea
We have rotated by a pure phase!
This implies that the measure
\be
dP' = |\Psi_{\Lambda'}[h,\tau']|^2 d\mu[h] = |\Psi_{\Lambda}[h,\tau]|^2 d\mu[h] = dP
\ee
is invariant.

\subsubsection{WdW and Holographic RG}

In this subsection we review the relationship between the holographic renormalization group
and the Wheeler--de-Witt equation. In fact, the holographic renormalization group is nothing but the Wheeler--de-Witt equation rewritten
for a particular slicing of spacetime. This works for general backgrounds including
cosmology. The formalism is crucially based on the existence of asymptotic spacetime data
and is in some sense a WKB-like version of some more general non-linear and
non-local formulation implied by the abstract structure of background independent Matrix theory.
Nevertheless, the following formulation summarizes some crucial points in the current thinking about
gauge theory/gravity duality and one of its trademarks: the UV/IR correspondence.

This formalism runs as follows~\cite{holobdbm,holorg,holorg1}. First we
fix the gauge so that the bulk metric can be written as
\begin{equation}
ds^2 = dr^2 + g_{ij}dx^i dx^j.
\end{equation}
This is just the ADM gauge discussed above:
the shift vector is set to zero and
the lapse to one.
Usually one envisions and holographic dual (of a non-gravitational nature,
as in AdS/CFT correspondence, {\em i.e.}, gauge theory/gravity duality) where the ultraviolet rescaling
in that dual corresponds to
the rescaling in the size of the extra dimension in the bulk spacetime, which in the chosen
gauge is nothing but the natural evolution parameter.
Given the fact that the bulk gravity theory is
reparameterization invariant, the local ultraviolet rescaling in the gauge theory
is encapsulated in the infrared by the four-dimensional Hamiltonian constraint
\begin{equation}
{\cal{H}} =0.
\end{equation}
More explicitly
\begin{equation}
{\cal{H}}= (\pi^{ij} \pi_{ij} - \frac{1}{2} \pi^{i}_{i} \pi^{j}_{j})
+\frac{1}{2} \pi_{I} G^{IJ}\pi_{J} + {\cal{L}}.
\end{equation}
Here $\pi_{ij}$ and $\pi_{I}$ are the canonical momenta conjugate to
$g^{ij}$ and $\phi^I$
\begin{equation}
\pi_{ij} = \frac{1}{\sqrt{-g}}\frac{\delta S}{\delta g^{ij}}, \quad
\pi_I = \frac{1}{\sqrt{-g}}\frac{\delta S}{\delta \phi^{I}}.
\end{equation}
Here $\phi^I$ denotes some background matter
fields coupled to $(3+1)$-dimensional gravity ---
for example, the Standard model fields;
${\cal{L}}$ is a local Lagrangian density, and
$G^{IJ}$ denotes the metric on the space of background matter fields.

As in the context of the AdS/CFT duality~\cite{holobdbm,holorg,holorg1}, the Hamiltonian
constraint can be formally rewritten as a renormalization group equation for
the dual renormalization group flow  \cite{holorg}.
In the Hamiltonian constraint
\begin{equation}
\frac{1}{\sqrt{-g}}\left( \frac{1}{2} \left(g^{ij}
{\frac{\delta S}{\delta g^{ij}}}\right)^2
-{\frac{\delta S}{\delta g^{ij}}}{\frac{\delta S}{\delta g_{ij}}}
-\frac{1}{2} G^{IJ}
\frac{\delta S}{\delta \phi^{I}} \frac{\delta S}{\delta \phi^{J}}\right)
= \sqrt{-g}\, {\cal{L}},
\end{equation}
assume that the local four-dimensional action $S$ can be separated into a
local and a non-local piece
\begin{equation}
S(g, \phi) = S_{loc}(g, \phi) + \Gamma (g, \phi).
\end{equation}
Given this rewriting of the four-dimensional action, the Hamiltonian constraint
can be formally rewritten as a Callan--Symanzik renormalization group
equation for the effective action~\cite{holorg} $\Gamma$ of the ultraviolet
theory at the scale $\Lambda$
\begin{equation}
\frac{1}{\sqrt{-g}}\left( g^{ij}
{\frac{\delta }{\delta g^{ij}}} - \beta^I \frac{\delta}{\delta
\phi^I}\right) \Gamma = HO,
\end{equation}
where $HO$ denotes higher derivative terms of the expression for the
four-dimensional conformal anomaly.
Here the ``$\beta$-function'' is defined (in analogy with the
AdS situation) to be $\beta^I = \partial_{\Lambda}
\phi^{I}$,
where $\Lambda$ denotes the cut-off of the defining ultraviolet theory.

In the context of the holographic renormalization group formalism developed in the
AdS/CFT correspondence, it is also possible to introduce a
holographic ``$c$-function'' which measures the number of accessible
degrees of freedom and which decreases during renormalization group flow. When the
spacetime is four-dimensional, one has \cite{holobdbm,holorg,holorg1}
\begin{equation}
c \sim \frac{1}{G \theta^2},
\end{equation}
where $\theta$ is the trace of the extrinsic curvature of the
boundary surface.
The trace of the quasi-local Brown--York
stress \cite{by1} tensor turns out to be
\be
\langle T^{i}_{i} \rangle \sim \theta
\ee
up to some terms constructed from local intrinsic curvature invariants
of the boundary.    Therefore the renormalization group equation of the defining ultraviolet theory is
given by
\be
\langle T^{i}_{i} \rangle = \beta^I \frac{\partial \Gamma}
{\partial \phi^I}.
\ee
Finally, in the context of the AdS/CFT correspondence the Raychauduri equation (discussed
in Sec. 3), that is,
gravitational focusing, implies monotonicity of the holographic ``$c$-function''
\begin{equation}
\frac{d \theta}{dt} \le 0,
\end{equation}
as long as a form of the weak positive energy condition is satisfied
by the background test matter fields.

\end{document}